\documentclass[11pt,dutch,british,openany,a4paper]{book}

\makeatletter
\renewenvironment{thebibliography}[1]
{\section*{\bibname}
	\@mkboth{\MakeUppercase\bibname}{\MakeUppercase\bibname}%
	\list{\@biblabel{\@arabic\c@enumiv}}%
	{\settowidth\labelwidth{\@biblabel{#1}}%
		\leftmargin\labelwidth
		\advance\leftmargin\labelsep
		\@openbib@code
		\usecounter{enumiv}%
		\let\p@enumiv\@empty
		\renewcommand\theenumiv{\@arabic\c@enumiv}}%
	\sloppy
	\clubpenalty4000
	\@clubpenalty \clubpenalty
	\widowpenalty4000%
	\sfcode`\.\@m}
{\def\@noitemerr
	{\@latex@warning{Empty `thebibliography' environment}}%
	\endlist}
\makeatother 

\usepackage{amsfonts,amsmath,amssymb}
\usepackage{comment} 
\excludecomment{comment}  

\usepackage{changepage}
\usepackage{mathtools} 
\usepackage{soul}
\usepackage{color} 
\usepackage{pslatex}
\usepackage[section]{placeins} 
\usepackage{float}
\usepackage{subfigure}
\usepackage{graphicx,wrapfig,hyperref}
\usepackage[small,bf]{caption}
\usepackage{slashed}
\usepackage{cite}
\usepackage{mciteplus}
\usepackage{mathrsfs}

\newcommand{\ali}[1]{\begin{align} #1 \end{align}}
\newcommand{\p}{\partial}
\newcommand{\ra}{\rightarrow}
\newcommand{\Ra}{\Rightarrow}
\newcommand{\vev}[1]{\langle #1 \rangle} 
\newcommand{\mn}{{\mu\nu}}
\newcommand{\ab}{{\alpha\beta}}
\newcommand{\al}{\alpha}

\DeclareMathOperator\tr{\text{tr}}
\definecolor{Darkgreen}{RGB}{0,100,0}
\definecolor{Forestgreen}{RGB}{34,139,34}
\definecolor{Mediumblue}{RGB}{0,0,205} 

\newdimen\mylength

\title{Introduction to holography}
\author{Nele Callebaut}
\date{}

\begin{document}

\maketitle

\tableofcontents

\chapter*{Introduction} 

These are course notes for the `Introduction to holography' Master level course at University of Cologne. 

The goal of the course is to give an introduction to holography. 
Holography is a popular approach to quantum gravity, in which a theory of gravity can be described by a lower-dimensional boundary theory that itself has no gravity. The most concrete known example of a holographic model is the AdS/CFT correspondence, where the gravitational theory has a negative cosmological constant (the universe is asymptotically Anti-de Sitter) and the boundary theory is a conformal field theory. Symmetry plays a very important role in this duality. We therefore start the course with a review of Poincar\'e symmetry in quantum field theory, before moving on in the second chapter to conformal symmetry in conformally invariant quantum field theories or CFT's. Then we move to the basics of AdS physics in chapters 3 and 4, which will already reveal hints to the existence of a duality with CFT. 
After gathering the basic ingredients (CFT and AdS), in the second half of the course we are ready to formulate the AdS/CFT correspondence (chapter 5), including finite temperature AdS/CFT (chapter 6), which involves black holes and their thermodynamics in the gravitational theory (chapter 7). We end the course with an introduction to entanglement in AdS/CFT and the origin of statements that `gravity emerges from entanglement' in holography.

\textit{Disclaimer:} This is a first version of the typed course, it will inevitably contain typos and mistakes. Comments and corrections are welcome at nele.callebaut@thp.uni-koeln.de. The most up-to-date version can be found at \url{https://www.thp.uni-koeln.de/callebaut/teaching.html}. 

\chapter{Poincar\'e invariance} 

In this chapter, we review Poincar\'e invariance in physics, in a notation that will be helpful for extending this to conformal invariance later.  

\section{Poincar\'e invariance in special relativity} \label{sect11}

In special relativity, we have a Minkowski metric 
\ali{
	\eta_\mn = \begin{pmatrix} -1 & & & \\ & 1 && \\ && 1 & \\ &&&1  
	\end{pmatrix}, \qquad ds^2 = \eta_\mn dx^\mu dx^\nu   \label{eta13}
}
with spacetime coordinates $x^\mu = (t,x^i)$, where we will assume 3 spacelike dimensions $i=1..3$ in this chapter. 
Under a coordinate transformation 
\ali{
	x^\mu \ra x'^\mu, 	
}
the metric transforms as a tensor 
\ali{
	\eta_\mn \ra \eta_\mn' = \eta_{\rho\sigma} \frac{\p x'^\rho}{\p x^\mu} \frac{\p x'^\sigma}{\p x^\nu}.  
}
Now consider specifically an infinitessimal coordinate transformation\footnote{The notation $\epsilon^\mu(x)$ is short for $\epsilon^\mu(x^\mu)$ - spacetime indices will often be left out to avoid heavy notation.}
\ali{
	x^\mu \ra x'^\mu = x^\mu + \epsilon^\mu(x)  \label{coordtransf}
}
where all terms in the function $\epsilon^\mu(x)$ scale with constants, let's already call them $\omega_a$, that are small $\omega_a << 1$, so that we are allowed to ignore $\mathcal O(\epsilon^2)$.  
We can ask, what are the infinitessimal coordinate transformations that leave the Minkowski metric invariant 
\ali{
	\eta'_\mn = \eta_\mn. \label{Poinccondition}
}
The answer will be the coordinate transformations \eqref{coordtransf} with $\epsilon^\mu$ satisfying the equation 
\ali{
	\p_\mu \epsilon_\nu + \p_\nu \epsilon_\mu = 0 \qquad \qquad  (\text{(flat) Killing equation}).  \label{flatKilling}
}
This is the \emph{Killing equation} for the Minkowski metric. If the coordinate transformation $\epsilon^\mu(x)$ satisfies this equation, it is a \emph{Killing vector} of the metric $\eta$, meaning the vector 
\ali{
	\epsilon \equiv \epsilon^\mu \p_\mu 
}
as a differential operator generates a symmetry of the geometry $\eta$, also called an \emph{isometry} of $\eta$. 
This is because we can write \eqref{coordtransf} as 
\ali{
	x'^\mu = x^\mu + \epsilon^\mu(x) = (1 + \epsilon) x^\mu = (1 + \omega_a t_a) x^\mu \label{inftransf}
}
with $\epsilon = \omega_a t_a$ expanded in sums\footnote{We are always using summation convention where double indices are summed over.} of small constants $\omega_a$ times generators $t_a$. The last expression in \eqref{inftransf} gives the standard definition of generators $t_a$, as giving rise to a transformation that is infinitessimally close to the identity operation. 

The solution to \eqref{flatKilling} is 
\ali{
	\epsilon_\mu = a_\mu + b_\mn x^\nu, \quad \text{with } b_\mn = -b_{\nu\mu}  \label{epsilonsol}
}
with $a$ and $b$ small constants, 
or 
\ali{
	\epsilon = a^\mu \p_\mu + \frac{1}{2} |b^\mn| (x_\nu \p_\mu - x_\mu \p_\nu).  
}
There are 10 constants in total that you can freely choose, 
corresponding to 10 independent coordinate transformations $\epsilon$: the $a^\mu$ correspond to 4 translations $x^\mu \ra x^\mu + a^\mu$ (in $t,x,y,z$ direction), the $b_\mn$ to 6 Lorentz transformations ($xy, yz, xz$ rotations and $tx,ty,tz$ boosts). The collection of those transformations are called \emph{Poincar\'e transformations}.  

We can read off $\omega_a$ and the Poincar\'e generators $t_a$ 
\ali{
	\omega_a &= a^\mu, \qquad t_a = \p_\mu \equiv p_\mu \qquad (``a" = \mu) \qquad \text{(translations)} \\ 
	\omega_a &= \frac{1}{2} |b^\mn|, \qquad t_a = x_\nu \p_\mu - x_\mu \p_\nu \equiv m_\mn \qquad (``a" = \mn) \qquad \text{(rotations and boosts)}. 
} 
These generators $t_a = (p_\mu , m_\mn)$ satisfy the Poincare algebra 
\ali{
	[m_\mn, p_\rho] &= -\eta_{\nu\rho} p_\mu + \eta_{\mu\rho} p_\nu \\
	[m_\mn,m_{\rho\sigma}] &= -\eta_{\nu\rho} m_{\mu\sigma} + \eta_{\mu\rho} m_{\nu\sigma}  - 
	\eta_{\nu\sigma} m_{\rho\mu} + \eta_{\mu\sigma} m_{\rho\nu}. 
}

The generators give you an infinitessimal transformation. To obtain an element of the Poincar\'e \emph{group}, we have to exponentiate the generators to a finite transformation. Following the notation of \cite{Weinberg}, we write the finite transformation as 
\ali{
	 x' &= T(\Lambda, a) x = \Lambda \, x + a, \label{finitetransfx}\\
	 \text{short for } \quad x'^\mu &= \Lambda^\mu_{\phantom{\mu}\nu} x^\nu + a^\mu. 
	}
Here $\Lambda$ (short for $\Lambda^\mu_{\phantom{\mu}\nu}$) are the Lorentz transformations, forming the Lorentz group, 
and $a$ the translations. 
Infinitely many 
infinitessimal transformations over $\omega_a = \frac{\alpha_a}{N}$ will give rise to a finite one over $\alpha_a$: 
\ali{
	e^{\alpha_a t_a} = \lim_{N \ra \infty} (1 + \frac{\alpha_a}{N} t_a)^N.  
}
Here, specifically, 
\ali{
	T(1,a) &= e^{a_\mu p^\mu} \\ 
	T(\Lambda,0) &= e^{\frac{1}{2} \theta_\mn m^{\mu\nu}} . 
}
The finite Poincar\'e transformations $T$ form a group, with composition rule 
\ali{
	T(\Lambda',a') T(\Lambda,a) = T(\Lambda' \Lambda, a' + \Lambda' a)  \label{compruleT}
}
and inverse 
\ali{
     T^{-1}(\Lambda,a) = T(\Lambda^{-1},-\Lambda^{-1} a). \label{inverseT}
}

\section{Poincar\'e invariance in quantum mechanics} \label{quantum}

How does Poincar\'e symmetry appear in a quantum setting? Before jumping into quantum field theory (QFT), let's consider quantum mechanics. 
(In this section we partly follow Chapter 2 of Weinberg's book \cite{Weinberg}, where he studies relativistic quantum mechanics as a first step in `deriving' QFT from reconciling quantum mechanics with special relativity.)  

\subsection{Conserved generators} 

Consider a quantum mechanical system with a Hilbert space $\mathcal H$ and its elements, wave-functions $\psi$. 
Under a Poincar\'e transformation $T$, like any symmetry transformation, the state-vector $\psi$ will transform with a unitary matrix $U$: 
\ali{
	\psi \ra U\, \psi.  
}
$\psi$ transforms in a different representation of the Poincar\'e group than the spacetime vector $x \ra T x$ in \eqref{finitetransfx}, but this representation will still satisfy the composition rule\footnote{This is true up to an important comment later on.} \eqref{compruleT}:  
\ali{
	U(\Lambda',a') U(\Lambda,a) = U(\Lambda' \Lambda, a' + \Lambda' a).   \label{compruleU}
}  
Like the generators $t_a$ are associated with infinitessimal $T(\omega_a)$, we can define generators $G_a$ associated with infinitessimal 
$U(\omega_a)$: 
\ali{
	U(\omega_a) = 1 - i \omega_a G_a.  \label{defGa}
}
From the unitarity of $U$, it follows immediately that the generators will be Hermitian, $G_a^+ = G_a$. This means they could be observables in the theory. In fact, they are even \emph{conserved observables}. This you can see as follows. $U$ will describe a symmetry of the quantum system if both $\psi$ and $U \psi$ are solutions of the Schr\"odinger equation $i \hbar \frac{d}{dt} \psi = H \psi$. This will be the case if $U H \psi = H U \psi$ or $[H,U]\psi = 0$, $\forall \psi$. We thus find that $U$ is a symmetry if 
\ali{
	[H,U] = 0.  
}
By \eqref{defGa}, this implies 
\ali{
	[H,G_a] = 0.  
}
In the Heisenberg picture, where operators are time-dependent with time evolution given by the Heisenberg equation $\frac{d \mathcal O}{dt} = \frac{i}{\hbar} [H,\mathcal O]$, 
the previous line expresses that the generator is a conserved quantity
\footnote{
	Including the possibility of explicit time dependence in $\mathcal O$, the Heisenberg equation is $\frac{d \mathcal O}{dt} = \frac{i}{\hbar} [H,\mathcal O] + \frac{\p \mathcal O}{\p t}$. As you have seen in Exercise 6, this is important to show that the boost generator is also conserved.    
}  
\ali{
	\frac{dG_a}{dt} = 0. 
	} 

Let us consider the translations $U(1,a) \equiv U(a)$ as an example. Writing $\psi'(x) = U(a) \psi(x)$ in bra ket notation, 
\ali{
	\langle x|\psi'\rangle = \langle x | U(a) | \psi \rangle,  
}
we can think of $U(a)$ working either actively to the right on $\psi$, or passively to the left on $x$. In the second viewpoint, using $U^{-1}(a) = U(-a)$, it follows that $\langle x|\psi'\rangle = \langle x-a | \psi \rangle$ or 
\ali{
	\psi'(x) = \psi(x-a).  
}
This we can Taylor expand around $x$ to find  
\ali{
	U(a) &= e^{- i a \cdot P} \approx 1 - i a \cdot P \\
	\text{with } P_\mu &= - i \p_\mu.  
} 
(with $a \cdot P$ notation for $a_\mu P^\mu$). 
We have thus identified the momentum operator $P_\mu$ as generator $G_{``a"=\mu}$ of translations from Taylor expanding $\psi'$. 
According to the general argument above, $P^\mu$ is the conserved operator associated with translation invariance. 


Assuming Poincar\'e symmetry of the system, with expectation values $\langle \psi | \mathcal O | \psi \rangle$ translation invariant, the action of $U$ on $\mathcal O$ should be 
\ali{
	\mathcal O \ra \mathcal O' = U(a) \mathcal O \, U^{-1}(a) 
} 
or infinitessimally, with notation $\delta \mathcal O \equiv \mathcal O' - \mathcal O$, 
\ali{
	\delta \mathcal O = - i a_\mu [P^\mu, \mathcal O] = -a_\mu \p^\mu \mathcal O. 
}

In more general notation, the finite and infinitessimal Poincar\'e transformations of $\mathcal O$ are respectively given by:  
\ali{
	\mathcal O' &= U \mathcal O \, U^{-1} \\
	\delta O &= - i \omega_a [G_a, \mathcal O].  
}

Similarly to the translations above, one can identify the generator and conserved operator for rotations and boosts as the angular momentum operator $M_\mn$, which appears in the expansion of an infinitessimal $U$ as 
\ali{
	U(\omega_a) \equiv U(1 + \omega, \epsilon) = 1 - i  \epsilon_\rho P^\rho + \frac{1}{2} i \omega_{\rho\sigma} M^{\rho\sigma}.  
}
The Poincar\'e generators $G_a = (P_\mu, M_\mn)$ (in the representation in which $|\psi\rangle$ transforms under the group) should satisfy the same Poincar\'e algebra as we found in section \ref{sect11}. Now this implies that the \emph{conserved observables} satisfy the Poincar\'e algebra. The Poincar\'e algebra for $P_\mu$ and $M_\mn$ is derived in \cite{Weinberg}. Here we only sketch the strategy. What you have to do is to look at the transformation properties of the generators themselves. One can derive what $U M U^{-1}$ and $U P U^{-1}$ have to be from the composition rule \eqref{compruleU} applied to $U(\Lambda, a) U(1 + \omega, \epsilon) U^{-1}(\Lambda,a)$. Next, taking $U$ infinitessimal and equating coefficients of $\omega$ and $\epsilon$ on each side, the Poincar\'e algebra is obtained (with a factor of $i$ due to the definition of $G_a$ compared to $t_a$): 
\ali{
	[P^\mu, P^\nu]&=0 \\ 
	i [P^\mu, M^{\rho\sigma}] &= \eta^{\mu\rho} P^\sigma - \eta^{\mu\sigma} P^\rho \\
	i [M^\mn,M^{\rho\sigma}] &= \eta^{\nu\rho} M^{\mu\sigma} - \eta^{\mu\rho} M^{\nu\sigma} - \eta^{\sigma\mu} M^{\rho\nu} + \eta^{\sigma\nu} M^{\rho\mu}. 
}

Here, a comment should be made about the possibility of \emph{central charges} in the algebra. Because the physical state is insensitive to an overall phase, there can be an extra phase factor in the composition rule \eqref{compruleU} for $U$: 
\ali{
	U(T') U(T) = e^{i \phi(T',T)} U(T'T). 
}
This corresponds to additional central charge terms in the algebra, i.e.~terms proportional to the identity operator. Schematically, $[G_b,G_c] = i c^a_{\phantom{a}bc} G_a + i c_{bc} 1$.  
It is shown in section 2.7 of \cite{Weinberg} that central charge terms in the $(P_\mu, M_\mn)$ algebra can always be eliminated by a redefinition of $P$ and $M$. 
We will encounter an important instance later where a central charge term will have to be taken into account. 
 
The Poincar\'e algebra for $(P_\mu, M_\mn)$ can be rewritten in a less compact but more familiar way by recombining the generators into $(H, \vec P, \vec J, \vec K)$ defined as 
\ali{
	H = P^0, \quad P^i, \quad  J_i = \frac{1}{2} \epsilon_{ijk} M^{jk}, \quad K^i = M^{0i} \qquad (i=1,2,3)
} 
respectively the Hamiltonian, momentum, angular momentum and boost operators. 
The angular momentum operator $\vec J = \vec x \times \vec P$ indeed satisfies the standard commutator 
\ali{
	[J_i, J_j] = i \, \epsilon_{ijk} J_k.  	\label{SO3alg}
}
A finite rotation $R_\theta$ by an angle $\theta$ over the direction of $\vec \theta$ can then be written as $U(R_\theta,0) = e^{i \vec \theta \cdot \vec J}$. 
Further we have 
\ali{
	[H,H] = [P_i,H]=[J_i,H]=0
}
but 
\ali{
	[H,K_i] = i \,P_i,  
}
and 
\ali{
	[J_i,K_j]=i \, \epsilon_{ijk}K_k, \quad [K_i,K_j]=-i \, \epsilon_{ijk}J_k, \quad [J_i,P_j]=i \, \epsilon_{ijk} P_k, \quad [K_i,P_j]=-iH \delta_{ij}, \quad [P_i,P_j]=0.  
}
The vanishing commutators indicate which operator eigenvalues can be used simultaneously to label a physical state (energy and momentum eigenvalues, or energy and spin eigenvalues). The fact that $\vec K$ does not commute with the Hamiltonian explains why we don't use the eigenvalue of the boost operator to label physical states. 


\subsection{Casimir and irreducible representations} \label{sectionCasimir}

Particularly useful operators to label and organize states in a QM system with symmetry $G$ are the \emph{casimir operators}. They have the properties that they 1) commute with all generators of the symmetry group $G$, and 2) give the same eigenvalue for all states in an irreducible representation of the group. This is best illustrated in the example of the SO(3) rotation group with Lie algebra \eqref{SO3alg} for the generators $J_i = (\vec x \times \vec P)_i$. The operator $J^2 = J_i J_i = J_1^2 + J_2^2 + J_3^2$ is a casimir operator, typically denoted $c_2$ (for `quadratic' casimir, quadratic in the generators)  
\ali{
	c_2^{rot} = J^2 
}
because $[J^2, J_i] = 0$.   
Since $J^2$ and $J_3$ commute, they can be diagonalized simultaneously with corresponding eigenvalues $\lambda$ and $m$ 
\ali{
	J^2 |\lambda,m \rangle &= \lambda |\lambda,m \rangle \\
	J_3 |\lambda,m \rangle &= m |\lambda,m \rangle. 
} 
We can regroup $J_1$ and $J_2$ into the operators 
\ali{
	J_\pm = J_1 \pm i J_2.  
} 
The states obtained by acting with $J_\pm$ on $|\lambda,m \rangle$ are again eigenstates of $J^2$ (with eigenvalue $\lambda$) and $J_3$ (with eigenvalue $m\pm 1$). In particular, $J_\pm$ act as ladder operators, raising or lowering the value of $m$: 
\ali{
	J_\pm |\lambda,m \rangle &= c_\pm(\lambda,m) |\lambda,m \pm 1 \rangle
}
where $c_\pm(\lambda,m)$ is a normalization constant that can be determined. 
From $J^2 - J_3^2 = J_1^2 + J_2^2 \geq 0$, we know the value of $m^2$ is bounded from above by $m^2 \leq \lambda$. Using the notation $j$ for the maximal value of $m$, you can show (Exercise) that $\lambda = j(j+1)$ and $m$ runs from $-j, -j+1, \cdots, j-1, j$. 
It is then natural to use $j$, which is called \emph{spin}, instead of $\lambda$ to label the eigenstates. 

The states 
\ali{
	|j,m\rangle, \qquad m=-j, -j+1, \cdots, j-1, j  \label{jmstates}
}
form a subspace in the Hilbert space that is left invariant by the action of the group $G=SO(3)$, in the sense that acting on such a state by a unitary group operator $U(g), g \in G$, will give back a linear combination of states in that subspace. The group does not take you outside of the subspace. This is true because the action of the group generators is 
\ali{
	J_3 |j,m \rangle &= m |j,m \rangle, \qquad J_\pm |j,m \rangle \sim |j,m \pm 1 \rangle .
}    
The states \eqref{jmstates} form an \emph{irreducible representation} of the group: symmetry transformations transform elements into other states in the irreducible representation (`irreducible' because it is a minimal such set of states, that cannot be reduced further). The eigenvalue of the casimir $j(j+1)$ is insensitive to those symmetry transformations. The casimir $c_2^{rot} = J^2$ therefore identifies the label $j$ for the irreducible representation as a physical organization label for the states, that does not distinguish between states that by symmetry are equivalent. 

Returning to the full Poincar\'e group, the quadratic casimirs are given by 
\ali{
	c_2^{\text{Poincar\'e}} = P^2 = P_\mu P^\mu, \qquad \tilde c_4^{\text{Poincar\'e}} = W^2 = W_\mu W^\mu  
}
with $W_\mu = -\frac{1}{2} \epsilon_{\mu\rho \sigma\lambda} M^{\rho\sigma} P^\lambda$. 
The first one measures the mass $P^2 = m^2$. For $P^2 = m^2 >0$, one can define a rest frame where $\vec P = 0$ and $W^2 = -m^2 \vec J^2$ with eigenvalue $-m^2 j(j+1)$ (the details are not important here). The casimirs of the Poincar\'e group thus suggest to use the mass $m$ and spin $j$ as Poincar\'e invariant 
labels that can be assigned to a physical state. In constructing a Poincar\'e invariant QM model for say electrons, which remain electrons as we translate or rotate around them, the \emph{particles} should be defined as \emph{irreducible representations} of the Poincar\'e group. 

\section{Poincar\'e invariance in classical field theory} 

Reference: Di Francesco ``Conformal field theory" Chapter 2 \cite{difran}, \cite{TongCFT}. 

Consider a classical field theory, with ingredients: a set of fields $\phi(x^\mu)$ (not necessarily only scalar fields) defined on a background Minkowski metric $\eta_\mn$. The field theory is described by an \emph{action} 
\ali{
	S[\phi, \p_\mu \phi] = \int d^d x \, \mathcal L[\phi, \p_\mu \phi] . 
}
Under a Poincar\'e transformation of the coordinates $x \ra x' = T(\Lambda,a) x$, the fields will transform 
\ali{
	\phi(x) \ra \phi'(x') = \mathcal F(\phi(x)) 
}
with two contributions (reflected in the notation $\phi'(x')$ with $\phi'$ and $x'$): a contribution from the induced transformation via the dependence on $x$, but also a contribution from the active transformation of the field $\phi$ itself, represented by the function $\mathcal F(\phi)$. 
The field theory is Poincar\'e invariant when the action is. 

Under a translation, $x' = x+a$, the field transforms as $\phi'(x+a) = \phi(x)$ (active point of view) or $\phi'(x) = \phi(x-a)$ (equivalent, passive point of view). The action is invariant under translations unless it explicitly depends on the position $x$. 

Under a Lorentz transformation, $x'^\mu = \Lambda^\mu_{\phantom{\mu}\nu} x^\nu$, the field transforms as 
\ali{
	\phi'(\Lambda x) = L_\Lambda \phi(x) 
}
where $L_\Lambda$ have to be matrices that form a representation of the Lorentz group. The action will be Lorentz invariant if all Lorentz indices are contracted. For example, for a scalar field $\varphi(x)$, a Lorentz invariant Lagrangian should be of the form $\mathcal L = f(\varphi) + g(\varphi) \p_\mu \varphi \p^\mu \varphi \, + $ higher order derivatives. 

We can read off the Poincar\'e generators $G_a$, defined as  
\ali{
	\phi'(x) = ( 1 - i \omega_a G_a ) \phi(x),  
}
from writing the infinitessimal Poincar\'e transformation 
\ali{
	x'^\mu &= x^\mu + \omega_a \, t_a \, x^\mu \\
	\phi'(x') &= \phi(x) + \omega_a \mathcal F_a \phi(x)  
}
for matrices $\mathcal F_a$, and Taylor expanding $\phi'(x') = \phi(x) + \omega_a \mathcal F_a \phi(x) = \phi(x') - \omega_a (t_a x^\mu) \p_\mu \phi(x') + \omega_a \mathcal F_a \phi(x')$. It follows that the generators are given by 
\ali{
	i G_a \phi = (t_a x^\mu) \p_\mu \phi - \mathcal F_a \phi. 
}   
Applying this to translations, with $t_a = \p_\nu$ and $\mathcal F_a = 0$, we find 
\ali{
	G_a^{transl} = \mathcal P_\nu = - i \p_\nu = - i p_\nu . \label{Gatransl}
}
For Lorentz transformations, with $t_a = x_\nu \p_\mu - x_\mu \p_\nu$ and $L_\Lambda \phi = (1 - i \omega_\mn S^\mn) \phi$, with $S^\mn$ a Hermitian matrix obeying the Lorentz algebra\footnote{
	The $S_\mn$ matrices act on the indices of the field $(S_\mn \phi)_\alpha = (S_\mn)_\alpha^{\phantom{\alpha}\beta} \phi_\beta$. They are called spin matrices and are determined by the spin-tensor structure of the field $\phi$. In particular \cite{Fradkin}, 
	\ali{
	S_\mn &= 0 \qquad \text{for the scalar field} \\
	(S_\mn)_\tau^{\phantom{\tau}\rho} &= i (\eta_{\mu\tau} \delta_\nu^\rho - \eta_{\nu\tau}\delta_\mu^\rho) 	\qquad \text{for the vector field} \\
	S_\mn &= \frac{i}{4}[\gamma_\mu,\gamma_\nu] \qquad \text{for the spinor field}  
	}
	with $\gamma_\mu$ the Dirac matrices $\{\gamma_\mu, \gamma_\nu \} = - 2 \eta_\mn$.  
}, 
\ali{
	G_a^{Lor} = L_{\mn} = i (x_\mu \p_\nu - x_\nu \p_\mu) + S_\mn = - i \, m_\mn + S_\mn .  \label{GaLor}   
}   
We see there is a new contribution compared to the $(p_\mu, m_\mn)$ generators, coming from the explicit field transformation $\mathcal F$. 

There is again a link to conserved quantities, which we will see at the quantum level. We can associate conserved charges in the classical theory thanks to Noether's theorem.  

\subsection{Noether theorem}  \label{sectionNoether}

Reference: Green Schwarz Witten (2.1.61) \cite{GSW}. 

The Noether theorem states that a global continuous symmetry of the action implies the existence of a conserved current, and gives you a procedure to explicitly construct the conserved current. Below we review a short derivation of the theorem, following e.g.~\cite{GSW}. 

Under the Poincar\'e symmetry, 
\ali{
	\phi(x) \ra 
	\phi(x) + \omega_a \mathcal F_a \phi(x), \qquad \text{($\omega_a$ constant)} 
}
with $\omega_a$ an infinitessimal constant. 
The action is invariant under these transformations
\ali{
	\delta_{global} S[\phi, \p \phi] = 0 
}
where the subscript on the variation is added to point out explicitly that the transformation is global (i.e.~$\omega_a$ are constant). 

Now consider the same transformation but with $x$-dependent (still infinitessimal) $\omega_a$
\ali{
	\phi(x) \ra \phi(x) + \omega_a(x) \mathcal F_a \phi(x), \qquad \text{(all $\omega_a(x)$)}.  
}
This is not a symmetry of the system, hence the corresponding variation of the action does not vanish $\delta_{all} S[\phi, \p \phi] \neq 0$. However, since we know it has to in the limit that $\omega_a$ becomes $x$-independent, we know $\delta_{all} S$ should be of the form 
\ali{
	\delta_{all} S[\phi, \p \phi] = \int d^d x \, j^\mu_a [\phi, \p \phi] \, \p_\mu \omega_a \,  . 
}
On a solution $\phi^*$ of the EOM, we do have that the variation of the action must vanish for all $\omega_a(x)$, 
\ali{
	\delta_{all}S[\phi^*, \p \phi^*] = 0 , \quad \forall \omega_a.  
}
This implies, making use of partial integration, that 
\ali{
	-\int d^d x \, \p_\mu j^\mu_a [\phi^*, \p \phi^*] \, \omega_a  = 0 , \quad \forall \omega_a 
}
and thus 
\ali{
	\p_\mu j^\mu_a [\phi^*, \p \phi^*] = 0.   
}
This expresses that $j^\mu_a$ is a conserved current 
and it is called the \emph{Noether current}. 
Di Francesco in \cite{difran} (2.141) derives the general form (for an action $S[\phi, \p \phi]$) to be 
\ali{
	j^\mu_a = \left( \frac{\p \mathcal L}{\p \p_\mu \phi} \p_\nu \phi - \delta_\mu^\nu \mathcal L \right) (t_a x^\nu) - \frac{\p \mathcal L}{\p \p_\mu \phi} \mathcal F_a \phi(x).  \label{Noethergen}
}

The corresponding conserved charge is given by integrating the time component of the Noether current over all space-like directions at a given instance of time 
\ali{
	Q_a = \int d^{d-1} x j_a^0 . 
}
The expected remaining time dependence $Q_a(t)$ has been left out of the notation because the charge is time-independent, i.e.~conserved: 
\ali{
	\frac{dQ_a}{dt} = \int d^{d-1} x \, \p_0 j_a^0 = -\int d^{d-1} x \, \p_i j_a^i = -\int_\infty j_a^i d\sigma^i = 0   \label{conservedQ}
}
by assuming the currents vanish at infinity. 

Let us identify the Noether current of translations. For translations, $\omega_a = a_\nu$ and the above argument defines the Noether current $T_\mn$, satisfying  
\ali{
	\p_\mu T^\mn = 0, \label{Econservation}
} 
via the variation of the action 
\ali{
	\delta S = \int d^d x \, T^\mn\,  \p_\mu a_\nu \, .  \label{this}
}
Applying \eqref{Noethergen} gives 
\ali{
	T^\mn = \frac{\p \mathcal L}{\p \p_\mu \phi} \p^\nu \phi - \eta^\mn \mathcal L.   
}
The conserved charge 
\ali{
	P^\nu = \int d^{d-1}x \, T^{0\nu}.  
}
In particular, $P^0 =\int d^{d-1}x \, T^{00} =  \int d^{d-1}x \, \left( \frac{\p \mathcal L}{\p \dot \phi} \dot \phi -\mathcal L \right) =  E$, allowing us to identify the Noether current for translations with the 
energy momentum tensor $T_\mn$. 
 
\eqref{this} counts as a \emph{definition} of the energy momentum tensor of the field theory. In a Lorentz invariant theory, a symmetric stress tensor can always be defined \cite{difran}. Then the expression \eqref{this} can be symmetrized, and written as   
\ali{
	\delta S = \int d^d x \, T^\mn\, \delta \eta_\mn \, . 
}
Allowing for more general background metrics, 
\ali{
	\delta S = \int d^d x \, \sqrt{|g|} \, T^\mn\, \delta g_\mn \, ,  \label{stresstensor}
}
with $g$ the determinant of the background metric $g_\mn$, leads to the \emph{stress tensor definition}  
\ali{
	T_\mn = \frac{2}{\sqrt{|g|}} \frac{\delta S}{\delta g^\mn}   \label{TmnQFTdef}  
 }
that we will use later in the course. 

To write the conserved current for the Lorentz transformations, instead of using the general $j^\mu_a(\mathcal F)$ formula, it is easy to see that the full conserved current has to take the form 
\ali{
	J^\mu = \epsilon_\nu T^\mn \label{Poinccurrent}
}
for the Poincar\'e transformations $\epsilon_\nu$ given in \eqref{epsilonsol}. Instead of showing that $J^\mu$ equals $\omega_a j^\mu_a$, we can just show that $J^\mu$ is the conserved current, by using conservation of energy \eqref{Econservation}, symmetry of $T^\mn$ in a Lorentz invariant theory, and the Killing equation \eqref{flatKilling}, 
\ali{
	\p_\mu J^\mu = 0. 
} 
The Poincar\'e conserved charges are then given in terms of the field theory stress tensor as 
\ali{
	Q_\epsilon = \int d^{d-1}x \,\,  \epsilon_\nu T^{0\nu} .  \label{Noetherdef}
}

Exercise: Write out the conserved current for Lorentz transformations, and the conserved charge.

\section{Poincar\'e invariance in quantum field theory}  \label{PoincQFT}

References: Di Francesco 2.4.3 \cite{difran}, Tong CFT lectures \cite{Tong}. 

The classical field theory of the previous section can be quantized to a quantum field theory. One can employ either the operator formalism for this,  leading to field operators $\hat \phi(x^\mu)$, or the path integral formalism.
An insertion $\phi(x^\mu)$ in the functional integral translates into an operator $\hat \phi(x^\mu)$ in the matrix element in the Hilbert space formalism [see (A.1.16) of Polchinski's chapter on path integrals \cite{Polchinski}]. The field is only strictly an operator in the operator formalism, but is still commonly referred to as `operator' also in the path integral formalism. We will leave out the hats indicating the operator nature in the operator formalism. 


In the \emph{quantum} field theory, the physics is described by the \emph{path integral}\footnote{
	The minus sign in the exponent signals that we are working in Euclidean signature in this section, following \cite{difran,Tong}. 
} 
\ali{
	Z = \int \mathcal D \phi \, e^{-S[\phi]}
	}
and \emph{correlation functions}  
\ali{
	\langle \phi(x_1) \cdots \phi(x_n) \rangle = \frac{1}{Z} \int \mathcal D \phi e^{-S[\phi]} \phi(x_1) \cdots \phi(x_n). 
	}


The quantum field theory will be invariant under a Poincar\'e transformation of the fields 
\ali{
	\phi(x) \ra \phi'(x) = \phi(x) - i \omega_a G_a \phi(x)  \label{symmtransf}
	}
if the path integral is. 
Both the action and the path integral measure will transform 
\ali{
	S[\phi] &\ra S[\phi'] \qquad \text{or} \quad S \ra  S' \\
	\mathcal D \phi &\ra \mathcal D \phi' . 
	} 
For a classical field theory with Poincar\'e invariance, $S' = S$ or $\delta_{global} S = 0$, it was shown in the previous section that there is a conserved Noether current $j^\mu_a$. 
For the corresponding quantum field theory to be Poincar\'e invariant, we assume additionally that $\mathcal D \phi' = \mathcal D \phi$ or $\delta_{global} (\mathcal D \phi) = 0$. Is there 
a corresponding quantum Noether theorem? 

To find the answer, we follow an analogue argument to the one in section \ref{sectionNoether}. That is, we start by considering the transformation with $x$-dependent (and still infinitessimal) $\omega_a$:  
\ali{
	\phi(x) \ra \phi'(x) = \phi(x) - i \omega_a(x) G_a \phi(x),    
}
under which the action and path integral measure will \emph{not} be invariant unless $\omega_a(x) \ra \omega_a$. The transformed path integral $Z'$ will take the form 
\ali{
	Z' &= \int \mathcal D \phi' e^{-S[\phi']} = \int \mathcal D \phi e^{-S[\phi] - \int d^d x \, j^\mu_a \p_\mu \omega_a} \approx \int \mathcal D \phi e^{-S[\phi]} \left( 1 + \int d^d x \, (\p_\mu j^\mu_a) \omega_a(x) \right),   
}
with the current $j^\mu_a$ not necessarily equal to its classical counterpart, because it can now contain contributions from the variation of the path integral measure. In the last line we expanded to first order in the infinitessimal $\omega_a(x)$. Now we look back at the expression $Z' = \int \mathcal D \phi' e^{-S[\phi']}$ and realize that the fields $\phi'$ are just dummy variables that are integrated over, and can be replaced by another integration dummy $\phi$. That means $Z'$ should be equal to $Z = \int \mathcal D \phi e^{-S[\phi]}$ (for all $\omega_a(x)$), which leads us to conclude that 
\ali{
	\vev{\p_\mu j^\mu_a} = 0  \qquad \text{(quantum Noether theorem)}.  
}

Similarly, we can use the same dummy variable argument to demand that the correlator 
\ali{
	\langle \phi(x_1) \cdots \phi(x_n) \rangle = \frac{1}{Z} \int \mathcal D \phi e^{-S[\phi]} \phi(x_1) \cdots \phi(x_n)
}  
equals 
\ali{
	\frac{1}{Z} \int \mathcal D \phi' e^{-S[\phi']} \phi'(x_1) \cdots \phi'(x_n),  
}
which can be further written out to first order in $\omega_a(x)$ as 
\ali{
\frac{1}{Z} \int \mathcal D \phi e^{-S[\phi]}\left( 1 + \int d^d x \, (\p_\mu j^\mu_a) \omega_a(x) \right) \left( \phi(x_1) \cdots \phi(x_n) - i \sum_{i=1}^n \int d^d x \, \omega_a(x) \delta(x-x_i) \phi(x_1) \cdots G_a \phi(x_i) \cdots \phi(x_n) \right). \nonumber 
	}
It follows that 
\ali{
	\p_\mu \langle j^\mu_a(x) \phi(x_1) \cdots \phi(x_n) \rangle = i \sum_{i=1}^n \delta(x-x_i) \langle \phi(x_1) \cdots G_a \phi(x_i) \cdots \phi(x_n) \rangle  \label{Wardid}
}
known as the \emph{Ward identity}. 

When $x \neq x_i$ in the Ward identity, it expresses that the current conservation is valid ``as an operator equation"\footnote{For a general explanation of this expression in the path integral formalism, see Polchinski's chapter on path integrals \cite{Polchinski} (A.1.19).} 
\ali{
	\p_\mu j^\mu_a = 0,  
}
meaning it holds (implicitly) within an expectation value $\vev{\p_\mu j^\mu_a \cdots} = 0$ with other possible operator insertions $\cdots$ present (at different locations).  

We will now use the Ward identity to study the action of the conserved Noether charge operator $Q_a$ on the field operator $\phi$. 
For this, we assume $x_1^0$ is different and larger than all the other time components $x_2^0 > \cdots > x_n^0$, and we integrate the Ward identity \eqref{Wardid} over a specific volume. 
The volume consists of time ranging from a value $t_-$ close to but lower than $x_1^0$, to a value $t_+$ close to but higher than $x_1^0$, over the whole spacelike range. Making use of the divergence theorem, the integrated Ward identity becomes 
\ali{
	\langle Q_a(t_+) \phi(x_1) Y \rangle - \langle Q_a(t_-) \phi(x_1) Y \rangle = i \langle G_a \phi(x_1) Y \rangle
} 
with the notation $Y = \phi(x_2) \cdots \phi(x_n)$. The expectation value is the vacuum expectation value of the time ordered product $\langle \phi(x_1) \cdots \phi(x_n) \rangle = \langle 0| \mathcal T \phi(x_1) \cdots \phi(x_n) \rangle |0 \rangle$, so that in the limit $t_- \ra t_+$:  
\ali{
	\langle 0| \, [Q_a,\phi(x_1)]\, Y |0\rangle = i \langle 0| G_a \phi(x_1) Y |0\rangle. 
}  
We find that from the Ward identity, we were able to identify the Noether conserved charge as the generator of the symmetry transformation of field operators 
\ali{
	\Aboxed{[Q_a,\phi] = i G_a \phi}\,. \label{Qagenerator}
}

This result confirms the intuition from the discussion in the QM system that the \emph{conserved charge operator becomes a generator}. 
For example, $[P_\mu,\phi] = - i \p_\mu \phi$ with $P^\mu = \int T^{0\mu} d^{d-1}x$. 

By showing that the $Q_a$ action generates the symmetry transformation, we have in a way implicitly showed that the 
conserved charges satisfy the Poincar\'e algebra 
\ali{
	[Q_{\epsilon_1},Q_{\epsilon_2}] = Q_{-[\epsilon_1,\epsilon_2]}.  \label{chargealgebra}
} 
To actually show this is non-trivial, and I refer to \cite{Boulware,SD} for it, where in particular it is shown that no central charge contributions appear in \eqref{chargealgebra}. (Remember this is a possibility in the quantum system, as discussed in section \ref{quantum}.)

\begin{figure}
	\centering \includegraphics[width=12cm]{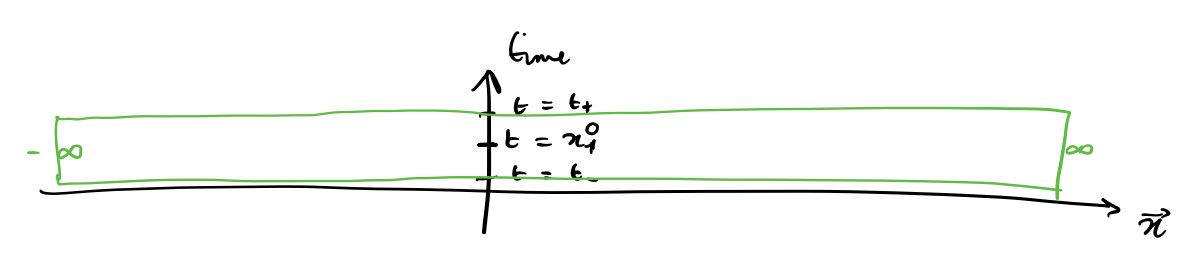}
	\caption{Integration volume.} 
\end{figure}

\newpage
\section*{Exercises}

\textbf{\underline{Exercise 1. Poincar\'e al\smash{g}ebra}} 

Check that $p_\mu$ and $m_\mn$, defined as 
\ali{
	p_\mu = \p_\mu, \quad m_\mn = x_\nu \p_\mu - x_\mu \p_\nu 
}	
satisfy the Poincar\'e algebra. \\

\textbf{\underline{Exercise 2. Finite Poincar\'e transformations}} 


In Weinberg's notation, we write the finite Poincar\'e transformation of $x$ as 
\ali{
	x' &= T(\Lambda, a) x = \Lambda \, x + a, \\
	\text{short for } \quad x'^\mu &= \Lambda^\mu_{\phantom{\mu}\nu} x^\nu + a^\mu   
}
with $\Lambda$ a Lorentz transformation and $a$ a translation. 

The finite transformations are defined as the exponential of the generator  
\ali{
	T(1,a) &= e^{a_\mu p^\mu}  \\
	T(\Lambda,0) &= e^{\frac{1}{2}\theta_\mn m^{\mn}} . 
}
Starting from the Poincar\'e generators $t_a = (p_\mu , m_\mn)$ given in Exercise 1, show that indeed 
\ali{
	x' &= T(1,a) x = x + a \\
	x' &= \Lambda \, x  \quad (\text{determine $\Lambda$ for one rotation or boost}) .  
}

\textbf{\underline{Exercise 3. Irreducible re\smash{p}resentation of rotation \smash{g}rou\smash{p}}} 

Consider the rotation group SO(3) with generators $J_i = (\vec x \times \vec P)_i$. In the notes on the course, the casimir of the rotation group SO(3) is introduced on p.7 (read that section for this exercise) as 
\ali{
	c_2^{rot} = J^2 
} 
with $J^2 = J_i J_i = J_1^2 + J_2^2 + J_3^2$. You can check that $[J^2, J_i] = 0$. 
The states $|\lambda,m \rangle$ satisfy $J^2 |\lambda,m \rangle = \lambda |\lambda,m \rangle$ and $J_3 |\lambda,m \rangle = m |\lambda,m \rangle$. 

a) Show that $J_\pm = J_1 \pm i J_2$ are ladder operators. For this, show that $J_\pm |\lambda,m \rangle$ are again eigenstates of $J^2$ and $J_3$, and determine their eigenvalues. 

b) Using the notation $j$ for the maximal value of $m$, show that $\lambda = j(j+1)$. Using the notation $j'$ for the minimal value of $m$, show that $\lambda = j'(j'-1)$. This has as a solution $j'=-j$ (why is $j'=j+1$ excluded?). You have shown that $m$ runs from $-j, -j+1, \cdots, j-1, j$.  \\

\textbf{\underline{Exercise 4. Poincar\'e casimir}} 

a. Given the Poincar\'e algebra for the momentum operator $P^\mu$ and angular momentum operator $M^\mn$, 
\ali{
	[P^\mu,P^\nu] &= 0, \\
	i [P^\mu, M^{\rho\sigma}] &= \eta^{\mu\rho} P^\sigma - \eta^{\mu\sigma} P^\rho \\ 
	i [M^\mn, M^{\rho\sigma}] &= \eta^{\nu\rho} M^{\mu \sigma} - \eta^{\mu\rho} M^{\nu \sigma} - \eta^{\sigma\mu} M^{\rho \nu} + \eta^{\sigma\nu} M^{\rho \mu}, 
}
check that $c_2 = P_\mu P^\mu$ is a casimir of the Poincar\'e group, i.e.~check that it commutes with all generators of the group.  	

b. What about $c_2 = \frac{1}{2} M_\mn M^\mn$, is it a casimir of the Poincar\'e group? \\

\textbf{\underline{Exercise 5. Poincar\'e action on o\smash{p}erator}} 	

Given the action of the angular momentum operator on a local operator $\mathcal O$ at the origin in the QFT acting on the vacuum state, $M_\mn \mathcal O(0) |0\rangle = S_\mn \mathcal O(0) |0\rangle$, with $S_\mn$ a matrix that satisfies the same algebra as $M_\mn$, calculate the action on the same operator $O(x)$ at the location $x$. \\


\textbf{\underline{\smash{Exercise 6. Boost conserved charge}}} 	

Using the general expression for the Poincar\'e current in a field theory,  
\ali{
	J^\mu = \epsilon_\nu T^\mn , 
}
write down the conserved charge $M^{\nu\rho}$ associated with Lorentz transformations in a field theory. (This is the same exercise as appears in the course.) 

Now specifically focus on the conserved charges associated with boosts. 
Use the notation of equation (1.41) in the course and rewrite your expression in terms of time $t=x^0$ and momentum $P^i$. 

In the quantum field theory, the Noether charges satisfy the Poincar\'e algebra, in particular (1.44) $[H,K_i] = i P_i$. Quantum mechanically, we interpreted the commutation relation $[H,G_a]=0$ as an expression of conservation of the charge $G_a$. Have we found an inconsistency regarding the conservation of the boost charge $K_i$? (Conserved by Noether but not by $[H,K_i] \stackrel{?}{=} 0$?) The hint for the resolution of the apparent inconsistency is in the previous paragraph.


\chapter{Conformal invariance} 

References: Ginsparg ``Applied conformal field theory" lectures \cite{Ginsparg}, Simmons-Duffin TASI lectures on conformal bootstrap \cite{SD}. 

Consider a $d$-dimensional flat spacetime $\mathbb{R}^d$ with metric $g_\mn$. We will use this general notation for the metric, so that we can easily switch between Euclidean signature 
\ali{
	g_\mn = \delta_\mn \qquad \text{(Euclidean)} 
}
or Lorentzian signature 
\ali{
	g_\mn = \eta_\mn \qquad \text{(Lorentzian)}.  
}
Here, we let $\eta_\mn$ be a more general Minkowski metric than the one in \eqref{eta13}, namely one with $p$ timelike and $q$ spacelike dimensions, $p+q=d$: 
\ali{
	\eta_\mn = (\underbrace{-1, \cdots, -1}_{p} \,,  \underbrace{1, \cdots, 1}_{q} ).  
}
Let us determine the infinitessimal coordinate transformations 
\ali{
	x^\mu \ra x'^\mu = x^\mu + \epsilon^\mu(x) 
}
for which the metric remains invariant \emph{up to an overall $x$-dependent scale factor} 
\ali{
	g'_\mn = \Omega(x) \, g_\mn .  \label{confcondition}
} 
The answer will be the coordinate transformations with $\epsilon_\mu$ satisfying the equation 
\ali{
	\p_\mu \epsilon_\nu + \p_\nu \epsilon_\mu = \frac{2}{d} (\p \cdot \epsilon) g_\mn \qquad \text{(conformal Killing equation)}. \label{confKillingeq}
}
The corresponding transformations are \emph{conformal transformations} that have the property of being `angle-preserving' in the sense that $\frac{v\cdot w}{\sqrt{v^2 w^2}}$ with $v \cdot w = g_\mn v^\mu w^\nu$ is left invariant. 
The solution to the conformal Killing equation is 
\ali{
	\epsilon_\mu = a_\mu + b_{[\mu\nu]} x^\nu + \lambda \, x_\mu + (b_\mu x^2 - 2 x_\mu (b \cdot x))  \label{epssolconf}
}
with $a_\mu, b_{[\mu\nu]}, \lambda$ and $b_\mu$ small constants, and $b_{[\mu\nu]}$ antisymmetric. This solution contains the Poincar\'e transformations (translations $a^\mu$ and Lorentz transformations $b_{[\mu\nu]}$) since they give a subset of the transformations satisfying \eqref{confcondition}, with $\Omega(x)=1$. The extra transformations that are allowed by relaxing the Poincar\'e condition \eqref{Poinccondition} to the conformal condition \eqref{confcondition} are the \emph{scale transformations or dilations} $\lambda$ and the \emph{special conformal transformations} $b_\mu$. 

The vector 
\ali{
	\epsilon = \epsilon^\mu \p_\mu  
}
is the generator of conformal transformations of functions 
\ali{
	f(x) \ra f(x') = f(x+\epsilon) = f(x) + \epsilon^\mu \p_\mu f(x) = (1 +\omega_a t_a) f(x) 
}
with in the last expression $\epsilon = \omega_a t_a$ expanded in sums of small constants $\omega_a$ times generators $t_a$. We use here the same notation as in section \ref{sect11}, but with the understanding that $t_a$ are now the conformal generators. In addition to the Poincar\'e generators $(p_\mu, m_\mn)$ they contain scale generators $\text{d}$ and special conformal generators $k_\mu$: 
\ali{
	t_a = (p_\mu, m_\mn, \text{d}, k_\mu) \qquad \text{(conformal generators)}  \label{confgen}
}
\ali{
	p_\mu &= \p_\mu \\ 
	m_\mn &= x_\nu \p_\mu - x_\mu \p_\nu \\ 
	\text{d} &= x^\mu \p_\mu \\ 
	k_\mu &= 2 x_\mu (x \cdot \p) - x^2 \p_\mu 
}
(here given in Euclidean signature). 
They satisfy the conformal algebra, which can be derived straightforwardly. For example, $[\text{d},p_\mu] = -p_\mu$. 

The finite conformal transformations 
are 
\ali{
	x' = x+a, \quad x' = \Lambda x, \quad x' = \lambda x, \quad x' = \frac{x + b x^2}{1 + 2 b\cdot x + b^2 x^2}.  \label{finiteglobaltransf}
}

\section{Conformal field theory} 

\subsection{Conformal charges and algebra}

We consider  
a QFT with conformal invariance, consisting of a set of fields $\{ \phi(x) \}$ on the background metric $g_\mn$. This is called a \emph{conformal field theory} or CFT. 
It can be observed that the conserved Noether current is still given by the same expression as the Poincar\'e current \eqref{Poinccurrent}
\ali{
	j^\mu_{conf} = \epsilon_\nu T^\mn .  \label{confcurrent}
}
This is so because the conformal theory has the defining property that the trace of the stress tensor vanishes 
\ali{
	T_\mu^\mu = 0.  \label{tracelessness}
} 
This statement is equivalent to the defining property of conformal transformations that $\delta g_\mn \sim g_\mn$ combined with the stress tensor definition \eqref{stresstensor}. 

The conformal charges are then given by 
\ali{
	Q_\epsilon = \int d^{d-1}x \,\,  \epsilon_\nu T^{0\nu}   \label{Qconf}
}
with $\epsilon_\mu$ given in \eqref{epssolconf}. We will use the notation \cite{SD} 
\ali{
	P_\mu = Q_{p_\mu}, \quad M_\mn = Q_{m_\mn}, \quad D = Q_{\text{d}}, \quad K_\mu = Q_{k_\mu}. 
}
The general argument in section \ref{PoincQFT} leading to the identification of the Noether charges as generators of symmetry transformations of the field operators \eqref{Qagenerator} applies to any symmetry transformation \eqref{symmtransf}, in particular for conformal symmetry transformations. This allows us to conclude that the conformal charges satisfy the conformal algebra $[Q_{\epsilon_1},Q_{\epsilon_2}] = Q_{-[\epsilon_1,\epsilon_2]}$ (although for an actual derivation, see \cite{SD} (30)-(31)). Written out: 
\ali{
	[M^\mn, P^\rho] &= g^{\nu\rho} P^\mu - g^{\mu\rho} P^\nu  \label{confalgfirst} \\
	[M^\mn,M^{\rho\sigma}] &= g^{\nu\rho} M^{\mu\sigma} - g^{\mu\rho} M^{\nu\sigma} - g^{\sigma\mu} M^{\rho\nu} + g^{\sigma\nu} M^{\rho\mu}  \label{Lorentzalg} \\
	[M^\mn, K^\rho] &= g^{\nu\rho} K^\mu - g^{\mu\rho} K^\nu \\ 
	[D,P^\mu] &= P^\mu, \qquad [D,K^\mu] = -K^\mu, \\
	[K^\mu, P^\nu] &= 2 g^\mn D - 2 M^\mn .  \label{confalglast}
}
By regrouping the generators in the following way 
\ali{
	J_\mn = M_\mn, \quad J_{0\mu} = \frac{1}{2}(P_\mu - K_\mu), \quad J_{0,d+1} = D, \quad J_{\mu, d+1} = \frac{1}{2}(P_\mu + K_\mu), 
	}
the full conformal algebra can be captured in one line: 
\ali{
	[J_{ab},J_{cd}] = g_{ad} J_{bc} + g_{bc} J_{ad} - g_{ac} J_{bd} - g_{bd} J_{ac} 
}
for indices $a,b \in \{0,1, \cdots, d, d+1 \}$.  
From comparison to the Lorentz algebra \eqref{Lorentzalg}, with corresponding finite elements the Lorentz matrices $\Lambda \in SO(d)$ of the Lorentz group, we can recognize the conformal group to be given by 
\ali{
	\text{conformal group:} \qquad SO(p+1,q+1)   \label{confgroup}
}
(or $SO(1,d+1)$ in Euclidean signature $g_\mn = \delta_\mn$). 
Indeed this group is bigger than the isometry group of $R^d = R^{p,q}$, as it should be. It is then natural to study the conformal symmetry of $R^d$ by embedding it in the bigger spacetime $R^{p+1,q+1}$ with Lorentz group $SO(p+1,q+1)$. This is called the `embedding space formalism'  \cite{Penedones, Rychkov}, it will not be discussed in this course.

\subsection{Field content of CFT: primaries and descendants} \label{CFTfieldcontent}

Let us consider the action of the conformal generators on a field $\phi(x)$. This will be given by \eqref{Qagenerator} with $Q_a$ the conformal generators \eqref{Qconf}. From \eqref{GaLor} and \eqref{Gatransl}, we already know the actions of the Poincar\'e generators to be  
\ali{
	[P_\mu, \phi(x)] = \p_\mu \phi(x)  
}
and 
\ali{
	[M_\mn, \phi(x)] = \left( m_\mn + S_\mn \right)\phi(x). \label{Monphi} 
}
(The $(-i)$ factors were dropped, working in Euclidean signature here.) 
Another way to arrive at the last expression, is the following. In a rotation or Lorentz invariant QFT, local operators at the origin should transform in irreducible representations of the rotation or Lorentz group. We therefore impose that such a QFT contains field operators $\phi$ that satisfy   
\ali{
	[M_\mn, \phi(0)] = S_\mn \phi(0) \label{primoprot}
}
with $S_\mn$ matrices (not operators) that satisfy the same algebra as $M_\mn$. One can then use the algebra to obtain the action in \eqref{Monphi} of $M_\mn$ on the translated field $\phi(x) = e^{x \cdot P} \phi(0) e^{-x \cdot P}$.  

In a scale invariant theory, similarly we impose that the theory contains field operators $\phi$ that diagonalize the dilation operator $D$ 
\ali{
	[D, \phi(0)] = \Delta \, \phi(0).  
} 
The eigenvalue $\Delta$ is called the \emph{conformal dimension} of the field $\phi(x)$. It follows from the conformal algebra that 
\ali{
	[D, \phi(x)] = (\text{d}+ \Delta) \phi(x) 
}
with $\text{d}= x^\mu \p_\mu$. 

The special conformal transformations $K_\mu$ can now be seen to act as \emph{lowering operators}. For this, consider $[D,[K_\mu,\phi(0)]]$. By letting this work on the vacuum state of the CFT, we can introduce a more readable notation for $[D,[K_\mu,\phi(0)]]$. 
The vacuum state $|0\rangle$ of the CFT is annihilated by all the conformal generators 
\ali{
	P_\mu |0 \rangle = M_\mn |0 \rangle = D |0 \rangle = K_\mu |0 \rangle = 0 \qquad \text{(vacuum state)} . 
} 
It follows that 
\ali{
	[Q_\epsilon, \phi] |0 \rangle = Q_\epsilon \phi |0 \rangle 
}
so that we can use the notation ``$[Q_\epsilon, \phi] = Q_\epsilon \phi$" 
when we implicitly let the expressions work on the vacuum state.  
Now considering 
\ali{
	[D,[K_\mu,\phi]] |0 \rangle  = D K_\mu \phi |0 \rangle,  
}
it is easy to show that 
\ali{
	D \,\, K_\mu \phi(0) |0\rangle = (\Delta-1) \, K_\mu \phi(0) |0\rangle 
}
or ``$D K_\mu \phi(0) = (\Delta-1) K_\mu \phi(0)$". That is, if $\phi$ is an eigenoperator of the dilation operator $D$ with eigenvalue $\Delta$, then $K_\mu \phi$ is an eigenoperator of $D$ with eigenvalue $\Delta-1$. Similarly, the translations $P_\mu$ are \emph{raising operators}\footnote{
	Indeed, this means that $K_\mu^\dagger = P_\mu$, shown in \cite{SD} eq.~(112). 
}, $P_\mu \phi$ being an eigenoperator of $D$ with eigenvalue $\Delta+1$
\ali{
	D \,\, P_\mu \phi(0) |0\rangle = (\Delta+1) \, P_\mu \phi(0) |0\rangle . 
}   

The eigenvalue of $D$ should be bounded from below, so there must exist eigenoperators of $D$ in the theory for which $\Delta$ can not be lowered further: 
\ali{
	[K_\mu, \phi(0)] = 0. 
	}
Such fields are called \emph{primary fields} of the CFT. To summarize, a primary field $\phi(x)$ satisfies 
\ali{
	[D, \phi(0)] = \Delta \, \phi(0) \quad \text{and} \quad [K_\mu, \phi(0)] = 0 \qquad \text{(primary field)}   \label{primop}
	}
as well as still $[M_\mn, \phi(0)] = S_\mn \phi(0)$ and $[P_\mu, \phi(x)] = \p_\mu \phi(x)$.  
Given a primary field $\phi$, we can construct operators $\varphi$ of higher conformal dimension by acting with momentum generators. Such operators are called \emph{descendants}, 
\ali{
	&\varphi(0) = [P_{\mu_1}, \cdots, [P_{\mu_n}, \phi(0)] ] \qquad \text{(descendant fields)}  \label{globaldesc} \\
	&[D,\varphi(0)] = (\Delta + n) \varphi(0).   
} 
Local operators in a CFT are divided into two types: primaries or descendants. 
The descendants can be written as (linear combinations of) derivatives of other local operators, the primaries cannot. 
It can be proven that any operator in a unitary CFT is a linear combination of primaries and descendants \cite{SD}. 
\begin{comment} 
For example, $\phi(x) = e^{x \cdot P} \phi(0) e^{-x \cdot P}$ is 
a linear combination of one primary and infinitely many descendant operators. 
Eigenlijk vind ik dit heel verwarrend! We noemen \phi(x) die in de origin voldoet aan de primary conditions een primary field, NIET een combinatie van primaries and descendants \emph{in the origin}. 
Kwam uit SD denk ik. 
\end{comment} 

The exponentiated charge 
\ali{
	U = e^{Q_\epsilon} 
}
implements a finite conformal transformation. More precisely, an infinitessimal conformal transformation $[Q_\epsilon, \phi]$ exponentiates to 
\ali{
	\phi' = U \phi U^{-1}.  
}
This follows from application of the Baker-Campbell-Hausdorff formula 
\ali{
	e^A B e^{-A} = e^{[A,\cdot]} B = B + [A,B] + \frac{1}{2!}[A,[A,B]]+ \cdots.  
}
The definition of a primary operator is often given in terms of its finite transformation: a primary operator $\phi(x)$ in the CFT on $\mathbb{R}^d$ is a field that under 
conformal transformations $x \ra x'$ transforms as 
\ali{
	\phi'(x') = \left|\frac{\p x'}{\p x}\right|^{-\Delta/d} \phi(x) \qquad \text{(primary field)}  \label{primaryfield}
}
with $\left|\frac{\p x'}{\p x}\right| = \frac{1}{\sqrt{\det g'}} = \Omega^{-d/2}$. 
For example, under a dilation $x' = \lambda x$, the transformation is $\phi'(x') = \phi'(\lambda x) = \lambda^{-\Delta} \phi(x)$ and $\left|\frac{\p x'}{\p x}\right| = \lambda^d$.

\subsection{Radial quantization and state-operator correspondence} 

Let us now consider a Euclidean CFT, and for simplicity start with a two-dimensional theory $d=2$. The background geometry is the plane $\mathbb{R}^2$, which can be parametrized by Euclidean coordinates $(x,y)$, complex coordinates $(z,\bar z)$ or polar coordinates $(r,\theta)$: 
\ali{
	ds^2_{plane} = dx^2 + dy^2 = dz d \bar z = dr^2 + r^2 d\theta^2 
}
with 
\ali{
	z &= x+ i y, \quad \bar z = x - i y \\
	z &= r e^{i \theta}, \quad \bar z = r e^{-i \theta}, \quad \theta \sim \theta + 2 \pi.   
}
To quantize the theory, a choice of time is made. This choice should respect the symmetries of the theory, but otherwise is quite arbitrary in the Euclidean theory. The typical choice 
is to use the vertical direction $y$ as time, with states $|\psi\rangle$ defined 
on each timeslice ($y = $ constant). Evolution in time is generated by the generator of time translations, i.e.~the Hamiltonian $P^{0} = H$. In this case, $P^y = H_y$, with the index on the Hamiltonian referring to the choice of time. 

\begin{figure}
	\centering \includegraphics[width=15cm]{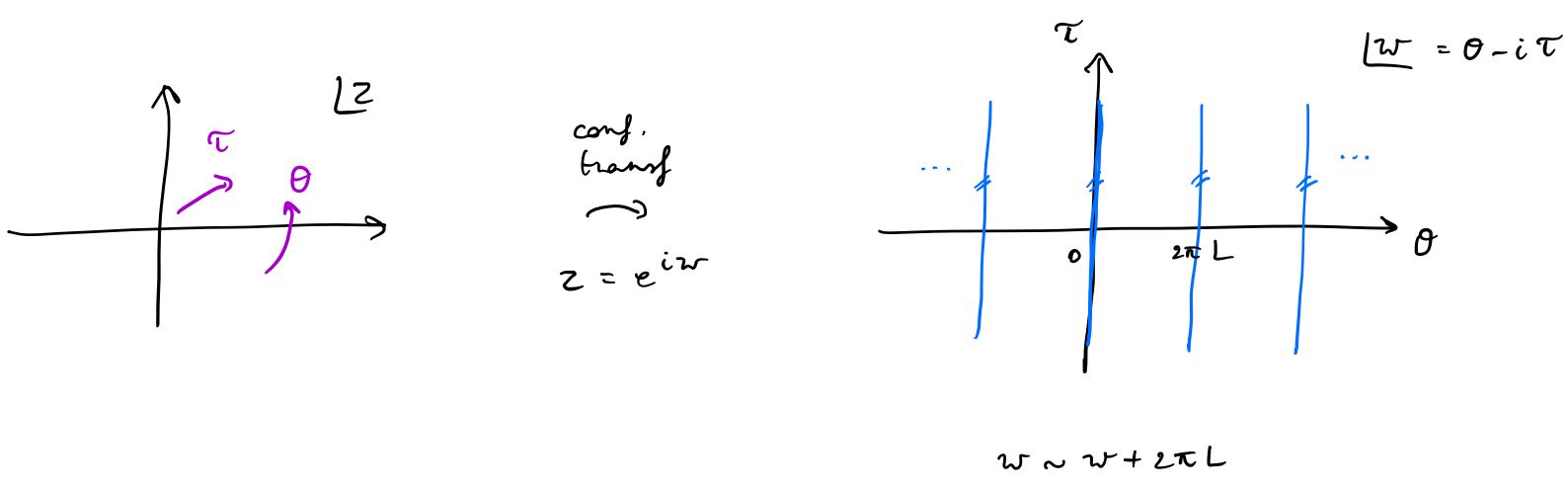} \\ 
	\vspace{0.5cm}
	\hspace{5cm} \qquad \includegraphics[width=3cm]{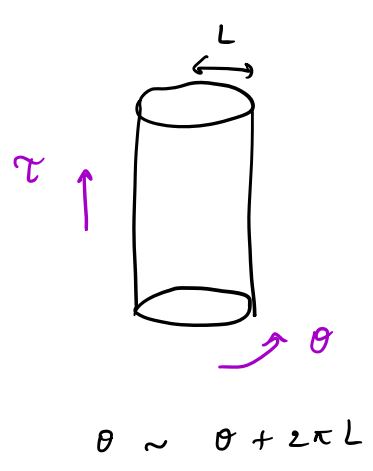}
	\caption{Conformal transformation from plane to cylinder.}  \label{figPlanetocyl}
\end{figure}

\begin{figure}
	\centering \includegraphics[width=12cm]{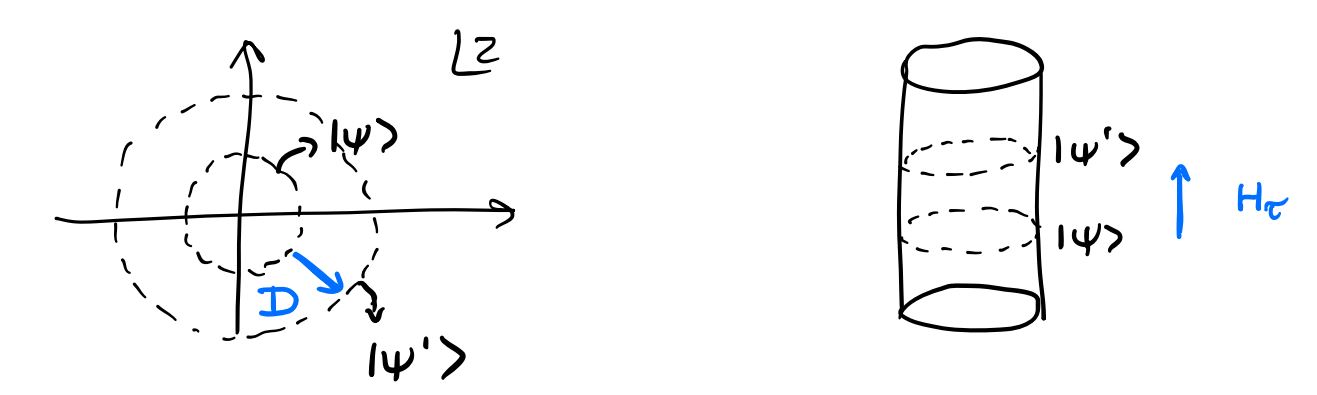}
\caption{Radial quantization on the $z$-plane: evolution in $\tau$ from one state $|\psi\rangle$ to another $|\psi'\rangle$ is generated by the scaling generator $D$. We can equivalently think of this as usual time evolution generated by $H_\tau = P^\tau$ in cylinder coordinates.}  \label{figRadialqu}
\end{figure}

In our scale invariant theory, it is also natural to use the radial direction as our choice of `time' $\tau$, with states living on 
circles (see figure \ref{figRadialqu}) 
and forward evolution in time generated by the dilation operator $D$:  
\ali{
	D &= H_\tau \qquad \qquad  \text{(radial quantization)}.  \label{radialqu}
}
Backwards evolution is stopped at the origin $r=0$, so that point should correspond to the infinite past $\tau \ra -\infty$. This suggests the definition of the radial time coordinate as 
\ali{
	r = L \, e^\tau,   \label{conftransf}
}
with $L$ an arbitrary length scale that is explicitly included for dimensional reasons. 
Under this coordinate transformation, the metric becomes  
\ali{
	ds^2_{plane} =r^2 (\frac{dr^2}{r^2} + d\theta^2) = e^{2\tau} \, L^2 (d\tau^2 + d\theta^2). \label{planetocyl}  
}
It is conformally equivalent to the metric of the cylinder of radius $L$ 
\ali{
	ds^2_{cyl} = d\tau^2 + d\theta^2 = dw d\bar w, \qquad \theta \sim \theta + 2 \pi L, \,\,\, w \sim w + 2 \pi L
}
where we also introduced the complex coordinates on the cylinder, $w = \theta - i \tau$ and $\bar w = \theta + i \tau$ (and rescaled $L \tau \ra \tau$ and $L \theta \ra \theta$ to coordinates that have the dimension of a length). The plane to cylinder conformal transformation \eqref{conftransf} (see also figure \ref{figPlanetocyl}) can then be written as 
\ali{
	z = e^{i w}, \qquad \bar z = e^{-i \bar w} \qquad \text{(plane to cylinder)}.   \label{planetocyltransf}
}
The conformal theory is insensitive to the conformal factor $\Omega = e^{2\tau}$ in \eqref{planetocyl}, so we can equally well think of the CFT living on the plane $ds^2_{plane}$ or the cylinder $ds^2_{cyl}$. The radial coordinate $\tau = \log(r/L)$ on the plane becomes the typical time coordinate $\tau$ on the cylinder, with dilations $r \ra \lambda r$ 
corresponding to cylinder time translations $\tau \ra \tau + \log \lambda$.  

The above generalizes 
for a Euclidean CFT on the $d$-dimensional plane $\mathbb{R}^d$, with the conformal transformation $r = L \, e^\tau$ to the (radius-$L$) `cylinder' $ds^2_{\mathbb R \times S^{d-1}}$:   
\ali{
	ds^2_{\mathbb R^d} = dr^2 + r^2 d\Omega^2_{S^{d-1}} = e^{2\tau} \, L^2 (d\tau^2 + d\Omega^2_{S^{d-1}}) = e^{2\tau} \, ds^2_{\mathbb R \times S^{d-1}}.   
	}
Radial quantization in flat space $\mathbb{R}^d$ is equivalent to usual quantization on the cylinder $ds^2_{\mathbb R \times S^{d-1}}$. States live on spheres $S^{d-1}$ and time evolution is generated by acting with $e^{-D \tau}$. This is summarized in figure \ref{figRadialqu}.

\begin{figure}
	\centering \includegraphics[width=13cm]{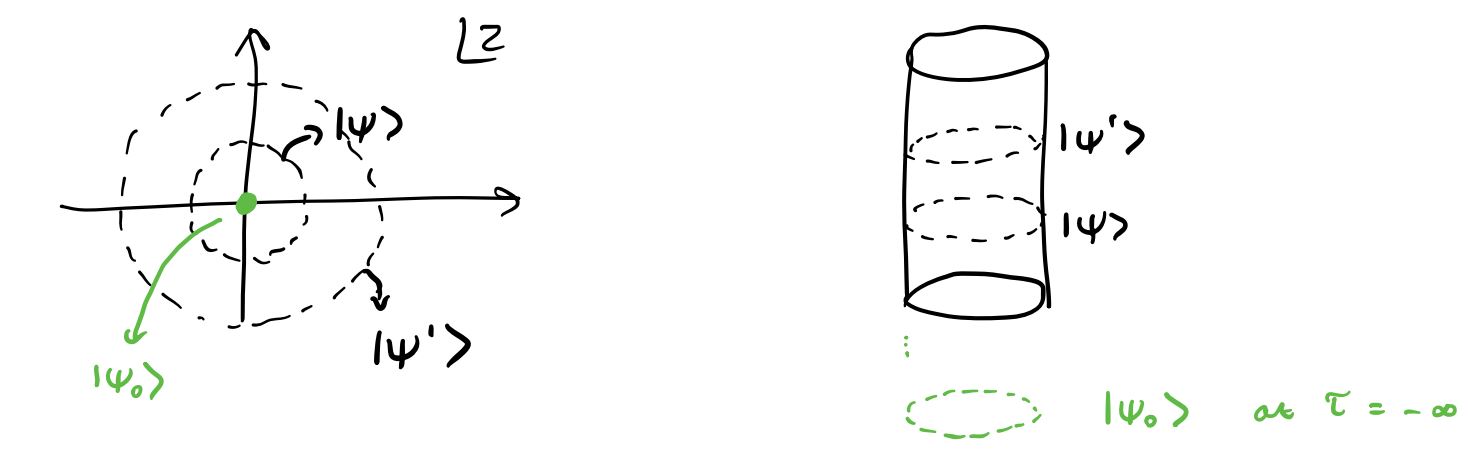}
	\caption{Asymptotic state $|\psi_0 \rangle = \mathcal O(0) |0\rangle$ at $\tau = -\infty$.} \label{figStateop}
\end{figure}

Consider the state $|\psi\rangle$ at a finite time $\tau$ in figure \ref{figRadialqu}. We can evolve that state backwards in time with $D$, all the way back to the point in the origin of the plane. The state that lives there at the origin, let's call it the initial or `asymptotic' state $|\psi_0\rangle$, is the state at the earliest possible time $\tau \ra -\infty$. The corresponding cylinder picture is given in figure \ref{figStateop}. 
As we can associate this state with a point, 
we can define the \emph{asymptotic state} $|\psi_0 \rangle$ to be created by the insertion of an \emph{operator} $\mathcal O$ at that point (the origin): $|\psi_0 \rangle = \mathcal O(0) |0\rangle$. This means there is a one-to-one correspondence between states and operators in a CFT. In a notation that makes that duality most clear, we define asymptotic states $|\mathcal O \rangle$ as  
\ali{
	\mathcal O(0) |0\rangle \equiv |\mathcal O \rangle \qquad \quad \text{(state-operator correspondence)}. \label{stateop}
}

This is the \emph{state-operator correspondence}. 
It is an axiom of CFT that 
becomes especially intuitive in radial quantization.  


Note that radial quantization \eqref{radialqu} also explains why it is so natural in a CFT to work with eigenoperators of the scaling generator $D$: this is as natural as working with energy eigenstates of the Hamiltonian $H_\tau$ in the cylinder picture. Said otherwise, because $D = H_\tau$, the conformal dimension $\Delta$ has the interpretation of an energy 
on the cylinder. 

\subsection{Primary and descendant states, conformal multiplet and OPE}  \label{sectconfcasimiroa}

Making use of the state-operator correspondence, the discussion of the field content of a CFT in section \ref{CFTfieldcontent} immediately 
applies to states of a CFT as well. Namely, we can define a \emph{primary state} $|\mathcal O \rangle$ as an eigenstate of the dilation operator and rotation/boost operator that is annihilated by special conformal transformations 
\ali{
	\begin{split} 
		K_\mu |\mathcal O \rangle &= 0 \\ 
		D |\mathcal O \rangle &= \Delta |\mathcal O \rangle \qquad \quad \text{(primary state)} \\
		M_\mn |\mathcal O \rangle &= S_\mn |\mathcal O \rangle. 
	\end{split} 
} 
This follows directly from letting equations \eqref{primop} and \eqref{primoprot} work on the vacuum, and using \eqref{stateop}.  

\emph{Descendant states} are obtained from successively working with momentum operators on a primary state. Such descendant states 
\ali{
	|\mathcal O \rangle, \quad P_\mu |\mathcal O \rangle, \quad P_\mu P_\nu |\mathcal O \rangle, \cdots  
}
form the basis for a \emph{conformal multiplet}. The state $\mathcal O(x)|0\rangle$ e.g.~can be expanded in this basis and thus belongs to the conformal multiplet. 

The states in the conformal multiplet transform into each other under conformal transformations, they form an irreducible representation of the conformal group \eqref{confgroup}. Each conformal multiplet is labeled by the quantum number $\Delta$, which is the conformal dimension of the operator $\mathcal O$ that creates the lowest weight state $|\mathcal O\rangle$ of the multiplet. Another notation of the basis of the multiplet makes this more clear:  
\ali{
	|\Delta\rangle, \quad |\Delta + 1\rangle, \quad |\Delta + 2\rangle, \cdots  \, .   
}
The conformal (quadratic) casimir 
\ali{
	c_2^{conf} &= \frac{1}{2} J_{ab} J^{ab}  \label{c2conf} \\
	&= \frac{1}{2} M_\mn M^\mn + D(D-d) - P_\mu K^\mu  
}
does not distinguish between states belonging to one conformal multiplet. Working on such a state it extracts an eigenvalue $\Delta(\Delta-d)$. This is for spinless fields. Including spin $l$, the casimir value is $\Delta(\Delta-d) + l(l+d-2)$. 

Just as in section \ref{sectionCasimir}, where the Poincar\'e casimir instructs us to label particles in a Poincar\'e invariant theory by their mass and spin, we see that the conformal casimir instructs us to label the field content of a conformal invariant theory by the conformal dimension $\Delta$ and spin $l$. 
Note that indeed we required a replacement of the notion of a mass, because a conformally invariant theory cannot contain massive particles. Any scale in the theory explicitly breaks conformal invariance. For example, the action of a massless boson $S = \int d^d x (\p_\mu \phi \p^\mu \phi)$ is conformal, but the one of a massive boson $S = \int d^d x (\p_\mu \phi \p^\mu \phi + \frac{1}{2} m^2 \phi^2)$ is not.

We can consider the state $O_i(x) O_j(0) |0 \rangle$. Because every state is a linear combination of primaries and descendants, it can be decomposed as 
\ali{
	O_i(x) O_j(0) |0 \rangle = \sum_{k \in \text{primaries}} c_{ijk}(x,P) O_k(0)|0\rangle 
}
where $c_{ijk}$ is an operator that packages together primaries and descendants in the $k$-th conformal multiplet. The state-operator correspondence allows to write an operator equation version of the above statement: 
\ali{
	O_i(x_1) O_j(x_2)  = \sum_{k \in \text{primaries}} c_{ijk}(x_{12},\p_2) O_k(x_2) \qquad \text{(OPE)}  \label{OPE}
}
(valid inside any correlation function where the other operators $O_n(x_n)$ have $|x_{2n}| > |x_{12}|$).  
This is called the `operator product expansion' or OPE. 
It expresses that two operators that come close together can be expanded in a complete set of local operators. The expansion coefficients $c_{ijk}$ as expected are singular. 
However, we will see later that this is perfectly fine, and the way in which they diverge encodes important information about the CFT.  

\section{2-dimensional conformal field theory} 

In two dimensions $d=2$, the structure of a conformal field theory is much richer than in higher dimensions. The results of the previous chapter all still apply, but only form part of the discussion. This is because the conformal Killing equation in two dimensions has infinitely more solutions than the ones we have discussed so far. 

We consider a 2D Euclidean CFT, $d=2$ and $g_\mn = \delta_\mn$ or 
\ali{
	ds^2 = dx^2 + dy^2 = dz d \bar z . 
}
Here we introduced complex coordinates 
\ali{
	z = x + i y, \qquad \bar z = x - i y. 
}
In a moment, we will also need the relation between the partial derivatives in both coordinates
\ali{
	\p_z = \frac{1}{2} (\p_x - i \p_y), \qquad \p_{\bar z} = \frac{1}{2} (\p_x + i \p_y).  
}

The conformal Killing equation \eqref{confKillingeq} takes the form 
\ali{
	\p_x \epsilon_x = \p_y \epsilon_y, \qquad \p_x \epsilon_y = - \p_y \epsilon_x 
}
which can be recognized as the Cauchy-Riemann equations of complex analysis. Rewritten in complex coordinates, 
\ali{
	\p_{\bar z} \epsilon^{z}(z,\bar z) = 0  \quad \text{and} \quad \p_z \epsilon^{\bar z}(z,\bar z) = 0, 
}
they express that the complex functions $\epsilon^z = \epsilon_x + i \epsilon_y$ and $\epsilon^{\bar z} = \epsilon_x - i \epsilon_y$ are respectively holomorphic (no $\bar z$ dependence) and antiholomorphic (no $z$ dependence) functions: 
\ali{
	\epsilon^{z}(z) , \qquad \epsilon^{\bar z}(\bar z) .  \label{epsilonz}
}
We have found that in two dimensions, not just the $\epsilon^\mu$ in \eqref{epssolconf} solve the conformal Killing equation, but \emph{all} $\epsilon^\mu$ of the form \eqref{epsilonz} do. 
The 2-dimensional conformal transformations are 
\ali{
	z \ra z' = z + \epsilon^z(z), \qquad  \bar z \ra \bar z' = \bar z + \epsilon^{\bar z}(\bar z). 
}
From this, we can read off the conformal generators $t_a$, using the notation introduced in \eqref{inftransf}.  We assume the functions $\epsilon^z(z)$ and $\epsilon^{\bar z}(\bar z)$ can be expanded in a Laurent series around the origin 
\ali{
	\epsilon^z(z) = \sum_{n=-\infty}^\infty c_n z^{n+1}, \qquad \epsilon^{\bar z}(\bar z) = \sum_{n=-\infty}^\infty \bar c_n \bar z^{n+1}. \label{epsilonLaurent}
}
Then the infinitessimal constants $\omega_a$ in \eqref{inftransf} are identified as $c_n$ and $\bar c_n$, and the generators $t_a$ as $l_n$ and $\bar l_n$, defined as 
\ali{
	l_n \equiv z^{n+1}\p_z, \qquad \bar l_n \equiv \bar z^{n+1} \p_{\bar z} \qquad (n \in \mathbb Z).  \label{lngen}
}
The algebra that they satisfy is the \emph{Witt algebra} 
\ali{
	[l_n,l_m] = (n-m) l_{n+m}, \qquad [\bar l_n,\bar l_m] = (n-m) \bar l_{n+m}, \qquad [l_n,\bar l_m] = 0 \qquad (n,m \in \mathbb Z).  \label{Wittalg}
	}
It consists of 
the direct sum of two infinite-dimensional algebras. These each contain a finite subalgebra generated by $(l_{-1},l_0,l_1)$ and $(\bar l_{-1},\bar l_0,\bar l_1)$ :
\ali{
	\begin{split} [l_0,l_1]=-l_1, \quad [l_1,l_{-1}] = 2l_0, \quad [l_{-1},l_0]=-l_{-1} \\
	[\bar l_0,\bar l_1]=-l_1, \quad [\bar l_1,\bar l_{-1}] = 2\bar l_0, \quad [\bar l_{-1},\bar l_0]=-\bar l_{-1} . \end{split}  
	} 
We can identify the corresponding conformal transformations as the ones discussed in the previous chapter: Poincar\'e transformations, dilations and special conformal transformations. Said otherwise, the generators $(l_{-1},l_0,l_1)$ and $(\bar l_{-1},\bar l_0,\bar l_1)$ map to the conformal generators $(p_\mu, m_\mn, \text{d}, k_\mu)$ in  \eqref{confgen}. Namely, $l_{-1} = \p_z = p_z$ and $l_1 = z^2 \p_z = k_{\bar z}$ (and $\bar l_{-1} = p_{\bar z}$ and $\bar l_1 = k_z$). Finally, $l_0 = z \p_z$ generates scale transformations and rotations. This is most clear from considering a finite such transformation $z \ra \zeta z$. Writing $z = r e^{i \theta}$, it is clear that $\zeta$ can affect either $r$ or $\theta$, specifically by a rescaling and a translation respectively. We can read off that dilations are generated by $\text{d}=l_0 + \bar l_0$ and rotations by $m_{xy} = i (l_0-\bar l_0)$. 
In summary, 
\ali{
	p_z = l_{-1}, \quad k_z = \bar l_1, \quad \text{d}=l_0 + \bar l_0, \quad m_{xy} = i (l_0-\bar l_0).  
}

What sets these generators apart from the other $l_n$? The answer is that $ l_{-1,0,1}$ (and $\bar l_{-1,0,1}$) are the ones that are globally well-defined on the Riemann sphere $S^2 = \mathbb C \cup \infty$. 
A conformal transformation (only considering the holomorphic part for the sake of the argument), 
\ali{
	\epsilon^z(z)\p_z = \sum_{-\infty}^\infty c_n z^{n+1} \p_z \quad  \stackrel{z=1/w}{=} \quad \sum_{-\infty}^\infty c_n \left(\frac{1}{w}\right)^{n-1} \p_w 
}
should be non-singular 
both in $z \ra 0$ and $z \ra \infty$ (or $w \ra 0$). The first condition requires $n \geq -1$, the second $n \leq 1$, so that combined $n = 0,\pm 1$ \cite{Ginsparg}.   

In 2 dimensions, the infinite set of conformal transformations generated by \eqref{lngen} are then referred to as \emph{local} conformal transformations, and the subset $n = 0,\pm 1$ as \emph{global} conformal transformations.  

Let us now consider the finite 
conformal transformations. 
The finite, local conformal transformations are given by \emph{any}  holomorphic resp.~antiholomorphic coordinate transformations 
\ali{
	z \ra f(z), \qquad \bar z \ra \bar f(\bar z) 
}
which are indeed immediately seen to take the metric to itself times a conformal factor $\Omega$: 
\ali{
 	ds^2 = dz d\bar z \ra \left|\frac{\p f}{\p z}\right|^2 dz d\bar z . 
}
The finite, global conformal transformations are the subset of these that take the form\footnote{
	Note that a pole is allowed. It is only the essential singularities that the function $f(z)$ should be free of \cite{difran}, which occur when the Laurent series expansion of the function has infinitely many powers of negative degree. See e.g.~\url{https://en.wikipedia.org/wiki/Singularity\_(mathematics)\#Complex\_analysis}. 
}
\ali{
	&z \ra \frac{a z + b}{c z + d}, \qquad \bar z \ra \frac{\bar a \bar z + \bar b}{\bar c \bar z + \bar d}  \label{Mobius} \\
	&a,b,c,d \in \mathbb C, \qquad ad - bc = 1. 
}
These are the \emph{M\"obius transformations}, which form the group 
\ali{
	SL(2,\mathbb C)/\mathbb Z_2 
}
(the quotient by $\mathbb Z_2$ because taking $a,b,c,d$ to $-a,-b,-c,-d$ does not alter the function $\frac{a z + b}{c z + d}$). This group is indeed isomorphic to $SO(3,1)$, consistent with the result \eqref{confgroup} of our discussion in general dimensions. If you're wondering how to write the finite global conformal transformations \eqref{finiteglobaltransf} in the M\"obius form \eqref{Mobius}, this can be found on p.8 in \cite{Ginsparg}.

\subsection{Conformal charges and algebra in 2 dimensions} 

Now let us discuss the generators at the quantum level. 
For a given choice of time, the Noether charge is given by 
\ali{
	Q_\epsilon = \int d^{d-1}x \, j^0_{conf}(x) \quad  \stackrel{\eqref{confcurrent}}{=} \quad \int d^{d-1}x \; \epsilon_\nu(x) T^{0\nu}(x)   
} 
with the conformal current integrated over a full time slice. 
In radial quantization, the time slices are origin-centered circles in the plane, and so the conformal charge in complex coordinates takes the form 
\ali{
	Q_\epsilon = \oint_0 \left(\frac{dz}{2\pi i} T_{zz}(z) \epsilon^z(z) + \frac{d\bar z}{2\pi i} T_{\bar z \bar z}(\bar z) \epsilon^{\bar z}(\bar z) \right). \label{Q2D}
} 
Here, we have 
written holomorphic and antiholomorphic stress tensor components. The (anti)holo\-mor\-phi\-ci\-ty follows from writing out stress tensor conservation and tracelessness of the stress tensor \eqref{tracelessness} in complex coordinates. 
  
\begin{figure}
	\centering \includegraphics[width=10cm]{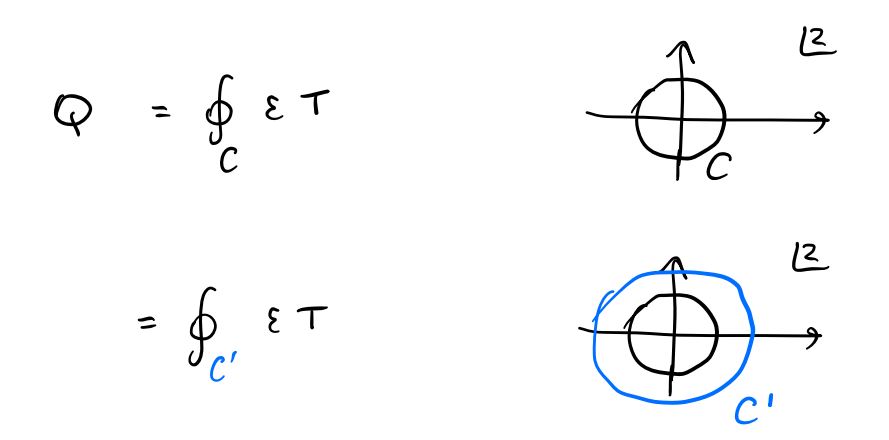}
	\caption{Topological charge conservation.} \label{Qcontour}
\end{figure}

When writing a contour integral $\oint_{\mathcal C}$, an integration contour $\mathcal C$ should be specified. In the case at hand, the contour is a radial time slice (see Figure \ref{Qcontour}). As long as there are no other operator insertions around, we are free to move the contour, i.e.~contract it to smaller or larger circles (or even deform it to another shape). This topological nature of $Q_\epsilon$ gives a very nice visualization of the conservation, i.e.~time-independence, of the charge \eqref{conservedQ}. In practical terms, it means it is sufficient to specify the contour in \eqref{Q2D} by its center $0$. We will soon encounter a situation where we are required to also specify the radius of the contour.  

Radial quantization is particularly useful in two dimensions because it unlocks the power of complex analysis, including the techniques of contour integration. For example, radial quantization allows us to `see' charge conservation (Figure \ref{Qcontour}).

We are interested in the conformal algebra $[Q_{\epsilon_1},Q_{\epsilon_2}]$. In previous instances (Poincar\'e invariant and global conformally invariant theories), we made use of the general QFT result \eqref{Qagenerator} that the conserved charge acts as a generator of the symmetry on fields $\phi$, 
\ali{
	\delta_\epsilon \phi = [Q_\epsilon, \phi],  \label{Qageneratorbis}
}
to argue that the charges satisfy the same algebra as the generators of the symmetry acting on functions. In two dimensions, there will be a correction to this claim. 

Let us first understand what is meant by the above commutator $[Q_\epsilon, \phi]$ when $Q_\epsilon$ is given by a contour integral. For this, we go back to how this commutator notation was introduced in the derivation of equation \eqref{Qagenerator}. Namely, this followed from integrating the Ward identity over a particular stretched volume such that 
\ali{
	\langle \delta \phi(t, \vec x) \cdots \rangle = \langle Q(t_+) \phi \cdots \rangle - \langle Q(t_-) \phi \cdots \rangle \, \stackrel{t_\pm \ra t}{=} \, \langle 0| [Q(t),\phi(t,\vec x)] \cdots |0\rangle 
} 
where the implicit time ordering in the correlators $\langle Q(t_\pm) \phi \cdots \rangle = \langle 0| \mathcal T \left( Q(t_\pm) \phi \cdots \right) |0\rangle$ leads to the equal time commutator of $Q$ and $\phi$ after taking the limit $t_\pm \ra t$ of the integration volume becoming infinitely stretched. 

\begin{figure}
	\centering 
	\includegraphics[width=6cm]{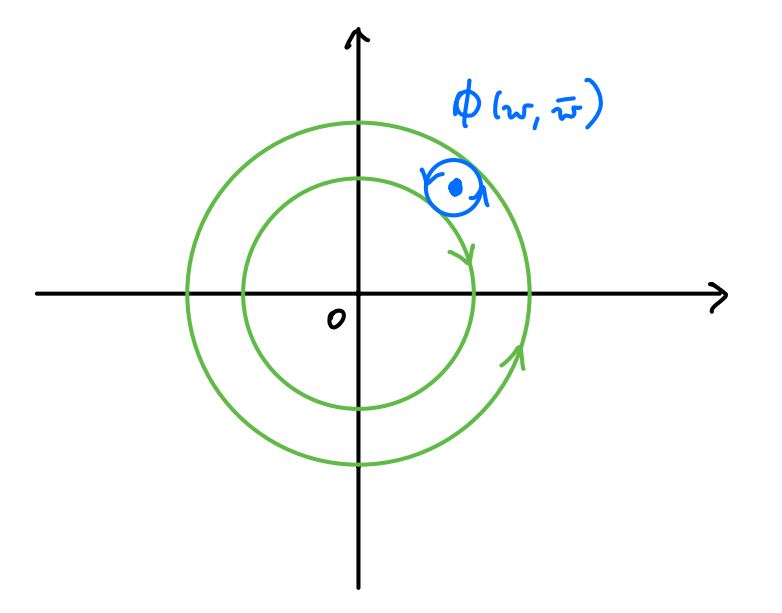} \\ 
	\includegraphics[width=5cm]{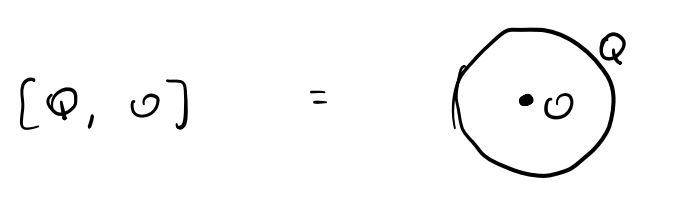}  
	\caption{Contour for $[Q_\epsilon,\phi]$ calculation, visualizing $\oint_{|z|>|w|} - \oint_{|z|>|w|} = \oint_{(w,\bar w)}$.  
		See e.g.~\cite{Flohr}. 
	} \label{Qphicontour} 
\end{figure}

Applying this to a 2D Euclidean CFT in radial quantization, with $Q_\epsilon$ given in \eqref{Q2D}: 
\ali{
	\langle \delta_\epsilon \phi(w,\bar w) \cdots \rangle &= \langle \left( \oint_{|z|>|w|} - \oint_{|z|<|w|} \right) \left(\frac{dz}{2\pi i}  T_{zz}(z) \epsilon^z(z) + \frac{d\bar z}{2\pi i} T_{\bar z \bar z}(\bar z) \epsilon^{\bar z}(\bar z) \right) \phi(w,\bar w) \cdots \rangle \, \nonumber \\ 
	& \stackrel{|z|\ra |w|}{=} \, \langle 0| [Q_\epsilon(|w|), 
	\phi(w,\bar w)] \cdots |0\rangle . 
}  
This time we had to specify the radius of the origin-centered contour in the definition of the charge with respect to the other operator insertion at location $(w,\bar w)$, because we can't simply deform the contour through that obstruction. The 
combination of contours appearing in $\delta_\epsilon \phi$ is illustrated in figure \ref{Qphicontour}. From the figure, we can see 
it can be deformed into a final contour that wraps the location of the operator tightly. It follows that 
\ali{
	\delta_\epsilon \phi(w,\bar w)  &= \oint_{(w,\bar w)}  \left(\frac{dz}{2\pi i} \epsilon^z(z) T_{zz}(z) \phi(w,\bar w)  + \frac{d\bar z}{2\pi i} \epsilon^{\bar z}(\bar z)  T_{\bar z \bar z}(\bar z) \phi(w,\bar w)  \right)  = [Q_\epsilon(|w|),\phi(w,\bar w)]. 
}

To work out the contour integral in $\delta_\epsilon \phi$ we need to make sense of the operator product $T_{zz} \phi$. This is done by the OPE \eqref{OPE}, and the operator product expansion $T_{zz}(z) \phi(w,\bar w)$ is in short referred to as the `$T \phi$ OPE'. Then the contour integrals can be worked out explicitly by making use of Cauchy's integral formulas 
\ali{
	\oint_a \frac{dz}{2\pi i} \, \frac{f(z)}{z-a} = f(a), \qquad \oint_a \frac{dz}{2\pi i} \, \frac{f(z)}{(z-a)^{n+1}} = \frac{\p^n f(a)}{n!} \qquad (n \geq 0).    
} 
The first formula can be remembered from working out that $\oint_0 \frac{dz}{z}$ in coordinates $z=e^{i \theta}$ 
gives $2\pi i$. The second can be remembered form the first Cauchy formula and partial integration. 

To summarize, knowledge of the conformal transformation behavior $\delta_\epsilon \phi$ of a field $\phi$ is equivalent (through Cauchy's integral formulas) with knowledge of the $T \phi$ OPE. 
Similarly, knowledge of the transformation behavior of the stress tensor $\delta_\epsilon T_{ww}(w)$ is equivalent to knowing the $T T$ OPE, short for the operator product expansion $T_{zz}(z)T_{ww}(w)$. We are specifically interested at this point in $\delta_\epsilon T_{ww}(w) = [Q_\epsilon, T_{ww}(w)]$ because it is an intermediate step in deriving the conformal algebra $[Q_{\epsilon_1}, Q_{\epsilon_2}]$.

We already know to a large extent what the conformal transformation behavior of the stress tensor should be. Namely, $T_\mn$ is a 2-tensor (an object with 2 spacetime indices) and transforms in the same way as the metric tensor,  
\ali{ 
	T_\mn(x) \ra T_\mn'(x) = T_{\rho\sigma}(x') \frac{\p x'^\rho}{\p x^\mu} \frac{\p x'^\sigma}{\p x^\nu} \quad \text{(active)}, \qquad  
	T_\mn(x) \ra T_\mn'(x') = T_{\rho\sigma}(x) \frac{\p x^\rho}{\p x'^\mu} \frac{\p x^\sigma}{\p x'^\nu} \quad \text{(passive)}
}
under finite, global 
conformal transformations, in respectively the active or passive picture. Written out in complex coordinates for the holomorphic stress tensor component (the discussion for the antiholomorphic component is completely analogous), 
\ali{
	T_{zz}'(z') = \left( \frac{\p z'}{\p z}\right)^{-2} T_{zz}(z). \qquad \text{(global)}
}
Indeed this gives the correct transformation behavior under Lorentz transformations of the stress tensor as a spin 2 field (2-tensor), 
and under scale transformations of the stress tensor as a $\Delta = 2$ field. The naive scaling dimension $\Delta = 2$ of the stress tensor follows from dimensional analysis: $\int d^{d-1}x \, T^{00} = E$ or $T^{00}$ is an energy density with dimension (energy/volume) or (1/length$^2$).   

The corresponding infinitessimal, global conformal transformation of $T_{zz}$ for $\delta z = \epsilon^z(z)$  is 
\ali{
	\delta_\epsilon T_{zz}(z) \equiv T'_{zz}(z) - T_{zz}(z) 
	= \epsilon^z(z) \p_z T_{zz}(z) + 2 \,  T_{zz}(z) \p_z \epsilon^z(z)  \label{CFTstresstensortransf}  
}
By $\delta_\epsilon T_{zz}(z) = [Q_\epsilon, T_{zz}(z)]$, yet another way to write the global conformal transformation behavior of $T_{zz}$ is as the $TT$ OPE 
\ali{
	T_{zz}(z) T_{ww}(w) = \frac{2 T_{ww}(w)}{(z-w)^2} + \frac{\p T_{ww}(w)}{z-w} + \cdots 
}
where $\cdots$ are higher order terms in $(z-w)$ that don't survive the Cauchy integral. It's only the divergent terms in the OPE that contain info about the transformation behavior.  The stress tensor components have dimension 1/length$^2$ so the left hand side of the $TT$ OPE has mass dimension $4$. You can check that the terms on the right hand side have the same mass dimension, as they of course should. However, there is one possible divergent term missing on the right hand side that is also allowed on dimensional grounds, namely 
\ali{
	\frac{c/2}{(z-w)^4}.  
}
The numerator should just be a dimensionless number, and it is by convention called $c/2$, with $c$ an important parameter in a 2D CFT called the \emph{central charge}. Including this allowed term in the stress tensor OPE, we have  
\ali{
	T_{zz}(z) T_{ww}(w) &= \frac{c/2}{(z-w)^4} + \frac{2 T_{ww}(w)}{(z-w)^2} + \frac{\p T_{ww}(w)}{z-w} + \cdots  \label{TTOPE} \\
	T_{\bar z\bar z}(\bar z) T_{\bar w\bar w}(\bar w) &= \frac{\bar c/2}{(\bar z-\bar w)^4} +  \frac{2 T_{\bar w\bar w}(\bar w)}{(\bar z-\bar w)^2} + \frac{\bar \p T_{\bar w\bar w}(\bar w)}{\bar z-\bar w} + \cdots 
} 
with central charges $c$ and $\bar c$. 

Working our way back, the central charge terms give contributions to the infinitessimal and finite transformation behavior 
\ali{
	\delta_\epsilon T_{zz}(z) &= \epsilon^z(z) \p_z T_{zz}(z) + 2 \,  T_{zz}(z) \p_z \epsilon^z(z) + \frac{c}{12} \p_z^3 \epsilon^z(z) \label{CFTstresstensortransffull} \\
	T_{z'z'}'(z) &= \left( \frac{\p z'}{\p z}\right)^{2} T_{zz}(z') + \frac{c}{12} \{z',z\}  \quad \text{(active)}, \qquad 
	T_{z'z'}'(z') = \left( \frac{\p z'}{\p z}\right)^{-2} \left( T_{zz}(z)  -  \frac{c}{12} \{z',z\} \right) \quad \text{(passive)} \\
	& \hspace{8.1cm}=   \left( \frac{\p z'}{\p z}\right)^{-2}  T_{zz}(z)  +  \frac{c}{12} \{z,z'\}   \quad \text{(passive)} \nonumber 
}
where the accolades are the standard notation for the \emph{Schwarzian derivative} 
\ali{
	\{f(z),z\} \, := \frac{\p_z^3 f}{\p_z f} - \frac{3}{2} \left( \frac{\p_z^2 f}{\p_z f} \right)^2 
}
which has the property $\{w,v\} = - \left(\frac{dw}{dv}\right)^2 \{v,w\}$.  
We have found the \emph{general transformation behavior of the stress tensor under conformal transformations}. Indeed, specifically for global conformal transformations \eqref{Mobius}, $f(z) = (a z + b)/(c z + d)$, the Schwarzian derivative vanishes $\{f(z),z\} = 0$ (check as an exercise), so that the general transformation behavior is still consistent with the transformation behavior under global conformal transformations we started the argument from.  

We were interested in the transformation behavior of the stress tensor in order to derive the conformal algebra of the conserved charges $Q_\epsilon$. Similarly as in the $d>2$ CFT case, where for generators $\epsilon = p,m,d,k$ we denoted the associated basis conserved charge generators $Q_\epsilon = P,M,D,K$ with capital letters, in the $d=2$ CFT case under consideration we use the notation $L_n, \bar L_n$ $(n \in \mathbb Z)$ for the basis conserved charges associated with the generators $\epsilon = l_n, \bar l_n$ $(n \in \mathbb Z)$. That is, we expand a general conserved charge as 
\ali{
	Q_\epsilon = \sum_n c_n L_n + \sum_n \bar c_n \bar L_n. 
}
Then from \eqref{Q2D} with \eqref{epsilonLaurent}, we can read off 
\ali{
	L_n = \oint_0 \frac{dz}{2\pi i} \, z^{n+1} T_{zz}(z), \qquad 	\bar L_n = \oint_0 \frac{d\bar z}{2\pi i} \, \bar z^{n+1} T_{\bar z\bar z}(\bar z). \label{Lnoint}
}
These are the \emph{Virasoro generators}. 
Inverting the contour integral by using the Cauchy formulas, we see that the Virasoro generators are the Laurent expansion modes 
of the stress tensor: 
\ali{
	T_{zz}(z) = \sum_{n=-\infty}^\infty \frac{L_n}{z^{n+2}}, \qquad T_{\bar z\bar z}(\bar z) = \sum_{n=-\infty}^\infty \frac{\bar L_n}{\bar z^{n+2}} . \label{TzzLaurent}
}
(Note that in cylinder coordinates this is just a Fourier expansion.) 

Using the $TT$ OPE, you can work out 
\ali{
	[L_n,L_m] &= \left( \oint \frac{dz}{2\pi i} \oint \frac{dw}{2\pi i} - \oint \frac{dw}{2\pi i} \oint \frac{dz}{2\pi i} \right) z^{n+1} T_{zz}(z) w^{m+1} T_{ww}(w) \\
	&= (n-m) L_{n+m} + 	\frac{c}{12} (n^3-n) \delta_{n+m,0} . 
}
In absence of the central charge, this would just be the classical Witt algebra \eqref{Wittalg}, which would express that the conserved charges $Q_\epsilon$  satisfy the same algebra as the conformal generators $\epsilon$ working on functions, $[Q_{\epsilon_1},Q_{\epsilon_2}] = Q_{-[\epsilon_1,\epsilon_2]}$, as we encountered for Poincar\'e symmetry and for global conformal symmetry. The second term is a \emph{central charge term} contribution to, or \emph{central extension} of, the conformal algebra at the \emph{quantum} level. The possibility of this occurring was first discussed in section \ref{quantum}. The central charge term has the property that it commutes with all elements of the algebra. 


We have found the famous \emph{Virasoro algebra} of 2D CFT, which is the unique central extension of the Witt algebra: 
\ali{
	\begin{split} 
	[L_n,L_m]&=(n-m)L_{n+m} + \frac{c}{12}(n^3-n)\delta_{n+m,0} \\
	[\bar L_n,\bar L_m]&=(n-m)\bar L_{n+m} + \frac{\bar c}{12}(n^3-n)\delta_{n+m,0} \qquad \text{(Virasoro algebra)} \\
	[L_n,\bar L_m]&=0	
	\end{split} \label{Virasoroalg} 
}
for $n,m \in \mathbb Z$. 
It has as a subalgebra the global algebra 
\ali{
	[L_n,L_m]=(n-m)L_{n+m}, \quad [\bar L_n,\bar L_m]=(n-m)\bar L_{n+m}, \quad [L_n,\bar L_m]=0 \qquad (n,m=-1,0,1) 
}
or 
\ali{
	[L_0,L_1]=-L_1, \qquad [L_1,L_{-1}]=2L_0, \qquad [L_{-1},L_0]=-L_{-1}  
}
so that we recover the global conformal algebra of conserved charges of the previous chapter, \eqref{confalgfirst}-\eqref{confalglast}.  

What is the physical interpretation of the central charge $c$ of the 2D CFT? (When we refer to \emph{the} central charge $c$, we are implicitly assuming that $\bar c = c$.)  
If you take as an example of a 2D CFT the theory of a free boson and calculate the $TT$ OPE, you would find \eqref{TTOPE} with $c=1$. If you take as an example the theory of $n$ free bosons, you would find \eqref{TTOPE} with $c=n$. This indicates that $c$ is a \emph{measure for the number of degrees of freedom} in the 2D CFT. 
It has additional physical interpretations, some of which will  appear in section \ref{sectCardyCFT}: $c$ for casimir, $c$ for Cardy, and others which will not be discussed in the course (Weyl anomaly, $c$-theorem, ...).  Reference: \cite{Tong} section 4.4.   

\subsection{Field content: Virasoro primaries and Virasoro descendants} 



The conformal symmetry now allows to organize the field content of the CFT. 
The strategy is completely analogous to the general $d (>2)$-dimensional case in section \ref{CFTfieldcontent}. There, we considered an eigenoperator $\phi^{(\Delta)}$ of the dilation operator $D$, 
\ali{
	[D,\phi^{(\Delta)}(0)] = \Delta \, \phi^{(\Delta)}(0).  
} 
It is labeled by its eigenvalue $\Delta$, called the conformal dimension of $\phi^{(\Delta)}$ or in radial quantization ($D = H_\tau$) the `energy'.  
(Note that it is therefore just as natural to consider eigenstates of $D$ in a CFT as it is to consider energy eigenstates in regular quantum mechanics.)  

For a 2-dimensional CFT, we write the dilation operator in terms of the Virasoro operators as $D = L_0 + \bar L_0$. Using this Virasoro operator notation, eigenoperators $\phi^{(h,\bar h)}$ 
\ali{
	[L_0,\phi^{(h,\bar h)}(0)] = h \, \phi^{(h,\bar h)}, \qquad [\bar L_0,\phi^{(h,\bar h)}(0)] = \bar h \, \phi^{(h,\bar h)}
}
are labeled by the eigenvalues $h$ and $\bar h$, called the \emph{conformal weights} of $\phi$. The field is implicitly or explicitly labeled by these values, and their sum gives the conformal dimension $\Delta = h + \bar h$. 

The remaining $L_n$ Virasoro operators are raising and lowering operators: $L_n, \bar L_n$ $(n>0)$ lower the eigenvalue by $n$ units 
\ali{
	L_0 L_n \phi(0) = (h-n) L_n \phi(0), \qquad \bar L_0 \bar L_n \phi(0) = (\bar h- n) \bar L_n \phi(0)
}
and $L_{-n}, \bar L_{-n}$ $(n>0)$ raise the eigenvalue by $n$ units 
\ali{
	L_0 L_{-n} \phi(0) = (h+n) L_n \phi(0), \qquad \bar L_0 \bar L_{-n} \phi(0) = (\bar h+ n) \bar L_{-n} \phi(0). 
}
That is, the $(L_n\phi^{(h,\bar h)})$ operator has an `energy' that is $n$ units smaller than $h$. The notation in the lines above assumes there is an implicit action left and right on the vacuum state $|0\rangle$ so that the action of $L_0$ applied after the action of $L_n$, namely $[L_0,[L_n,\phi]]$, can be written as $L_0 L_n \phi$. (See a bit further below for the definition of the vacuum state. At this point it is just a matter of shortening the notation.) 

We use the lowering operator to define a \emph{Virasoro primary} and the raising operators to define \emph{Virasoro descendants}. The idea is that the Virasoro primary is a field whose `energy' eigenvalue cannot be lowered furthered, hence it is annihilated by all the lowering operators 
\ali{
	\text{Virasoro primary } \phi: \qquad [L_n,\phi(0)] = 0, \qquad [\bar L_n,\phi(0)] = 0 \qquad (n>0).  \label{Virprim}
}
In particular, these conditions include the conditions $[L_1,\phi(0)] = 0 = [\bar L_1,\phi(0)]$ and thus $[K_\mu, \phi(0)]=0$, which is the defining condition for a primary field in \eqref{primop}. That is, a \emph{Virasoro primary} is also a primary. But a primary is not a Virasoro primary, because the number of conditions in \eqref{Virprim} is much larger. The distinction between the $d=2$ concept of Virasoro primary field and global conformal concept of primary field is sometimes made by referring to what we call the primary as `global primary field' or `quasi-primary'. 

Now we can build operators that have higher dimension by acting with raising operators. This constructs the descendants 
\ali{
	\text{Virasoro descendants of }\phi: \qquad  (L_{-1})^n \phi(0), \quad L_{-n} \phi(0)  
}
having an additional `energy' $n$. For example, first descendants $L_{-1}\phi, \bar L_{-1}\phi$ have conformal weights $h+1, \bar h+1$ thus conformal dimension $\Delta +1$. Second descendants $(L_{-1})^2 \phi, (\bar L_{-1})^2 \phi, L_{-1} \bar L_{-1} \phi, L_{-2}\phi, \bar L_{-2}\phi$ have $\Delta +2$, etc. The Virasoro descendants contain the global descendants as defined in \eqref{globaldesc} (constructed by the action of $L_{-1},\bar L_{-1}$ or the momentum operator) but also new types of descendants (constructed by the action of other $L_{-n}$).

For completeness, we give some other (equivalent) definitions of the Virasoro primary. One is in terms of its finite conformal transformation behavior 
\ali{
	\phi'(z',\bar z') = \left(\frac{\p z'}{\p z}\right)^{-h} \left(\frac{\p \bar z'}{\p \bar z}\right)^{-\bar h} \phi(z,\bar z) 
} 
under all $z \ra f(z), \bar z \ra \bar f(\bar z)$. This includes the global conformal transformations $z \ra \frac{a z+b}{c z + d}, \bar z \ra \frac{\bar a \bar z + \bar b}{\bar c \bar z + \bar d}$ as a subset. The Virasoro primary has conformal weights $h = \frac{\Delta +s}{2}$, $\bar h = \frac{\Delta -s}{2}$, and conformal dimension or `energy' $\Delta = h + \bar h$ (the eigenvalue of $D = L_0 + \bar L_0$) and spin $s = h - \bar h$ (the eigenvalue of $M_{xy} = i (L_0 - \bar L_0)$, for $z=x+i y$).     
 
Finally, the Virasoro primary can be defined as a field $\phi^{(h,\bar h)}$  with the following $T \phi $ OPE: 
\ali{
	T_{zz}(z) \phi^{(h,\bar h)}(w,\bar w) = \frac{h \, \phi^{(h,\bar h)}(w,\bar w)}{(z-w)^2} + \frac{\p_w \phi^{(h,\bar h)}(w,\bar w)}{z-w} + \cdots \\
	T_{\bar z\bar z}(\bar z) \phi^{(h,\bar h)}(w,\bar w) = \frac{\bar h \, \phi^{(h,\bar h)}(w,\bar w)}{(\bar z-\bar w)^2} + \frac{\p_{\bar w} \phi^{(h,\bar h)}(w,\bar w)}{\bar z-\bar w} + \cdots .  
} 
Indeed, the $T \phi$ OPE by the Cauchy formulas contains the same information as the transformation properties of $\phi$ under conformal transformations. Therefore $\delta_\epsilon \phi = [Q_\epsilon, \phi] = \oint \cdots T \phi$ all provide different ways of packaging the same info (infinitesimal transformation behavior).

\subsection{Virasoro primary and descendant states, Virasoro multiplet} 

By the state-operator correspondence, which is CFT-specific, the organizational structure of the field operator content translates directly into a corresponding organizational structure of the states. 

The \emph{vacuum state} $|0\rangle$ is defined as 
\ali{
	L_n |0\rangle = 0 = \bar L_n |0\rangle \qquad (n \geq 1) .  
}
This includes the global conformal vacuum condition $L_{\pm 1,0}|0\rangle = 0$ as well as other conditions that ensure zero energy $\vev{T_{zz}} = 0 = \vev{T_{\bar z\bar z}}$ in the vacuum state, with $T_{zz}(z)|0\rangle = \sum_{n=-\infty}^\infty \frac{L_n}{z^{n+2}} |0\rangle$ and $T_{\bar z\bar z}(\bar z)|0\rangle = \sum_{n=-\infty}^\infty \frac{\bar L_n}{\bar z^{n+2}} |0\rangle$ well-defined, 
using that the definition also implies $\langle 0|L_n^\dagger = \langle 0|L_{-n} = 0$ $(n \geq -1)$.  

An \emph{asymptotic state} is defined as the state associated with a Virasoro primary operator (state-operator correspondence) as 
\ali{
	\phi^{(h,\bar h)}(0)|0\rangle = |\phi^{(h,\bar h)}\rangle.  
}
This state is alternatively denoted as $|h,\bar h\rangle$, in a closer notation to $|\Delta \rangle$ in section \ref{sectconfcasimiroa} before: 
\ali{
	|h,\bar h\rangle = \lim_{z,\bar z \ra 0} \phi^{(h,\bar h)}(z,\bar z) |0\rangle . 
} 
The asymptotic state is an eigenstate 
\ali{
	L_0 |h,\bar h\rangle = h |h,\bar h\rangle, \qquad \bar L_0 |h,\bar h\rangle = \bar h |h,\bar h\rangle. 
}

The definition of the Virasoro primary operator \eqref{Virprim}, acting on the vacuum state, together with the definition of an asymptotic state, naturally imposes 
the definition of a \emph{Virasoro primary state} 
\ali{
	L_n |\psi \rangle = 0 = \bar L_n |\psi \rangle \qquad (n > 0).  
}
A given Virasoro primary state $|\psi \rangle$ carries with it a collection of Virasoro descendant states, e.g.~$L_{-1}|\psi \rangle = |\p \psi \rangle$ with $h+1$ eigenvalue, $(L_{-1})^2|\psi \rangle, L_{-2}|\psi \rangle$ with $h+2$, etc. This collection of states is called the Virasoro multiplet or Verma module of the state, labeled by $(h,\bar h)$.


\begin{comment} 
\ali{
	T'_\mn(x) = T_{\alpha\beta}(x'(x)) \frac{\p x'^\alpha}{\p x^\mu}(x) \frac{\p x'^\beta}{\p x^\nu}(x) 
}
in an active picture, or 
\ali{
	T'_\mn(x') = T_{\alpha\beta}(x(x')) \frac{\p x^\alpha}{\p x'^\mu}(x') \frac{\p x^\beta}{\p x'^\nu}(x') 
}
in a passive picture. 
is a spin $2$ field because it is a 2-tensor (having 2 spacetime indices)
\end{comment}


\newpage
\section*{Exercises}

	\textbf{\underline{\smash{Exercise 1. Finite special conformal transformations}}} 

a) Given the generator $k_\mu = 2 x_\mu (x \cdot \p) - x^2 \p_\mu $ of special conformal transformations, calculate the action of a finite special conformal transformation. 

\textit{Tip: Perform an inversion $y^\mu = x^\mu/x^2$.}	

\begin{comment}

SOLUTION: Seminar slides Julian p.14/40, but with notation small $k$ for the generator. 

\end{comment}

b) 
Using $y^\mu$, what is the interpretation of a finite special  conformal transformation in terms of more familiar transformations? \\

\begin{comment}

SOLUTION: SCT = translation, preceded and followed by an inversion. [Di Fran (4.17)] 

\end{comment}

\textbf{\underline{\smash{Exercise 2. Conformal group}}} 

Count the number of conformal generators ($t_a$ in the notes). This should be equal to the amount of generators of the group $SO(p+1,q+1)$. \\

\begin{comment}

SOLUTION: Ginsparg p.5 under point d). 
\frac{1}{2} (p+q+1)(p+q+2)  

\end{comment} 

\textbf{\underline{\smash{Exercise 3. Conformal Killing equation}}} 

Determine the function $\cdots$ in the conformal Killing equation 
\ali{ 
	\p_\mu \epsilon_\nu + \p_\nu \epsilon_\mu = \cdots  g_\mn . 
} 

Determine $\Omega(x)$ for the different conformal transformations $\epsilon$. 
\\*
\\*
\\*
\textbf{\underline{\smash{Exercise 4. Conformal casimir}}} 

What is the quadratic casimir of the conformal group? 

\textit{Tip: Use a solution of the previous Exercise sheet on Poincar\'e invariance.}	
\\*
\\*
\\*
\textbf{\underline{\smash{Exercise 5. Conformal multiplets $=$ irreducible representations of conformal group}}} 

The conformal casimir can be written as 
\ali{
	c_2^{conf} = \frac{1}{2} M_{\mn} M^\mn + D(D-d) - P_\mu K^\mu.   
	} 
Use this form of the casimir to read off the action of $c_2^{conf}$ on a primary, scalar field $\mathcal O$: 
\ali{
	[c_2^{conf}, \mathcal O(0)]|0\rangle = ? 
} 
Use the state-operator correspondence 
to rewrite your result in terms of the casimir working on the corresponding \emph{state}: 
\ali{
	c_2^{conf} |\Delta \rangle = ? 
}
Now act with the raising operator $P_\mu$ on $|\Delta \rangle$ to create the state $|\Delta + 1 \rangle$. Work out the action of the casimir on this state. 
You could have found this result without explicitly doing the calculation by using which property of the casimir? 

We can create the basis of a conformal multiplet 
\ali{
	\{ |\Delta \rangle, |\Delta + 1\rangle, |\Delta + 2 \rangle, \cdots \}  
}
by successive action of the momentum operator. 
What is the action of $c_2^{conf}$ on any state in this conformal multiplet?  
Does the state corresponding to the operator $\mathcal O(x)$ belong to this multiplet? What is $[c_2^{conf}, \mathcal O(x)]$? 

We have built here a lowest-weight representation of the conformal group in a way that is very analogous to the construction of the highest-weight representation of the rotation group in Ex.3 on Poincar\'e invariance. In this analogy, the conformal generators $D, P_\mu $ and $K_\mu$ take the role of which rotation generators? 
\\*
\\*
\\*
\textbf{\underline{\smash{Exercise 6. Equivalent definitions of Virasoro primary operator.}}}

Show that a field $\phi^{(h,\bar h)}(w,\bar w)$ that satisfies the following $T \phi$ OPE 
\ali{
	T_{zz}(z) \phi^{(h,\bar h)}(w,\bar w) &= \frac{h \, \phi^{(h,\bar h)}(w,\bar w)}{(z-w)^2} + \frac{\p_w \phi^{(h,\bar h)}(w,\bar w)}{z-w} + \cdots \nonumber \\ 
	T_{\bar z\bar z}(\bar z) \phi^{(h,\bar h)}(w,\bar w) &= \frac{\bar h \,  \phi^{(h,\bar h)}(w,\bar w)}{(\bar z-\bar w)^2} + \frac{\p_w \phi^{(h,\bar h)}(w,\bar w)}{\bar z-\bar w} + \cdots \nonumber 
	}
transforms under conformal transformations as 
\ali{
	[L_n, \phi^{(h,\bar h)}(w,\bar w)] = \left(w^{n+1} \p_w + (n+1) w^n h \right) \phi^{(h,\bar h)}(w,\bar w), 
}
which is the transformation behavior of a Virasoro primary field \\   $\phi^{(h,\bar h)}(w,\bar w) = e^{w L_{-1} + \bar w \bar L_{-1}} \phi^{(h,\bar h)}(0,0) e^{-w L_{-1} - \bar w \bar L_{-1}}$.  
\\*
\\*
\\*
\textbf{\underline{\smash{Exercise 7. Action of conformal generators on descendant}}} 

In a 2D Euclidean CFT, calculate the action of $L_1$ on the first descendant of the primary field $\phi(z)$, i.e. calculate $[L_1,\partial_z\phi(z)]$.
\\*
\\*
\\*
\textbf{\underline{\smash{Exercise 8. Descendant of the identity operator.}}} 

Use \eqref{Lnoint} to calculate the descendant of the unit operator $L_{-2} \hat 1$. Formulate a conclusion. 
\\*
\\*
\\*
\textbf{\underline{\smash{Exercise 9. Schwarzian derivative.}}} 

Use Mathematica to check that the Schwarzian derivative of a M\"obius transformation vanishes.

\chapter{Anti-de Sitter geometry} 

We have only considered flat spaces so far, $g_\mn = \eta_\mn$ or $\delta_\mn$ of $\mathbb R^d$. In this chapter, we discuss more general, curved geometries and their symmetries. We start with the sphere $S^2$.

\section{Sphere $S^2$} 

The sphere $S^2$ is a 
2-dimensional curved geometry that can be defined as a particular surface in $\mathbb R^3$. 
Namely, the points in $\mathbb R^3$ that satisfy 
\ali{
	X^2 + Y^2 + Z^2 = L^2   \label{embS2}
}
lie on a 2D surface embedded in $\mathbb R^3$, which is the sphere of radius $L$. We will refer to this as an `embedding definition' of $S^2$, and $\mathbb R^3$ as the embedding space. 

The flat metric on embedding space is 
\ali{ 
	ds^2_{emb} = dX^2 + dY^2 + dZ^2 = g_{ab}^{emb} dX^a dX^b  \label{ds2emb}
} 
in terms of flat space coordinates 
\ali{
	X^a = (X,Y,Z) \qquad (a = 1,2,3). 
}
The scalar product or dot product in different notations is 
\ali{
	X^2 = X \cdot X = X_a X^a = g_{ab}^{emb} X^a X^b. 
}
The dot product is per definition invariant under rotations: 
$X'_b = R_{ba} X_a$, then $X'_b X'^b = X_a X^a$ if $R^T R = \mathbf 1$ (with transpose $(R^T)_{ab} = R_{ba}$), i.e.~if $R \in O(3)$ is a $3\times 3$ orthogonal matrix, element of the orthogonal group ($(\det R)^2 = 1$), and $R \in $ rotation group $SO(3)$ if $\det R = 1$. 

The embedding definition of the sphere can be written in shorter notation as 
\ali{
	X_a X^a = L^2. 
}
$SO(3)$ transformations leave the $\mathbb R^3$ dot product and thus the $S^2$ definition invariant. It follows that $SO(3)$ is the symmetry group or isometry group of $S^2$, or said otherwise that the three $SO(3)$ generators 
\ali{ 
	m_{ab} = X_b \p_a - X_a \p_b \qquad  (a,b=1,2,3) 
}
are \emph{Killing vectors} of the sphere. 


We'd prefer to use only 2 intrinsic coordinates to describe the 2-dimensional sphere, rather than 3 embedding coordinates satisfying the additional constraint \eqref{embS2}. To this end we find a parametrization for which \eqref{embS2} is automatically satisfied 
\ali{
	X &= L \sin \theta \cos \phi \\
	Y &= L \sin \theta \sin \phi \\ 
	Z &= L \cos \phi . 
} 
This defines the intrinsic coordinates 
\ali{
	x^\alpha = (\theta,\phi)  \qquad (\alpha=1,2) . 
}
These are of course just the polar and azimuthal angle. The parametrization $X^a(x^\alpha)$, with $X^2 + Y^2 = L^2 \sin^2 \theta$, is such that $\theta$ and $\phi$ have the interpretation of describing orbits of radius $L \sin \theta$ and angle $\phi$ in the $XY$ plane of $\mathbb R^3$.  

The induced metric on the sphere is then 
\ali{
	ds^2 &= L^2 (d\theta^2 + \sin^2 \theta \,  d\phi^2) = g_{\alpha\beta}(x) dx^\alpha dx^\beta  \label{metricS2}
}
with range $\theta \in (0,\pi)$ and $\phi \in (0,2\pi)$. The metric of the sphere $g_{\alpha\beta}(x)$ depends on the coordinates, indicating it is curved. With this metric given, you can compute properties of the geometry such as its curvature. 

\paragraph{Some general relativity concepts.}

Using the general relativity formulas\footnote{Here and further in the course, the go to reference for general relativity concepts and definitions is \cite{Poisson}.} for respectively the Ricci scalar $R$, Ricci tensor $R_\ab$, Riemann tensor $R^\alpha_{\phantom{\mu}\beta \gamma\delta}$ and connection $\Gamma^\alpha_{\beta\gamma}$, 
\ali{
	R &= R^\alpha_{\phantom{\alpha}\alpha}, \qquad R_{\alpha\beta} = R^\mu_{\phantom{\mu}\alpha \mu\beta} \\
	R^\alpha_{\phantom{\mu}\beta \gamma\delta} &= \Gamma^\alpha_{\beta \delta,\gamma} -  \Gamma^\alpha_{\beta \gamma,\delta} + \Gamma^\alpha_{\mu\gamma}\Gamma^\mu_{\beta \delta} - \Gamma^\alpha_{\mu\delta}\Gamma^\mu_{\beta \gamma} \\
	\Gamma^\alpha_{\beta\gamma}&= \frac{1}{2} g^{\alpha\mu} \left(g_{\mu\beta,\gamma} + g_{\mu\gamma,\beta} - g_{\beta\gamma,\mu} \right) , 
}
you can check that the sphere 
has constant positive curvature 
\ali{
	R = \frac{2}{L^2}
}
and is a solution of the vacuum Einstein equations 
\ali{ 
	G_\ab \equiv R_\ab - \frac{1}{2} R g_\ab = 0 
	}
with $G_\ab$ the Einstein tensor. 

To find the symmetry of a given geometry, we ask what are the coordinate transformations 
\ali{
	x^\al \ra x'^\al = x^\al + \epsilon^\al(x) 
}
that keep the metric invariant 
\ali{ 
	g'_\ab(x) = g_\ab(x)
} 
with the transformed metric given by 
\ali{
	g'_\ab(x) = g_{\rho\sigma}(x'(x)) \frac{\p x'^\rho}{\p x^\al}(x) \frac{\p x'^\sigma}{\p x^\beta}(x) = g_\ab(x) + \delta_\epsilon g_\ab(x) + \mathcal O(\epsilon^2) 
} 
with
\ali{
	\delta_\epsilon g_\ab = \nabla_\alpha \epsilon_\beta + \nabla_\beta \epsilon_\al . 
}
Here, the covariant derivative is defined as $\nabla_\rho p_\sigma = \p_\rho p_\sigma - \Gamma^\alpha_{\rho\sigma} p_\alpha$. 
Solutions of the Killing equation 
\ali{
	\nabla_\alpha \epsilon_\beta + \nabla_\beta \epsilon_\al = 0 \qquad \text{(Killing equation)}
}
are the Killing vectors $\epsilon^\al(x)$. They generate the isometries of $g_\ab(x)$. We know the Killing vectors of the sphere in embedding coordinates and can also write them in terms of the intrinsic coordinates (see \textit{Exercise}). Could there have been more intrinsic symmetries that the embedding definition missed? (There cannot be less Killing vectors than $m_{XY}, m_{XZ}$ and $m_{YZ}$, but could there be more?) The answer is no, because a 2D manifold can have maximum 3 isometries. 

More generally, an $n$D manifold can have maximum $\frac{n(n+1)}{2}$ isometries. For example, flat space $\mathbb R^n$ has $\frac{n(n+1)}{2}$ isometries: $n$ translations and $\frac{1}{2}n(n-1)$ rotations. A manifold with maximum possible symmetry is a \emph{maximally symmetric space MSS}. All the geometries discussed in this chapter will be MSS's. Their properties are constant curvature and $R_\mn = R g_\mn/n$. 

\section{Hyperboloid} 

The 2D surface defined by the embedding definition 
\ali{
	X^2 + Y^2 - Z^2 = L^2 
}
is a 1-sheeted hyperboloid,  
and 
\ali{
	X^2 + Y^2 - Z^2 = -L^2, \qquad Z > 0  
}
is one half $H_2$ of a 2-sheeted hyperboloid. 

This time we cannot rewrite the left hand side of these definitions as a dot product in the embedding space \eqref{ds2emb}, which was very helpful in the case of the sphere to discuss the isometry group. Therefore, we can instead consider a Lorentzian embedding space 
\ali{ 
	ds^2_{emb} = dX^2 + dY^2 - dT^2 = g_{ab}^{emb} dX^a dX^b  \label{ds2embLor}
} 
and define the hyperboloids as 
\ali{
	X^2 + Y^2 - T^2 &= L^2  \qquad (\text{1-sheeted hyperboloid}) \\
	X^2 + Y^2 - T^2 &= -L^2, \qquad T > 0 \qquad (H_2) 
}
which can be rewritten 
\ali{
	X_a X^a  &= L^2  \qquad (\text{1-sheeted hyperboloid}) \\
	X_a X^a &= -L^2, \qquad T > 0 \qquad (H_2).  
}
It can be shown that the hyperboloid surfaces defined in this way are still Euclidean surfaces. 
Using the Lorentzian embedding space now allows to argue that the isometry group of the hyperboloid is the group that leaves the dot product of the embedding space \eqref{ds2embLor} invariant: the isometry group of $H_2$ is $SO(2,1)$, or of $H_n$ is $SO(n,1)$. 

The hyperboloids have constant negative curvature.

\section{$S^n$, $H_n$, dS$_n$ and AdS$_n$}

Summarizing the previous sections, and generalizing at the same time to higher dimensions, an $n$-sphere $S^n$ and hyperboloid $H_n$ are defined respectively as: 
\ali{
	S^n: \qquad &X_1^2 + X_2^2 + \cdots + X_{n+1}^2 = L^2 \\ 
	\text{in  } &\mathbb R^{n+1} \text{  with  } ds^2_{emb} = dX_1^2 + dX_2^2 + \cdots + dX_{n+1}^2 
}
with the isometry group of $S^n$ equal to the rotation group $SO(n+1)$ of $\mathbb R^{n+1}$; 
and 
\ali{
	H_n: \qquad &X_1^2 + \cdots + X_{n}^2 - U^2 = -L^2, \quad U > 0  \\ 
	\text{in  } &\mathbb R^{n,1} \text{  with  } ds^2_{emb} = dX_1^2  + \cdots + dX_{n}^2 - dU^2 
}
with the isometry group of $H_n$ equal to the Lorentz group $SO(n,1)$ of $\mathbb R^{n,1}$. 

An $n$-dimensional \emph{de Sitter} space dS$_n$ is now defined as the Lorentzian version of a sphere, meaning the Lorentzian spacetime of constant positive curvature. Its embedding definition is  
\ali{
	\text{dS}_n: \qquad &X_1^2 + \cdots + X_{n}^2 - U^2 = L^2 \\ 
	\text{in  } &\mathbb R^{n,1} \text{  with  } ds^2_{emb} = dX_1^2  + \cdots + dX_{n}^2 - dU^2 
}
with the isometry group of dS$_n$ equal to the Lorentz group $SO(n,1)$ of $\mathbb R^{n,1}$. De Sitter space has constant positive curvature. 

An $n$-dimensional \emph{Anti de Sitter} space AdS$_n$ is defined as the Lorentzian version of a hyperboloid, meaning the Lorentzian spacetime of constant negative curvature. Its embedding definition is 
\ali{
	\text{AdS}_n: \qquad &X_1^2 + \cdots + X_{n-1}^2 - U^2 - V^2 = -L^2 \label{AdSembdef} \\ 
	\text{in  } &\mathbb R^{n-1,2} \text{  with  } ds^2_{emb} = dX_1^2  + \cdots + dX_{n-1}^2 - dU^2 - dV^2 
}
with the isometry group of AdS$_n$ equal to the Lorentz group $SO(n-1,2)$ of $\mathbb R^{n-1,2}$. Anti de Sitter space has constant negative curvature.

\section{Symmetry of AdS$_{d+1}$} \label{sectionAdSsymm}

Let us repeat the embedding space definition of $(d+1)$-dimensional AdS$_{d+1}$ with AdS radius $\ell$ in embedding space $\mathbb R^{d,2}$: 
\ali{
	\text{AdS}_{d+1}: \qquad &X_1^2 + \cdots + X_{d}^2 - U^2 - V^2 = -\ell^2. 
}
The embedding space definition is kept invariant under $SO(d,2)$ transformations, and therefore the isometry group of AdS$_{d+1}$ is $SO(d,2)$. 

Euclidean AdS space EAdS$_{d+1}$ is defined as (in embedding space $\mathbb R^{d+1,1}$)
\ali{
	\text{EAdS}_{d+1}: \qquad &X_1^2 + \cdots + X_{d}^2 + X_{d+1}^2 - U^2 = -\ell^2, \qquad U > 0 \label{EAdS3embedding}
}
(or EAdS$_n = H_n$). The isometry group of EAdS$_{d+1}$ is $SO(d+1,1)$. 

\fbox{ \parbox{0.97\textwidth}{ We find that the isometry group $SO(d,2)$ of AdS$_{d+1}$ equals the conformal group of a $d$-dimensional CFT, CFT$_d$. And the isometry group $SO(d+1,1)$ of EAdS$_{d+1}$ equals the conformal group of a $d$-dimensional Euclidean CFT, ECFT$_d$. This is a first hint at the relation between  AdS$_{d+1}$ and CFT$_d$. } }

\section{AdS$_3$} \label{sectAdS3}

References: Kaplan lectures \cite{Kaplan lectures}. 

\paragraph{Intrinsic coordinates: global or Poincar\'e}

Embedding definition in embedding space $\mathbb R^{2,2}$ 
\ali{
	-U^2 - V^2 + X^2 + Y^2 = -\ell^2.  \label{embdefAdS3}
}

Intrinsic coordinates: global AdS coordinates $(\tilde \rho, t,\phi)$ or Poincar\'e AdS coordinates $(Z, t_P,x)$: 
\ali{
	U &= \ell \cosh \frac{\tilde \rho}{\ell} \cos \frac{t}{\ell} = \frac{\ell^2 + Z^2 + x^2 - t^2_P}{2Z}\\ 
	V &= \ell \cosh \frac{\tilde \rho}{\ell} \sin \frac{t}{\ell} = \ell \frac{t_P}{Z}\\
	X &= \ell \sinh \frac{\tilde \rho}{\ell} \cos \phi = \ell \frac{x}{Z} \\
	Y &= \ell \sinh \frac{\tilde \rho}{\ell} \sin \phi =  -\frac{\ell^2 - Z^2 - x^2 + t^2_P}{2Z} 
}
give induced metric on AdS$_3$ (often we will set the AdS radius $\ell$ to one for convenience) 
\ali{
	ds^2 = d\tilde \rho^2 - \cosh^2 \frac{\tilde \rho}{\ell} dt^2 + \ell^2 \sinh^2 \frac{\tilde \rho}{\ell} \,  d\phi^2 
}
or 
\ali{
	ds^2 = \frac{\ell^2}{Z^2} (dZ^2 + dx^2 - dt_P^2).  \label{AdS3Poinc}
}
The Poincar\'e parametrization should be thought of as an analytic continuation from the Euclidean parametrization of the EAdS$_3$ embedding coordinates in terms of Poincar\'e coordinates $(\tau_P, x,Z)$: 
\ali{
	X_1 &= \ell \frac{x}{Z} \\
	X_2 &= \ell \frac{\tau_P}{Z} \\ 
	X_3 &= \frac{\ell^2 - Z^2 - x^2 - \tau_P^2}{2Z} \\
	U &= \frac{\ell^2 + Z^2 + x^2 + \tau_P^2}{2Z} . \label{UPoincareEucl}
}

\begin{figure}
	\centering 
	\includegraphics[width=8cm]{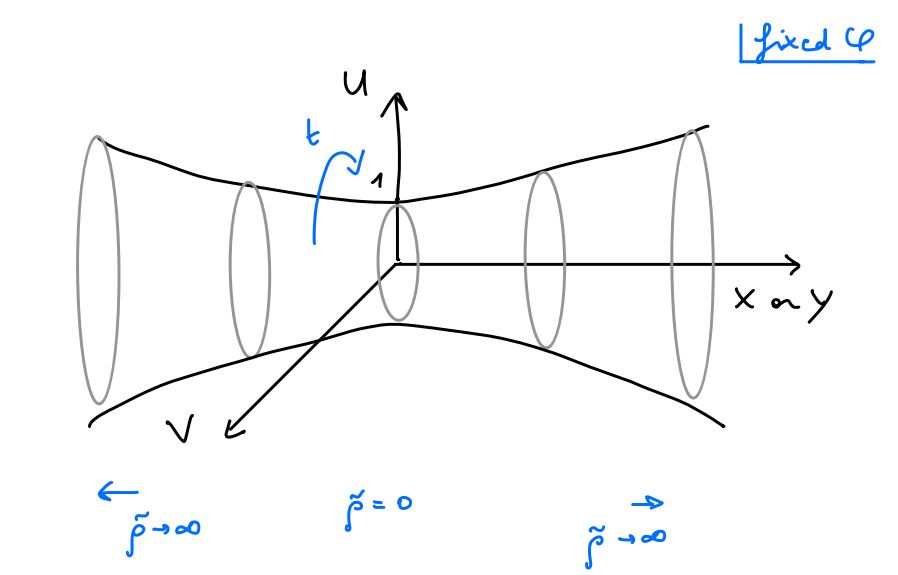} \qquad \qquad
	\includegraphics[width=6cm]{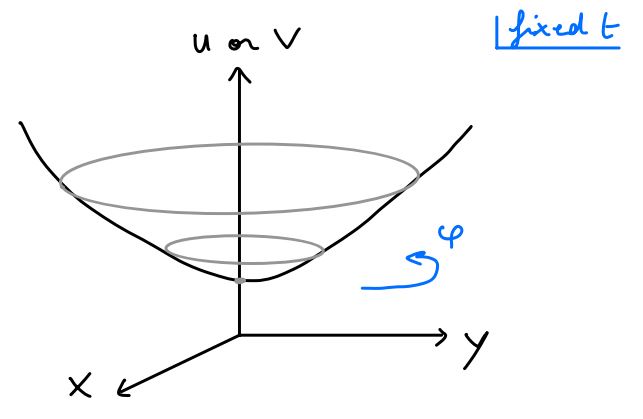}  
	\caption{$UV$ orbits (left) and $XY$ orbits (right) span hyperbolic AdS$_3$ surface in embedding space, showing resp.~$\p_t$ and $\p_\phi$ symmetries.  
	} \label{figAdS3emb} 
\end{figure}

Global coordinates describe $UV$ orbits of (exponentially growing) radius $\cosh \tilde \rho/\ell$ and angle $t/\ell$, and $XY$ orbits of (exponentially growing) radius $\sinh \tilde \rho/\ell$ and angle $\phi$. Another way of stating this is that 
these orbits are generated respectively by the rotation generators $m_{UV} = -\ell \, \p_t$ and $m_{XY} = -\p_\phi$. As such, the global coordinates reveal an $SO(2) \times SO(d)$ subgroup of the $SO(d=2,2)$ symmetry group of AdS$_3$. This is illustrated in figure \ref{figAdS3emb}. 

From the metric in Poincar\'e coordinates \eqref{AdS3Poinc}, we see that the Poincar\'e coordinates $x$ and $t_P$ span a flat space metric at any constant value of the Poincar\'e coordinate $Z$. In this way, these coordinates reveal a specific subgroup of the AdS$_3$ symmetry group $SO(d=2,2)$, namely the $d$-dimensional Poincar\'e group $SO(d)$ $\cup$ translations for $d=2$, with respective generators $m$ and $p$.   

We thus see that \emph{different choices of coordinates make different sets of AdS isometries manifest}.

\paragraph{AdS$_{d+1}$}
 
In general dimensions, AdS$_{d+1}$ in global vs Poincar\'e coordinates is given by 
\ali{
	ds^2 = d\tilde \rho^2 - \cosh^2 \frac{\tilde \rho}{\ell} dt^2 + \ell^2 \sinh^2 \frac{\tilde \rho}{\ell} \,  d\Omega^2_{d-1}  
}
and
\ali{
	ds^2 = \frac{\ell^2}{Z^2} (dZ^2 + \sum_{i=1}^{d-1} (dx^i)^2 - dt_P^2).  
}

AdS$_{d+1}$ is a maximally symmetric solution of Einstein's equations 
\ali{
	R_\ab - \frac{1}{2} R g_\ab + \Lambda g_\ab = 0 
}
with negative cosmological constant 
\ali{
	\Lambda = -\frac{d(d-1)}{2 \ell^2}.  
}
It has constant negative curvature 
\ali{
	R = -\frac{d(d+1)}{\ell^2}, 
}
Ricci tensor $R_\ab = -\frac{d}{\ell^2} g_\ab$ and $R_{\mn\ab} = -\frac{1}{\ell^2} (g_{\mu\al}g_{\nu\beta} - g_{\mu\beta} g_{\nu\al})$.

\paragraph{Ranges of the intrinsic coordinates} 

The parametrization of embedding in terms of intrinsic coordinates is such that the global coordinates describe $UV$ orbits or radius $\ell \cosh \frac{\tilde \rho}{\ell}$ and angle $\frac{t}{\ell}$, and $XY$ orbits of radius $\ell \sinh \frac{\tilde \rho}{\ell}$ and angle $\phi$. 
For the range 
\ali{
	0 < \tilde \rho < \infty 
	} 
the respective radii or the orbits are larger than $\ell$ ($UV$) and positive ($XY$). Further, $\phi$ is an angle with range 
\ali{
	0 < \phi < 2\pi 
}
and $t/\ell$ is an angle as well, 
\ali{
	0 < t < 2 \pi \ell.  
}
However, to avoid closed timelike curves, AdS is defined as the \emph{universal covering space} of the surface defined by \eqref{embdefAdS3}: you let $t$ run from $-\infty$ to $\infty$, and each time $t$ becomes a multiple of $2 \pi \ell$, you go to a new sheet or a new covering of the hyperbolic surface \eqref{embdefAdS3}. That is, AdS in global coordinates is 
\ali{
	ds^2 = d\tilde \rho^2 - \cosh^2 \frac{\tilde \rho}{\ell} dt^2 + \ell^2 \sinh^2 \frac{\tilde \rho}{\ell} \,  d\phi^2, \\
	0 < \tilde \rho < \infty, \quad 0 < \phi < 2\pi, \quad -\infty < t <  \infty. 
}
For the range of the Poincar\'e coordinates, in Euclidean signature we have $-\infty < x,y < \infty$ and $Z > 0$ so that $U$ in \eqref{UPoincareEucl} is positive \eqref{EAdS3embedding}. In Lorentzian signature, 
\ali{
	ds^2 = \frac{\ell^2}{Z^2} (dZ^2 + dx^2 - dt_P^2) \\ 
	-\infty < x, t_P < \infty, \qquad Z > 0.  
}
While in Euclidean signature, Poincar\'e and global coordinates cover the 
full AdS spacetime, it is important to note that in Lorentzian signature, Poincar\'e coordinates only cover a patch of the full AdS spacetime, called the \emph{Poincar\'e patch}.  
To see this, we can consider the coordinate transformation between global and Poincar\'e coordinates: 
\ali{
	t_P &= \frac{\sin t}{\cos t - \sin \phi \tanh \tilde \rho} \\
	Z &= \frac{1}{\cosh \tilde \rho \, (\cos t - \sin \phi \tanh \tilde \rho)}  \\ 
	x &= \frac{\cos \phi \, \tanh \tilde \rho}{\cos t - \sin \phi \tanh \tilde \rho} 
} 
From the expression $t_P(t,\phi,\rho)$ it is clear that $t_P$ will reach $\pm \infty$ at the surface $\cos t = \sin \phi \tanh \tilde \rho$ of finite global time $t$ points. This surface is called the \emph{Poincar\'e horizon} of the Poincar\'e patch of AdS.

Indeed, in Euclidean Poincar\'e coordinates 
\ali{
	\tau_P &= \frac{\sinh \tau}{\cosh \tau - \sin \phi \tanh \tilde \rho} \\
	Z &= \frac{1}{\cosh \tilde \rho \, (\cosh \tau - \sin \phi \tanh \tilde \rho)}  \\ 
	x &= \frac{\cos \phi \, \tanh \tilde \rho}{\cosh \tau - \sin \phi \tanh \tilde \rho}  
}
the denominator $\cosh \tau - \sin \phi \tanh \tilde \rho$ 
in terms of global Euclidean time $\tau$ does not have any zeros, so that the infinite range of Euclidean Poincar\'e time $\tau_P$ maps to an infinite range of Euclidean global time $\tau$ and there is no Poincar\'e horizon in Euclidean signature.  

An alternative Euclidean Poincar\'e to global transformation (with boundary metric in complex coordinates $d\tau_P^2 + dx^2 = dz d\bar z$) is given by \cite{Guica}  
\ali{
	Z = \frac{e^\tau}{\cosh \tilde \rho}, \qquad z = \tanh \tilde \rho \, e^{\tau + i \phi}, \qquad \bar z = \tanh \tilde \rho \, e^{\tau - i \phi} . 
}
It makes more clear 
that Poincar\'e (or also called `planar') to global reduces to the plane to cylinder transformation on the AdS boundary. 

Penrose diagram of AdS$_3$: 
\ali{ 
	\cosh \tilde \rho = \frac{1}{\cos \rho}
}
with finite range for $\rho$.

\section{Visualization of AdS$_3$: Penrose diagrams} 

The causal structure of a spacetime ($ds^2 >=<0$) is captured in a locally Minkowski spacetime by local lightcones of $45^{\circ}$ light rays. Light propagates along $ds^2=0$ so an overall conformal factor in $ds^2$ is irrelevant to light, and thus to the causal structure. This is exploited in constructing Penrose diagrams of spacetimes. The idea of Penrose diagrams is  
to represent the causal structure of a given, infinite spacetime in a \emph{finite} diagram by using a particular conformal coordinate transformation that brings infinity to a finite distance. It has the property that light rays are at $45^{\circ}$ paths. 

\paragraph{Penrose diagram of flat space $\mathbb R^{1,1}$} 

\ali{
	ds^2 &= -dt^2 + dx^2 \qquad (-\infty < t,x < \infty) \quad \text{(original coordinates)} \\
	&= \frac{-d\tilde t^2 + d\tilde x^2}{f(\tilde t,\tilde x)} \qquad (-\frac{\pi}{2} < \tilde t \pm \tilde x < \frac{\pi}{2}) \quad \text{(Penrose coordinates)} 
}
via the conformal transformation 
\ali{
	\tan (\tilde t \pm \tilde x) = t \pm x 
	}
(and thus 
$f(\tilde t,\tilde x) = \cos^2(\tilde t + \tilde x) \cos^2(\tilde t - \tilde x)$). The Penrose diagram of 2D Minkowski spacetime is a diamond.

\begin{figure}[h!!]
	\centering 
	\includegraphics[width=8cm]{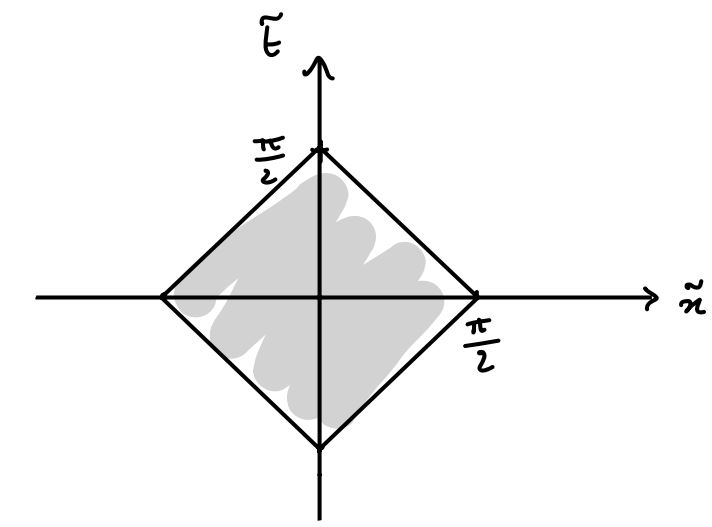} \qquad \qquad
	\includegraphics[width=6cm]{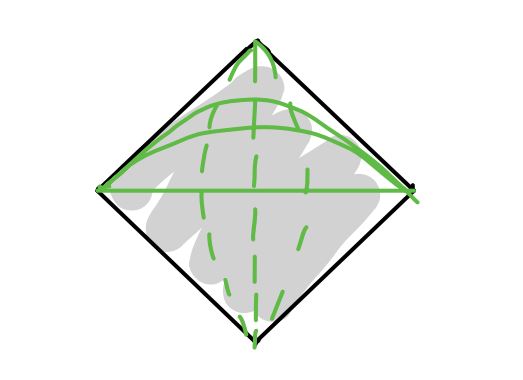}  \\
	\includegraphics[width=3cm]{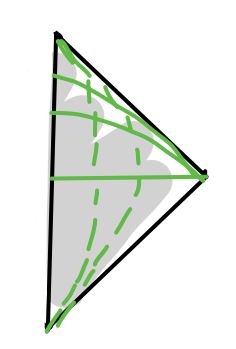} \\
	\includegraphics[width=7.5cm]{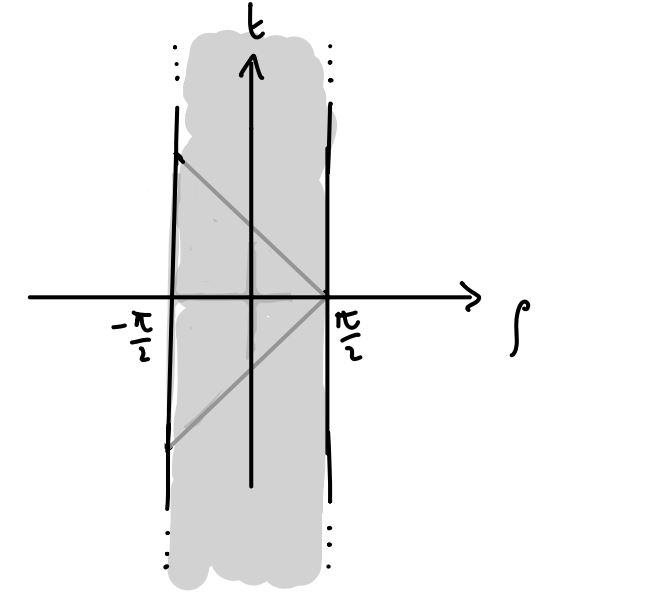} 
	\qquad \qquad  \qquad \includegraphics[width=2.7cm]{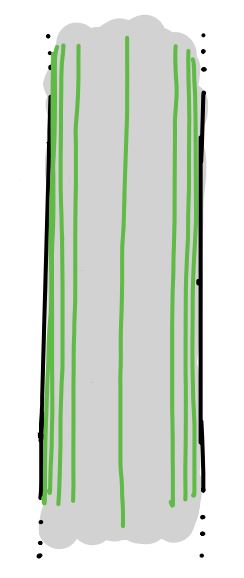} 
	\caption{
		First line: $\mathbb R^{1,1}$ Penrose diagram, first including Penrose coordinates as `help' coordinates, then more typically without axes because the spacetime is the grey region alone, and including constant $t$ lines (full green) and constant $x$ lines (dashed green). Middle line: Poincar\'e AdS$_2$ Penrose diagram, with constant $t_P$ (full green) and constant $Z$ (dashed green) lines and Poincar\'e horizon at $t_P \ra \infty$. It forms the Poincar\'e patch (marked with grey lines) of the full global AdS$_2$ Penrose diagram in the third line (with constant $\tilde \rho$ lines in full green).  } 
	\label{figPenrose}  
\end{figure}

\paragraph{Penrose diagram of Poincar\'e-AdS$_2$}

\ali{
	ds^2 &= \frac{1}{Z^2} (dZ^2 - dt_P^2) \qquad (Z>0, \, -\infty < t_P < \infty) 
}
is conformal to half of $R^{1,1}$ and thus it immediately follows that the Penrose diagram of Poincar\'e-AdS$_2$ is half a diamond. 

\paragraph{Penrose diagram of global-AdS$_2$} 

\ali{
	ds^2 &= d\tilde \rho^2 - \cosh^2 \tilde \rho dt^2 \qquad (-\infty < t,\tilde \rho < \infty) \quad \text{(original coordinates)} \\ 
	&= \frac{1}{\cos^2 \rho} (-dt^2 + d\rho^2) \qquad (-\infty < t < \infty, -\frac{\pi}{2} < \rho < \frac{\pi}{2}) \quad \text{(Penrose coordinates)} 
}
via the conformal transformation 
\ali{
	\cos \rho = \frac{1}{\cosh \tilde \rho}  \label{Penrosetransf}
}
($\sin \rho = \tanh \tilde \rho$ and $\tan \rho= \sinh \tilde \rho$). 

The Penrose diagram is then a strip. 
Note that in this case, only the radial coordinate is compactified by a conformal transformation, but the range of time $t$ is still infinite. 

\paragraph{Penrose diagram of global-AdS$_3$}

\ali{
	ds^2 &= d\tilde \rho^2 - \cosh^2 \tilde \rho dt^2 + \sinh^2 \tilde \rho \, d\phi^2 \qquad (-\infty < t < \infty, \, 0 < \tilde \rho < \infty, \, 0<\phi<2\pi) \quad \text{(original coordinates)} \\ 
	&= \frac{1}{\cos^2 \rho} (-dt^2 + d\rho^2 + \sin^2 \rho \,  d\phi^2) \qquad (-\infty < t < \infty, \, 0 < \rho < \frac{\pi}{2}, \, 0<\phi<2\pi) \quad \text{(Penrose coordinates)} 
}
via the conformal transformation \eqref{Penrosetransf}. 

\begin{figure}
	\centering \includegraphics[width=3.2cm]{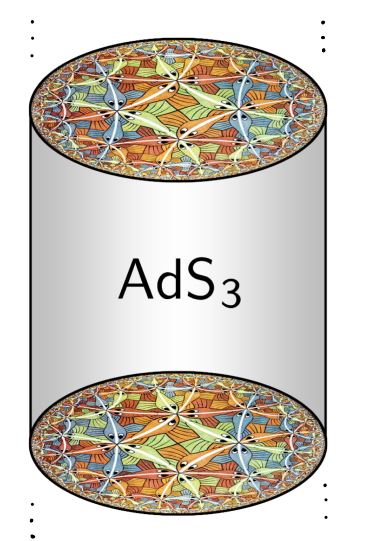} 
	 \qquad\qquad\qquad \includegraphics[width=2.8cm]{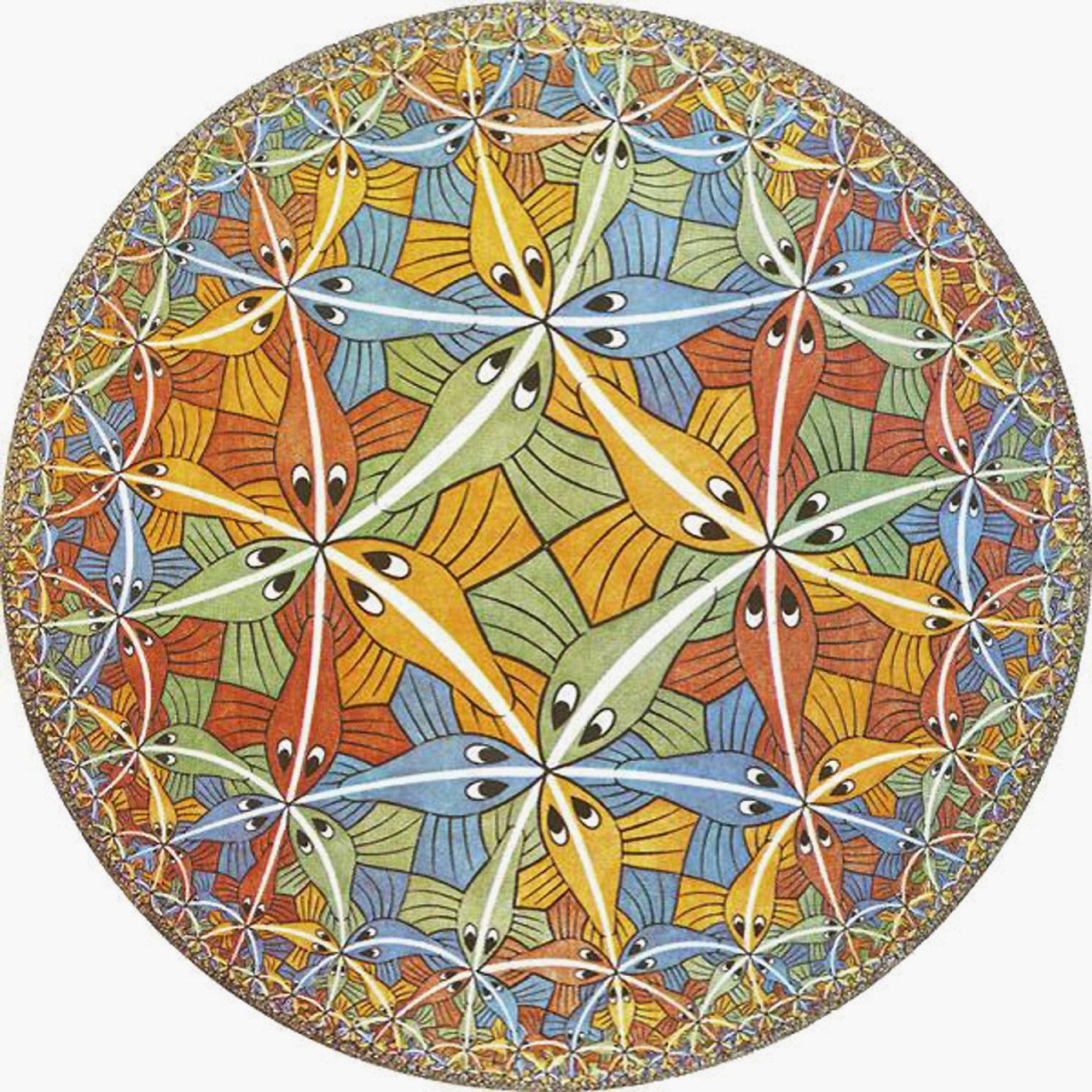}  \caption{a) Penrose diagram of global AdS$_3$ is a solid (infinite) cylinder b) Poincar\'e disk visualization of constant time slice.} \label{fig-AdScylinder} 
\end{figure}

The Penrose diagram is the solid cylinder traced out by a revolving global-AdS$_2$ Penrose strip, see figure \ref{fig-AdScylinder}a. 

\subsection{$H_2$ time slice}

A constant time slice of global-AdS$_3$ has the metric 
\ali{
	ds^2 = d\tilde \rho^2 + \sinh^2 \tilde \rho \, d\phi^2 = \frac{1}{\cos^2 \rho} (d\rho^2 + \sin^2 \rho \, d\phi^2) 
}
of a \emph{Poincar\'e disk}. 
The geometry of a Poincar\'e disk is famously pictured in the prints of Escher. One example is given in figure \ref{fig-AdScylinder}b. The fish getting ever smaller and closer together near the boundary is a reflection of the constant $\tilde \rho$ lines in the Penrose diagram. 
You should think of all the fish being of the same size, in a hyperbolic spacetime.  

A constant time slice of Poincar\'e-AdS$_3$ has the metric 
\ali{
	ds^2 = \frac{1}{Z^2}(dZ^2 + dx^2), \qquad (Z>0)
}
of an \emph{upper half-plane} $H_2$.

\subsection{Conformal boundary}  

The boundary of AdS$_3$ is at $\tilde \rho \ra \infty$, $\rho \ra \pi/2$ or $Z \ra 0$. The 2-dimensional metric at a constant value of the AdS radial direction approaching the boundary takes the form 
\ali{
	ds^2 = \underbrace{\frac{1}{\cos^2 \rho}}_{\text{infinite conformal factor}} \underbrace{(-dt^2 + d\phi^2)}_{\text{conformal boundary $\mathbb R \times S^1$}}, \qquad (-\infty< t < \infty, \, 0 < \phi < 2\pi)  
} 
or 
\ali{
	ds^2 = \underbrace{\frac{1}{Z^2}}_{\text{infinite conformal factor}} \underbrace{(-dt_P^2 + dx^2)}_{\text{conformal boundary $\mathbb R^{1,1}$}}, \qquad  (-\infty< t_P, x < \infty). 
}
Dropping the infinite conformal factor in the boundary metric, we obtain what is called the \emph{conformal boundary} of AdS. It has a flat metric. In particular, the conformal boundary of global AdS$_{d+1}$ is a cylinder $\mathbb R \times S_{d-1}$, while the conformal boundary of Poincar\'e AdS$_{d+1}$ is a plane $\mathbb R^{1,d-1}$.

\newpage
\section*{Exercises}

\textbf{\underline{\smash{Exercise 1. Curved space with Mathematica: sphere and AdS$_3$}}} 

Given the metric of the sphere $ds^2 = L^2 (d\theta^2 + \sin^2 \theta \, d\phi^2)$, 
use the RGTC package in Mathematica to compute: the scalar curvature $R$ and the Einstein tensor (i.e.~check which Einstein equation is satisfied by the sphere solution).  

Same question for AdS$_3$, with metric $ds^2 = \frac{\ell^2}{Z^2} (dZ^2 + dz d\bar z)$.  	\\

\textbf{\underline{\smash{Exercise 2. Killing vectors of sphere}}} 

In class, we wrote the Killing vector $\epsilon$ for the sphere in embedding coordinates $X^a = (X,Y,Z)$,  
\ali{
	\epsilon = c_{XY} \, m_{XY} + c_{XZ} \, m_{XZ} + c_{YZ} \, m_{YZ}.  
}	
Write it in terms of the intrinsic sphere coordinates $x^\alpha = (\theta,\phi)$, and then check that it indeed solves the Killing equation 
\ali{
	\nabla_\alpha \epsilon_\beta + \nabla_\beta \epsilon_\alpha = 0.  
} 

Use Mathematica (RGTC) to check your results. \\


\textbf{\underline{\smash{Exercise 3. Conformal compactification}}} 

In the 2D CFT chapter, we often used the Riemann sphere $S^2$ to describe the complex plane $\mathbb C$ with the point at infinity included, $\mathbb C \cup \infty = S^2$. 
(We could think of the inclusion of infinity as a result of stereographic projection.) In fact, this gives us an example of the \emph{conformal compactification} we use in Penrose diagrams, as we will see in this exercise. 

We start from the plane $\mathbb R^{p+1}$ (compared to the introductory paragraph, we look at the general dimensional case, in real signature now) 
\ali{
	ds^2 = dr^2 + r^2 d\Omega_p^2 
} 	
with $d\Omega_p^2$ the metric on the $p$-sphere. Define a new coordinate $\theta$ through the transformation 
\ali{
	r = \frac{1}{\tan \frac{\theta}{2}}. 
}
Work out the coordinate transformation, what is the new metric $ds'^2$? 

In what sense is the new metric compactified \`a la Penrose? Discuss the range of $r$ and $\theta$. \\

\textbf{\underline{\smash{Exercise 4. Penrose diagram of flat space}}} 

Use Mathematica to draw some constant $t$ and $x$ lines in the Penrose diagram of flat space $ds^2 = -dt^2 + dx^2$. (Use the command ParametricPlot.) \\	

\newpage

\textbf{\underline{\smash{Exercise 5. AdS$_3$ metrics}}} 

Use the coordinate transformations between embedding, global and Poincar\'e AdS coordinates to verify the different metric representations of AdS$_3$. You can use Mathematica for this exercise. 	
\\*

\textbf{\underline{\smash{Exercise 6. Poincar\'e to global from the boundary to the bulk}}}

Find the Poincar\'e to global coordinate transformation of Euclidean AdS$_3$ 

\ali{
	ds^2_{Poincare} &= \frac{1}{Z^2} (dZ^2 + dz d \bar z)  \nonumber \\
	ds^2_{global} &= \frac{1}{\cos^2 \rho} \left(d\rho^2 + d\tau^2 + \sin^2 \rho \, d\phi^2 \right)  \nonumber 
}

by starting from the plane to cylinder transformation in 2D CFT, and extending it into the bulk. 
Start by comparing the conformal factors in the AdS boundaries in Poincar\'e and global coordinates. 
\\*

\textbf{\underline{\smash{Exercise 7. Killing vectors of AdS$_3$}}} 

Consider AdS$_3$ in embedding coordinates $(U,V,X,Y)$ and global coordinates $(\tilde \rho, t, \phi)$. 
Show that the generator of rotations in the $UV$ plane is equal to the Hamiltonian in global coordinates. 

Next, consider Euclidean AdS$_3$ in global coordinates $(\tilde \rho, \tau, \phi)$ and Poincar\'e coordinates $(Z, \tau_P, x)$, related by the transformation obtained in the previous exercise.  
Rewrite the generator of global Euclidean time translations $\p_\tau$ in Poincar\'e coordinates. Shortly comment on the interpretation of your result from the point of view of the conformal boundary coordinates $(\tau_P, x)$. \\

\chapter{Anti-de Sitter: add matter and gravity} 

In this chapter we will add matter to the AdS background geometry of the previous chapter, as well as consider gravity in AdS. 

\section{Classical massless and massive particles} 

Consider a single massless particle at the center of global AdS. Such a particle will travel on lightlike paths. 
From the Penrose diagram we can read off that it takes a finite 
time $\Delta t = \frac{\pi}{2}$ for the particle to reach the boundary along a light ray. At the timelike boundary of AdS, a boundary condition needs to be imposed to consistently describe the evolution of the system.\footnote{
	This statement is related to the fact that AdS is not globally hyperbolic, i.e.~ there are no Cauchy hypersurfaces. 
	This led to statements in '70s general relativity literature that the ``Cauchy problem is not well-posed in AdS". The Cauchy problem being: given initial conditions in a spacelike slice, 
	determine evolution of the system. The resolution is that the evolution of an AdS system requires, because of the presence of the timelike boundary, boundary conditions to be imposed at the boundary. Including the boundary conditions, the evolution problem is perfectly well-defined, and there is no issue \cite{bengtsson, McGreevy}.    
} 
Typically, reflective boundary conditions will be imposed. The photon will then bounce back at the boundary and follow a periodic path of period $\Delta t = 2\pi$ in global time, bouncing back and forth. 

\begin{figure}
	\centering \includegraphics[width=6cm]{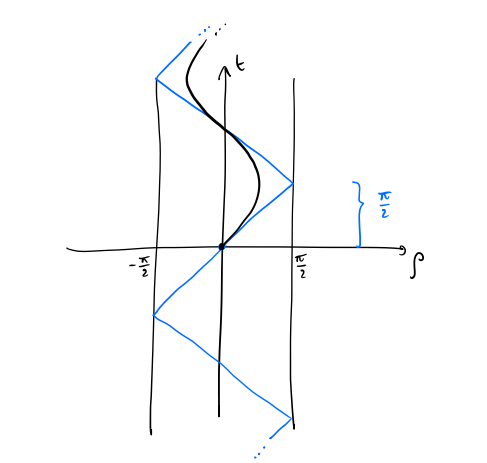} 
	\caption{Lightlike (blue) and timelike (black) path of particles in AdS.} \label{fig-AdSbox}
\end{figure}

A massive particle starting at the center of AdS will follow a timelike path of 
the same period $\Delta t = 2\pi$, bouncing back and forth between regions close to the boundary. This behavior, illustrated in figure \ref{fig-AdSbox}, can be summarized as: \emph{AdS acts like a box that confines particles}.  

In the rest of this section, we discuss how to obtain the path of classical particles mathematically. 

A classical single particle in AdS$_3$ will trace out a worldline in spacetime. 
This path or worldline can be described by 
\ali{
	x^\alpha(\lambda) \quad \text{or} \quad X^a(\lambda) 
}
using either intrinsic AdS coordinates $x^\alpha$ ($\alpha=1,2,3$) or embedding coordinates $X^a$ ($a=1,2,3,4$), and where $\lambda$ is a wordline coordinate. It will be easiest to work with embedding coordinates. 

First consider the path of a particle in the flat embedding space. It will be given by a geodesic, namely a path of extremal 
distance in the spacetime. It is obtained by extremizing 
\ali{
	S &= \int ds = \int d\lambda \sqrt{g_{ab}(X) \dot X^a(\lambda) \dot X^b(\lambda)} = \int d\lambda \sqrt{\dot X^2} 
}
The above is equivalent to extremizing (see e.g. \cite{Nakahara} (7.58) and further) 
\ali{
	S =  \int d \lambda  \frac{\dot X^2}{2} .  
}
The equivalence comes down to the fact that geodesics are also curves on which tangent vectors are parallel transported. 

Now a particle in AdS $=$ particle in embedding space constrained to live on the AdS surface defined by $X_a X^a = -1$ (this is short notation for \eqref{AdSembdef}). We can impose this constraint by using a Lagrange multiplier $\kappa$ in the ``action" $S$: an AdS geodesic extremizes 
\ali{
	S = \int d\lambda \left ( \frac{1}{2} \dot X^2 -\frac{1}{2} \kappa (X^2 + 1) \right).  
} 
The Euler-Lagrange equations for $S$ are 
\ali{
	\ddot X_a(\lambda) = - \kappa X_a(\lambda) \label{EOMforX}
	}
and the constraint 
\ali{
	X_a X^a = -1. 
}
There are 3 classes of choices for $\kappa$: positive, zero or negative. Because we can rescale the unphysical wordline coordinate $\lambda$, it is sufficient to consider one representative for each of the 3 classes: $\kappa = 1, 0, -1$. These choices correspond to respectively timelike, null or spacelike solutions $X_a(\lambda)$. 

A massive particle will follow a timelike path $(\lambda = t)$ and hence will have to satisfy 
\ali{
	\ddot X_a(t) = - X_a(t), \qquad X_a X^a = -1.  
}
The solution is simply given by 
\ali{
	X_a = A_a \cos t + B_a \sin t   \label{Xasol}
}
with 
\ali{
	A\cdot B = 0, \quad A^2 = B^2 = -1. 
}
From the solution \eqref{Xasol}, it follows indeed (as expected on pictorial grounds from the AdS Penrose diagram in figure \ref{fig-AdSbox}) that all timelike paths of the massive particle have the same frequency with respect to AdS time $t$, i.e.~a periodicity $\Delta t = 2 \pi$. (Note this is not so surprising from the embedding coordinates point of view, since $t$ was introduced as an angle of $UV$-orbits, before we imposed the universal covering interpretation.) 

The most trivial solution is 
\ali{
	U =  \cos t, \quad V =  \sin t, \quad X = Y = 0. \label{trivialsol}
}
Comparison to the parametrization of AdS$_3$ embedding coordinates in terms of global Penrose coordinates 
\ali{
	U =  \frac{\cos t}{\cos \rho}, \quad V = \frac{\sin t}{\cos \rho}, \quad X = \tan \rho \cos \phi, \quad Y = \tan \rho \sin \phi,  
	}
indicates that the trivial solution \eqref{trivialsol} represents a massive particle at rest at the center $\rho = 0$ of AdS. 

An example of a non-trivial solution \cite{Kaplan lectures} is 
\ali{
	U =  \frac{\cos t}{\cos \rho_*}, \quad V = \frac{\sin t}{\cos \rho_*}, \quad X = \tan \rho_* \cos t, \quad Y = \tan \rho_* \sin t
}
describing a particle at fixed $\rho = \rho_*$ making a circular orbit in the XY plane. 



\section{Classical gravity in AdS$_3$} \label{sectionAdSgrav}

\paragraph{Gravitational action} 
Next, let's consider gravity in AdS$_{d+1}$. For this, we consider the gravitational action 
\ali{
	\begin{split} 
	S_{grav}[g_\ab,\psi] &= -\frac{1}{16 \pi G} \int_M d^{d+1} x \sqrt{-g} ( R - 2 \Lambda) + 
	S_M[g_\ab,\psi] - \frac{1}{8 \pi G} \int_{\p M} d^d x \sqrt{-\gamma} \, K    \\
	&= S_{EH}
	+ S_{GH}
	\end{split}  \label{Sgrav} 
}
Here, $G$ is the coupling constant of gravity or Newton constant. 
The first term is the Einstein-Hilbert (EH) action $S_{EH}$ for the $(d+1)$-dimensional metric $g_\ab$ on a manifold $M$. It is given in terms of the square root of the determinant of the metric $\sqrt{-g}$, the Ricci scalar $R$ of the metric, a negative cosmological constant 
\ali{ 
	\Lambda = -\frac{d(d-1)}{2 \ell^2}   
}
and a matter 
action $S_M$ for matter fields $\psi$ that we will leave out for most of the discussion. The second term is the Gibbons-Hawking action which depends on the induced metric $\gamma_\mn$ on the boundary $\p M$ of the manifold. 
It is given in terms of $K$, which is the trace $K \equiv \gamma_\mn K^\mn$ of the extrinsic curvature $K_\mn \equiv -\frac{1}{2} \mathcal L_n \gamma_\mn$ of the hypersurface $\p M$, defined as the Lie derivative of $\gamma_\mn$ in the direction of the outward pointing unit normal $n^\mu$ to $\p M$. (The definition of the Lie derivative will be reviewed in the paragraph below.) The Gibbons-Hawking term is there to insure a well-defined variational principle exists, despite the EH action containing second order derivatives of the metric field (which is atypical compared to previously discussed field theories $S[\phi, \p_\mu \phi]$). For details, see e.g.~\cite{HartmanQG}. 


\paragraph{Diffeomorphism invariance} This action is diffeomorphism invariant, in short diff invariant, or ``general relativity is diff invariant" 
\ali{
	S_{grav}[g_{\ab}(x),\psi(x)] \quad \stackrel{x \ra f(x)}{=} \quad S_{grav}[f_*g_\ab(x), f_*\psi(x)].  
}
In GR (general relativity), we're working with a manifold $M$ with metric $g_\ab$ and matter fields $\psi$, or a system $(M, g_\ab, \psi)$. A mapping $f: M \ra M$ is a diff if $(M, g_\ab, \psi)$ and $(M, f_* g_\ab, f_* \psi)$ represent the same physical situation, with $f_*$ the pullback defined below. Diffeomorphisms are just another name for active coordinate transformations (see e.g.~\cite{carroll-link} and \cite{difran} section 2.4).  
In the passive picture, $x \ra x'$, a field transforms $\phi(x) \ra \phi'(x')$. In the active picture, $x \ra x$ and $\phi(x) \ra \phi'(x)$. 
The transformed field evaluated at the original position $x$ is called the pullback. If $x'=f(x)$ in the passive picture, the pullback is denoted as $\phi'(x) = f_* \phi(x)$.  The pullback is also used to define the Lie derivative of $\phi$ along $X$ as $\mathcal L_X \phi = \lim_{\epsilon \ra 0} \frac{\phi'(x) - \phi(x)}{\epsilon}$ 
where $\phi'(x)$ is the pullback $f_* \phi(x)$ for $x \ra f(x)$ a small 
displacement of the point $x$ along the flow lines of the vector field $X$.

Because the metric is a dynamical variable, there is no prior geometry in GR and thus no \emph{preferred} coordinate system for spacetime (in contradistinction to the discussions of field theory on a fixed background in the first chapters of the course). The 
diff invariance of GR is a gauge invariance: not a physical symmetry, but an expression of a redundancy in the description of the system. 
To describe the gauge invariant physics, 
the diff invariance should be fixed by a gauge choice, i.e.~a choice of coordinates.

\paragraph{Variational problem} 
Let's consider the variational problem. We vary the action while imposing boundary conditions that fix the metric and matter fields at the boundary of the manifold, $\delta g_\ab|_{\p M} = 0$ and $\delta \psi|_{\p M} = 0$  (Dirichlet boundary conditions or `gravity in a box'). 
This imposes the Einstein equations as Euler-Lagrange equations for the metric 
\ali{
	R_\ab - \frac{1}{2} R \, g_\ab + \Lambda g_\ab = 8 \pi G \, T_\ab^M(\psi) ,  \label{ELeom}
}
with $T_\ab^M(\psi)$ the matter stress tensor defined as $T_\ab^M = \frac{-2}{\sqrt{-g}} \frac{\delta S_M}{\delta g^\ab}$,  
and the Euler-Lagrange equations for the matter fields, e.g.~ $\Box \phi - m^2 \phi = 0$ for the example matter action $S_M = \frac{1}{2} \int d^{d+1} x \sqrt{-g} \left(  (\p\phi)^2 + m^2 \phi^2 \right)$. 

\paragraph{Fefferman-Graham gauge}
One solution of this variational problem in the absence of matter fields, i.e.~of the vacuum Einstein equations 
\ali{
	R_\ab - \frac{1}{2} R g_\ab + \Lambda g_\ab = 0 , 
}
is the AdS$_{d+1}$ solution  
\ali{
	ds^2 = g_\ab dx^\alpha dx^\beta = \frac{\ell^2}{Z^2} (dZ^2 + \eta_\mn dx^\mu dx^\nu), \qquad (\mu,\nu = 1 \cdots d) 
}
with the AdS bulk direction $Z$ going to zero at the boundary. 
This solution suggests an ansatz for the most general solution in a particular choice of coordinates: 
\ali{
	ds^2 = \frac{\ell^2}{Z^2} (dZ^2 + \tilde \gamma_\mn(Z,x) dx^\mu dx^\nu), \qquad (\mu,\nu = 1 \cdots d).  \label{FGgaugemetric}
}
It was shown by Fefferman and Graham that $\tilde \gamma_\mn$ 
satisfy the following 
expansion in the coordinate $Z$: 
\ali{
	\tilde \gamma_\mn(Z,x) = \tilde \gamma_\mn^{(0)}(x) + \cdots + Z^d \tilde \gamma^{(d)}_\mn(x) + \cdots  + Z^d \log(Z^2) h^{(d)}_\mn(x)  + \cdots  \qquad (\text{FG expansion})   \label{FGexp}
}
in terms of $\tilde \gamma^{(d)}_\mn$ and $h^{(d)}_\mn$ that only depend on the boundary coordinates $x^\mu$. The logarithmic piece appears only for even $d$, and the expansion is in $Z^2$ (which is the reason it is more standardly given as an expansion in $\rho \equiv Z^2$ \cite{Skenderis00}).   
It is an \emph{asymptotic}, i.e.~near-boundary $Z \ra 0$, expansion, called the \emph{Fefferman-Graham (FG) expansion}. 
Writing a solution in the form \eqref{FGgaugemetric}, with the asymptotic behavior given by \eqref{FGexp}, is then called the \emph{FG gauge}. 
In this gauge, 
\ali{
	\tilde \gamma_{ZZ} = \tilde \gamma_{Z\mu} = 0 \qquad (\text{FG gauge}). 
	\label{FGgauge}
	} 

\paragraph{Banados solution and BTZ solution} 

In 3 (bulk) dimensions, a general solution to the vacuum Einstein equations with negative cosmological constant and Dirichlet boundary conditions $\gamma^{(0)} = dz d\bar z$ can be written down. It is known as the Banados solution 
and takes the form of a truncated 
FG expansion to order $Z^4 \tilde \gamma_\mn^{(4)}$,   
\ali{
	ds^2 = \frac{1}{Z^2} \left[ (dZ^2 + dz d\bar z)  + Z^2 \left( L(z) dz^2 + \bar L(\bar z) d\bar z^2 \right) + 2 \, Z^4 L(z) \bar L(\bar z) dz d\bar z \right]   \qquad \text{(Banados)}  \label{ds2Banados}
	} 
where $L(z)$ and $\bar L(\bar z)$ are general holomorphic resp.~anti-holomorphic functions.  
The conformal boundary is given by a flat metric $\tilde \gamma^{(0)}$. 
The functions $L(z)$ and $\bar L(\bar z)$ describe fluctuations away from the pure AdS$_3$ solution (in Poincar\'e coordinates) $L = 0 = \bar L$. These $L$-functions only change the geometry in the bulk, not the leading order in $Z \ra 0$ behavior or `asymptotics'. The Banados solution is therefore called an \emph{asymptotically AdS$_3$ solution}.  

One important example is the black hole solution called BTZ \footnote{To be precise, planar BTZ, because of the planar conformal boundary.} (after the authors Banados Teitelboim Zanelli), which is the Banados solution for constant $L$-functions, $L(z) = L$ and $\bar L(\bar z) = \bar L$
\ali{
	ds^2 = \frac{1}{Z^2} (dZ^2 + dz d\bar z)  + 
	\left( L \, dz^2 + \bar L \, d\bar z^2 \right) + 2 \, 
	Z^2 L \, \bar L \, dz d\bar z. \qquad \text{(BTZ)} 
} 
It describes a 3-dimensional AdS black hole with mass and angular momentum 
\ali{
	M = \frac{L + \bar L}{4G}, \qquad J = \frac{L - \bar L}{4G}. 
	 \label{BTZmass}
}
One way to calculate the mass of the black hole solution is to use the definition of the Brown-York stress tensor.

\paragraph{Brown-York stress tensor} 

In a gravitational theory, given a metric solution $g_\ab$, we can define a quasi-local stress tensor associated with a spacetime region $\mathcal N$  
as the variation of the gravitational action evaluated on the solution with respect to the intrinsic metric $\gamma_\mn$ 
of the boundary $\p \mathcal N$ of the region,   
\ali{
	T^\mn_{BY} := \frac{2}{\sqrt{- \gamma}} \frac{\delta S_{grav}^{on-shell}[\gamma_{\mn}]}{\delta \gamma_\mn}.   \label{TBYdef}
}
This gravitational stress tensor is called the Brown-York stress tensor. It measures the 
energy-momentum carried by gravitational degrees of freedom (metric and matter fields) and allows to calculate gravitational mass. 
In particular, we will consider the Brown-York stress tensor for $\mathcal N$ the asymptotically AdS$_3$ geometry or Banados geometry, and $\p \mathcal N$ the AdS boundary $\gamma_\mn = \frac{\ell^2}{Z^2} \tilde \gamma_\mn$ at $Z \ra 0$. 

The definition \eqref{TBYdef} is reminiscent of the QFT stress tensor definition \eqref{TmnQFTdef} in 
field theory on a fixed background geometry. In that context, conserved Noether charges $Q_\epsilon$ 
were associated with Killing fields $\epsilon$ of the background geometry via \eqref{Noetherdef}. 
Similarly, the Brown-York stress tensor definition allows to construct conserved Noether charges in gravity. They are associated with diffeomorphisms $\xi^\alpha$ that 
preserve $\gamma_\mn$:  
\ali{
	Q_\xi^{BY} = \int d\phi \, j_\xi^0 = \int d\phi \, \, \xi_\nu  T_{BY}^{0\nu} 
}
with $\phi$ and $t$ the boundary coordinates of AdS$_3$.\footnote{In more general notation, the integration $\int d\phi$ would be replaced by $\int_{\text{time slice $\Sigma$ of $\p \mathcal N \,$}} d^{d-1} x \sqrt{\sigma}$, with $\sigma_{ab}$ the metric of the spacelike surface $\Sigma$ in the boundary $\p \mathcal N$, and the integrand replaced by $u^\mu T_\mn^{BY} \xi^\nu$ with $u^\mu$ a timelike unit normal (defining local flow of time in $\p \mathcal N$) to $\Sigma$.}   

The AdS boundary geometry is independent of $\phi$ and $t$, and thus preserved under the diffeomorphisms $\xi = \p_t$ and $\p_\phi$. The corresponding gravitational Noether charges are respectively the mass and angular momentum of the asymptotically AdS geometry, 
\ali{
	Q^{BY}_{\p_t} = M \qquad \text{and} \qquad Q^{BY}_{\p_\phi} = J. \label{QBY}	
}

The Brown-York tensor can be calculated from the definition to be 
\ali{
	T_{BY}^\mn = \frac{1}{8\pi G} \left[ K^\mn - K  \gamma^\mn  + \frac{2}{\sqrt{-\gamma}} \frac{\delta S_{ct}}{\delta \gamma_\mn} \right]_{Z \ra 0}.   
	}
The last term is a counterterm that needs to be added to renormalize the divergences from the first two terms as $Z \ra 0$. A counterterm $S_{ct}$ can be added to the gravitational action: it does not alter the EOM and can be uniquely fixed\footnote{
	This is not in general true. Brown and York proposed a subtraction in the stress tensor that is derived from embedding the boundary with the same intrinsic metric $\gamma_\mn$ in a reference spacetime, e.g.~flat space. It is not always possible to embed a boundary with an arbitrary intrinsic metric in the reference spacetime and therefore the Brown-York procedure is not always well-defined. The procedure in \cite{BalasubKraus} is applied to asymptotically AdS in particular. 
} to $-\frac{1}{\ell} \int \sqrt{-\gamma}$ after imposing that $S_{ct}$ only depends on $\gamma_\mn$ \cite{BalasubKraus}.   
Then $T_{BY}^\mn = \frac{1}{8\pi G} \left[ K^\mn - K  \gamma^\mn  - \frac{1}{\ell} \gamma^\mn \right]_{Z \ra 0}$ 
and in particular, for the Banados geometry, 
\ali{
	T^{BY}_{zz} = -\frac{ L(z)}{4G}, \qquad T^{BY}_{\bar z \bar z} = -\frac{ \bar L(\bar z)}{4G}.  \label{BanadosBY}
}
From this result (for $\ell=1$), you can derive (\textit{Exercise}) that the gravitational mass of the BTZ black hole is indeed given by \eqref{BTZmass}. 
More generally, this result gives a \emph{physical interpretation to the $L$-functions} in the Banados geometry \eqref{ds2Banados} for an asymptotically AdS$_3$ background. Namely, $L(z)$ and $\bar L(z)$ are the holomorphic and anti-holomorphic Brown-York stress tensor components (up to a proportionality constant) of the Banados geometry. We will use this in the next section to discuss the famous Brown-Henneaux symmetry of AdS$_3$ gravity.

\subsection{Hints of AdS$_3$/CFT$_2$: Virasoro algebra} \label{subsVirasoro}

We first write the Banados geometry \eqref{ds2Banados} in terms of an alternative choice of bulk coordinate $r$, with $Z = \ell^2/r$, 
\ali{
	ds^2 = g_\ab(x; L,\bar L) dx^\alpha dx^\beta = 
	\frac{1}{r^2} dr^2 + r^2 dz d\bar z  +  L(z) dz^2 + \bar L(\bar z) d\bar z^2  + \frac{2}{r^2} L(z) \bar L(\bar z) dz d\bar z   
	\label{ds2Banados2}
}
The AdS boundary is now at $r \ra \infty$. (The coordinate change from $Z$ to $r$ is just so we can follow \cite{HartmanQG} for this section.)

We have seen that we can associate conserved gravitational charges with bulk diffeomorphisms $\xi^\alpha$ that preserve the boundary metric. We already mentioned the examples of translations in the boundary coordinates. But there is a much bigger, in fact infinitely bigger, set of bulk diffs $x^\alpha \ra x^\alpha + \xi^\alpha(x)$ that preserve the boundary metric. They are given by 
\ali{
		\begin{split} 
	z \ra z + \epsilon(z) - \frac{\ell^4}{2 r^2} \bar \epsilon''(\bar z) \\
	\bar z \ra \bar z + \bar\epsilon(\bar z) - \frac{\ell^4}{2 r^2}  \epsilon''(z) \\
	r \ra r - \frac{r}{2} \epsilon'(z) - \frac{r}{2} \bar \epsilon'(\bar z) 	. 
	    \end{split} \label{BrHdiffs}
} 
These are the \emph{Brown-Henneaux} diffs of asymptotically AdS$_3$. They have the property that they preserve the asymptotics of the asymptotically AdS$_3$ 
metric \eqref{FGgaugemetric} (with flat $\tilde \gamma^{(0)}_\mn$). In Banados notation \eqref{ds2Banados} or \eqref{ds2Banados2}, the Brown-Henneaux diffs' only effect is to change the $L$-functions, 
i.e.~change the geometry in the bulk. 
The Brown-Henneaux diffs are not Killing vectors but \emph{asymptotic Killing vectors} of the asymptotically AdS$_3$ solution. 

Indeed you can check that under the diff $\xi^\alpha$ given in \eqref{BrHdiffs}, the metric solution $g_\ab$ transforms such that 
\ali{
	ds^2 \ra ds^2 + \left( - \epsilon \, \p L -2 L \, \p \epsilon + \frac{\ell^2}{2} \p^3\epsilon \right) dz^2 + \left( - \bar \epsilon \, \bar \p \bar L -2 \bar L \, \bar \p \bar \epsilon + \frac{\ell^2}{2} \bar \p^3 \bar \epsilon \right) d\bar z^2  + \mathcal O(\frac{1}{r^2}).  
}
With the interpretation \eqref{BanadosBY} of the $L$-functions as the gravitational stress tensor components, we see that the effect of the Brown-Henneaux diffs is 
\ali{
	\begin{split} 
	\delta_\xi T_{zz}^{BY} &= - \epsilon(z) \p T_{zz}^{BY} - 2 T_{zz}^{BY} \p \epsilon - \frac{\ell}{8 G} \p^3 \epsilon(z) \\
		\delta_\xi T_{\bar z \bar z}^{BY} &= - \bar \epsilon(\bar z) \bar \p T_{\bar z\bar z}^{BY} - 2 T_{\bar z\bar z}^{BY} \bar \p \bar \epsilon - \frac{\ell}{8 G} \bar \p^3 \bar \epsilon(\bar z). 
		\end{split}  \label{BrHTtransf}
}
We recognize this as the transformation behavior of a 2D \emph{CFT} stress tensor \eqref{CFTstresstensortransffull}, in a CFT with a specific central charge given by 
\ali{
	\Aboxed{ c = \frac{3 \ell}{2G} } \qquad \text{(Brown-Henneaux central charge)} .  \label{BrHcentralcharge}
}
Indeed, the Brown-Henneaux diffs can be seen to take the form of 2D conformal transformations of the boundary coordinates $z$ and $\bar z$ at the boundary $r \ra \infty$. We see that symmetry of the asymptotic \emph{geometry} of AdS$_3$ coincides with conformal symmetry of a 2D \emph{CFT} at the conformal boundary of AdS$_3$!   

The more concrete statement is that the conserved charges associated with the asymptotic symmetry of an asymptotically AdS$_3$ geometry obey a Virasoro algebra with a central charge $c = \frac{3 \ell}{2G}$.  Namely, if we expand the Brown-York stress tensor components into modes 
\ali{
	T_{zz}^{BY}(z) = \sum_n \frac{L_n^{BY}}{z^{n+2}}, \qquad T_{\bar z\bar z}^{BY}(\bar z) = \sum_n \frac{\bar L_n^{BY}}{\bar z^{n+2}}
}
then the conserved charges $Q^{BY}_\xi$ associated with the Brown-Henneaux diffs $\xi$ take the form $Q^{BY}_\xi = \sum_n c_n L_n^{BY} + \sum_n \bar c_n \bar L_n^{BY}$. They are classical charges, with a Poisson bracket 
algebra $\{ Q^{BY}_{\xi_1}, Q^{BY}_{\xi_2} \}_{PB}$. The resulting algebra (given here without derivation) 
is the Virasoro algebra 
\ali{
	\begin{split} 
	\{ L_n^{BY}, L_m^{BY} \}_{PB} &= (n-m) L^{BY}_{n+m} + \frac{\ell}{8G} (n^3 - n) \delta_{n+m,0} \\
	\{ \bar L_n^{BY}, \bar L_m^{BY} \}_{PB} &= (n-m) \bar L^{BY}_{n+m} + \frac{\ell}{8G} (n^3 - n) \delta_{n+m,0} \\
	\{ L_n^{BY}, \bar L_m^{BY} \}_{PB} &=  0 .   
	\end{split}  \label{BrHalg}
	} 
As we have seen before, the algebra of conserved charges \eqref{BrHalg} contains the same information as the transformation behavior of the stress tensor \eqref{BrHTtransf}. From both points of view a Virasoro symmetry with central charge \eqref{BrHcentralcharge} emerges. This result was obtained by Brown and Henneaux in 1986. It is a pre-AdS/CFT result that nowadays is interpreted as an AdS/CFT statement about the duality between AdS$_3$ gravity (with gravitational parameters $\ell$ and $G$)  and a 2D CFT (with central charge parameter $c$ related to the dual parameters by \eqref{BrHcentralcharge}).  


\paragraph{Emergent connection to 2D CFT} 

Before addressing some comments about the Brown-Henneaux result, let us first summarize the emergent connection to 2D CFT. 
We have seen that the asymptotic symmetry transformations of AdS$_3$ at the AdS boundary equal the (local) conformal transformations of a 2D CFT with $c = 3 \ell/2G$. In higher dimensions, the asymptotic isometry group of AdS$_{d+1}$ ($d>2$) is just the same as the isometry group $SO(d,2)$ of AdS$_{d+1}$. Combined with the boxed result of section \ref{sectionAdSsymm} (which also applies to $d=2$ but as a statement about the global conformal transformations of the CFT), we conclude that: 

\fbox{ \parbox{0.97\textwidth}{ The asymptotic isometry transformations of AdS$_{d+1}$ geometry at the AdS conformal boundary 
		equal the conformal transformations of a CFT$_d$. } }  

Another consequence 
of the Brown-Henneaux result is that: 

\fbox{ \parbox{0.97\textwidth}{  The (classical) Brown-York stress tensor of asymptotically AdS$_3$ gravity has the properties of (the expectation value of) a 2D CFT stress tensor. } }


\paragraph{Central charge appears at classical level?} 
In section \ref{quantum} we first discussed the possibility of a central charge appearing in the algebra of conserved charges at the \emph{quantum level}, reflecting the phase factor ambiguity of a quantum state. How does the physics of \emph{classical} asymptotic AdS$_3$ gravity give rise to a central charge usually associated with quantum indeterminacy? The answer is in the presence of the boundary, which leads to an ambiguity of the classical charge and a corresponding central charge term in the algebra. 
For more details, including the relation to the comment discussed in the next paragraph, see \cite{Carlip05} (section 3.1 and 2). 

\paragraph{Physical diffs?} 
Another question concerns the interpretation of diffs in gravity. In the beginning of section \ref{sectionAdSgrav} we paused at the interpretation of diff invariance in gravity as an unphysical or `gauge' redundancy. But the Brown-Henneaux diffs give rise to the physics described by the Virasoro algebra. How is it that these diffs do have a physical interpretation? Again, it is the presence of the boundary of AdS that provides the answer. Namely, GR is locally diff invariant, but not invariant under diffs that reach the boundary. Some of the diffs that reach infinity are actual symmetries. This is best illustrated by an example. A time reparametrization with compact support $t \ra t'(t,x)$ such that $t' \ra t$ at $r \ra \infty$ is a local diff (no physics). In contrast, a global time shift $t \ra t+1$ acts at infinity, describing true time evolution (physics). Roughly, ``local diffs are fake, global diffs are real" \cite{HartmanQG}.

\subsection{Hints of AdS$_3$/CFT$_2$: Ryu-Takayanagi formula} 

As we will see later in more detail, there is an important relation between the classical gravitational concept of geodesics in AdS$_3$ and the quantum mechanical concept of entanglement entropy in a 2D CFT. It is called the Ryu-Takayanagi relation. 

In AdS$_3$, $ds^2 = \frac{dZ^2 - dt^2 + dx^2}{Z^2}$, boundary-anchored geodesics are given by half-circles $x^2 + Z^2 = \left(\frac{L}{2}\right)^2$.  
Such a geodesic has a length 
\ali{
	\int ds = 2 \ell \int_{\epsilon}^{L/2} \frac{dZ}{Z} \sqrt{x'(Z)^2 + 1} = 2 \ell \log \frac{L}{\epsilon} + \mathcal O(\epsilon) .  
} 
The Ryu-Takayanagi formula identifies $\frac{1}{4G}$ times the length of the boundary-anchored AdS$_3$ geodesic with the entanglement entropy of the interval at the conformal boundary that the geodesic anchors to. Indeed, using the Brown-Henneaux central charge we find 
\ali{
	\frac{\text{length geodesic}}{4G} = \frac{c}{3} \log \frac{L}{\epsilon}.   \label{RTfirstencounter}
}
The right hand side can be identified as the expression for the entanglement entropy for an interval of length $L$ in a 2D CFT. (We have not derived this expression here. It can be derived using for example the replica trick \cite{CardyCalabrese}.) 

\section{Conformal matter on conformal boundary of AdS}

In this chapter we have considered adding matter and gravity to the AdS geometry 
\ali{
	ds^2 = \frac{dZ^2 - dt^2 + d\vec x^2}{Z^2} 
}
introduced in the previous chapter. The logical procedure is to add dynamical degrees of freedom in the bulk of the AdS spacetime. But it is also perfectly well-defined to take some conformal matter and place it on the AdS conformal boundary 
\ali{
	ds^2 = -dt^2 + d\vec x^2   
} 
(while leaving the bulk of AdS untouched).  
The result is a $d$-dimensional CFT living at the conformal boundary of AdS$_{d+1}$. 

\paragraph{AdS/CFT duality} The statement of AdS/CFT duality is that in fact adding conformal matter to the AdS conformal boundary is physically equivalent to adding gravity to the AdS bulk. Indeed we have already encountered some hints for this duality in the previous section when discussing gravity in AdS$_3$. 
It was found that the same physics appear in the asymptotics of AdS$_3$ gravity and 2D CFT, more precisely in the Virasoro algebra and in the description of bipartitioning the system (across a geodesic in the bulk or across an interval in the CFT). In both instances, the physical effect (i.e.~central charge extension of Virasoro algebra or appearance of $\frac{c}{3} \log \frac{L}{\epsilon}$ in bipartioning of system) has a classical origin in the gravity theory but a quantum origin in the dual CFT.  

\fbox{ \parbox{0.97\textwidth}{  Classical gravity in asymptotically AdS$_{d+1}$ geometry is dual to a quantum theory without gravity but with conformal symmetry at the $d$-dimensional conformal boundary. } } 
 
This is a very bold statement, claiming that theories that we usually consider to be fundamentally different are in fact describing the same physics. The rest of the course is devoted to further investigating  
this claimed duality. 



\newpage
\section*{Exercises}

\textbf{\underline{\smash{Exercise 1. Diff invariant actions}}}

In this exercise you will show the diff invariance 
\ali{
	S[\varphi'(x), g_\mn'(x)] = S[\varphi(x), g_\mn(x)] 
}
of actions of the form 
\ali{
	S = \int_{\mathcal M} d^{d+1} x \sqrt{|g|} \,\, \psi(\varphi,g_\mn) 
}
with $\psi$ any scalar combination of the dynamical fields $g_\mn(x)$ and $\varphi(x)$. 

1) Show that the metric field transforms as 
\ali{ 
	g'_\mn(x) = g_\mn(x) + \delta_\xi g_\mn, \qquad \delta_\xi g_\mn = \nabla_\mu \xi_\nu + \nabla_\nu \xi_\mu 
} 
under diffs or active coordinate transformations $x'(x) = x + \xi(x)$. For this, start from the passive coordinate transformation behavior $g_\mn'(x') = g_\ab(x) \frac{\p x^\alpha}{\p x'^\mu} \frac{\p x^\beta}{\p x'^\nu}$ imposed by invariance of $ds^2$. Write down the active coordinate transformation behavior effectively by switching $x$ and $x'$. 

2) Show that a scalar field transforms as 
\ali{ 
	\phi'(x) = \phi(x) + \delta_\xi \phi, \qquad \delta_\xi \phi = \xi^\mu \p_\mu \phi  = \xi^\mu \nabla_\mu \phi 
} 
under diffs or active coordinate transformations $x'(x) = x + \xi(x)$.

3) Write down $\delta_\xi \sqrt{|g|}$ using the result of 1). You can use for this (without derivation) that $\delta \sqrt{|g|} = -\frac{1}{2} \sqrt{|g|} g_\mn \delta g^\mn = \frac{1}{2} \sqrt{|g|} g^\mn \delta g_\mn$ (the sign is not a typo but comes from the sign in the derivative of the inverse matrix).  

4) Use these results to write $\delta_\xi S$ and show that it vanishes, but only for diffs $\xi^\mu$ that vanish on the boundary of the spacetime $\xi^\mu|_{\p \mathcal M} = 0$. This means diffs that reach the boundary are physical transformations whereas diffs in the bulk are gauge transformations!

\chapter{AdS/CFT dictionary}

\section{Quantum gravity (QG)} 

\paragraph{Planck scale} 
In $D$-dimensional gravity, the Newton constant has the dimension of length to the power $D-2$, 
\ali{
	[G] = L^{D-2}.  \label{Gdim}
}
Usually gravity is weak, but at the Planck scale 
\ali{
	\ell_P = (\hbar G)^{\frac{1}{D-2}}  
}
QM interactions $(\hbar)$ and gravity $(G)$ become comparably strong, and a QM treatment of gravity forces itself. In $D=4$ for example, $\ell_P= \sqrt{\hbar G} \approx 10^{-33} \text{cm}$. (We only sporadically reintroduce $\hbar$, using natural units $\hbar = 1$ through most of the course.) 
The Planck mass is 
\ali{
	M_P = \frac{1}{\ell_P}   
}
and is a very high energy scale. 

In a gravitational theory, we have an action $S_{grav}[g_\ab, \psi]$ for dynamical fields $\psi$ and $g_\ab$. 
In contrast to a field theory, the metric $g_\ab$ is itself dynamical. Still, you could attempt to quantize gravity \eqref{Sgrav} using a background field method 
\ali{
	g_\ab = \bar g_\ab + \sqrt{G} \,\, h_\ab = \bar g_\ab + \left(\frac{1}{M_P}\right)^{\frac{D-2}{2}} h_\ab  \label{weakfield}
	}
where only the metric fluctuations $h_\ab$ are dynamical. 
The $\sqrt{G}$ prefactor is included to ensure a canonical kinetic term for the metric fluctuations $\int \p h \p h$ in the $\frac{1}{M_P}$ expansion of the gravitational action 
around $\bar g_\ab$. 
The expanded action will contain non-linear interactions to all orders in $h_\ab$, with a dimensionful coupling constant $G$ \eqref{Gdim}. 
This means the theory is \emph{non-renormalizable} for $D>2$. 

One can still make sense of the non-renormalizable theory in an \emph{effective field theory} (EFT) sense. 

\paragraph{Gravity as an EFT} 


References for more details: \cite{HartmanQG} p.8 and further, Donoghue gr-qc/9512024 and \url{http://www.scholarpedia.org/article/Quantum_gravity_as_a_low_energy_effective_field_theory}. 

To treat gravity as an EFT, one must first write down the most general possible action consistent with the symmetries, and organize the terms in increasing order of derivatives. The starting assumption is that nature has a graviton, i.e.~a massless spin-2 field. Consistency (diff invariance) 
imposes that the action takes the form of the EH action of GR plus an infinite number of terms which are higher order in curvatures $R, R_\ab, \cdots$ or derivatives (remember $R \sim \p \p g$) 
\ali{
	S_{grav}[g_\ab,\psi] = -\frac{1}{16 \pi G} \int d^D x \sqrt{-g} \left(-2 \Lambda + R + c_1 R^2 + c_2 R_\ab R^\ab + c_3 R_{\ab\gamma\delta}R^{\ab\gamma\delta} + \cdots \right) + S_M. 
}
Choose a fixed order in derivatives up to which you will keep the terms in the action. You are allowed to do quantum field theory with this action, including loops, in a perturbative expansion \eqref{weakfield}. Quantizing the metric fluctuations $h_\ab$ (which describe gravitational waves at the classical gravity level) into gravitons is similar in spirit to quantizing sound waves into phonons, i.e.~quantized effective particles.  
However, the answer you get can only be trusted if the neglected terms in the derivative expansion of the action are much smaller than the terms you kept. Because the theory is non-renormalizable, the perturbative expansion will break down at high energies. The point is that \emph{at low energies} $E < M_P$, the quantized EFT is a perfectly good quantum theory of gravity. It describes the low-energy phenomenology of a more fundamental theory, which 
is expected to involve new degrees of freedom at or above the Planck energy scale. The more fundamental UV theory is often referred to as the \emph{UV-completion} of gravity.   

We will not be concerned with the higher curvature corrections to the EH action in this course. We are allowed to neglect them when the AdS length $\ell$ is large compared to the length scale $\ell_s$ of new physics 
\ali{
	\ell_s \ll \ell.  
}
This is because the coefficients $c_{1,2,3,...}$ scale with powers of $\ell_s$, by dimensional analysis and the rules of EFT.  The length scale of new physics is smaller or equal than the Planck scale $l_s \leq l_P$. 
The notation $\ell_s$ refers to the fundamental scale of string theory, which is one proposed UV completion of quantized gravity. Another proposed UV completion is a lower-dimensional CFT, through (the strong version of) the AdS/CFT correspondence. 

\paragraph{Gravitational path integral} 

In section \eqref{PoincQFT}, we reviewed how a field theory with action $S[\psi]$ on a fixed spacetime manifold $M$ can be quantized in the path integral formalism by integrating over fields defined on $M$. In quantum gravity, we must also integrate over the dynamical metric, giving rise to the gravitational path integral (in Euclidean signature) 
\ali{
	Z &= \int \mathcal D g \mathcal D \psi \, e^{-S_{grav}^E[g_\ab,\psi]} 
}
with 
\ali{ 
	S_{grav}^E[g_\ab,\psi] &= -\frac{1}{16 \pi G} \int_M d^{d+1} x \sqrt{g} ( R - 2 \Lambda) + 
	S_M[g_\ab,\psi] - \frac{1}{8 \pi G} \int_{\p M} d^d x \sqrt{\gamma} \, K    	. 	 \label{SEgrav}
} 
Boundary conditions are specified at infinity $\p M$, for the matter fields as well as the geometry. 

The gravitational path integral is in most cases only a formal expression that we don't know how to actually evaluate. What we can do is approximate it by expanding it around a saddle point 
\ali{
	Z \approx e^{-S_{grav}^E[g_\ab^*,\psi^*] + S^{(1)} + \cdots}.    
}
Here, $g_\ab^*$ and $\psi^*$ are solutions to the classical equations of motion forming a `classical saddlepoint' for the path integral. It gives an order $1/G$ contribution to the exponent.  
The next term $S^{(1)}$ is the one-loop term of order $G^0$, and the dots are higher-loop contributions.  

Reinstating $\hbar$ as the unit in which the action is measured in the path integral exponent $Z = \int \mathcal D g \mathcal D \psi$ $\, e^{-\frac{1}{\hbar} S_{grav}^E[g_\ab,\psi]}$, it is clear that the expansion in $G$ from the previous paragraph is in fact an expansion in $\hbar G$. The classical limit is 
\ali{
	\hbar G \ra 0 \qquad \text{(classical)}    
} 
and quantum corrections are higher order in $\hbar G$. 
The leading $Z$ is often referred to as the \emph{semi-classical path integral} 
\ali{
	Z \approx e^{-\frac{1}{\hbar} S_{grav}^E[g_\ab^*,\psi^*]}    
	}
	(and sometimes as the `classical path integral'). 
It describes a 	\emph{semi-classical approximation} of the theory, valid for small $\hbar G$. 
Since $\hbar G$ has dimensions, of length to the power $D-2$ to be precise (when $\hbar = 1$), the limit of small $\hbar G$ more correctly is phrased as the limit where the Planck length is small compared to the curvature length scales 
\ali{
	\ell_P \ll \ell \qquad \text{(semi-classical gravity)}. \label{semicllP}
}

\section{String theory background} 

\emph{Not for exam. Only mentioning the take-away result for the progression to the next section.} 


The historical route to AdS/CFT proceeded through a string theory derivation involving different interpretations of string theoretic objects called D-branes. It lead Maldacena to formulate his AdS/CFT conjecture in 1997: 

\hphantom{~~~~} \emph{Moderate AdS/CFT} \\ 
\hphantom{~~~~} Classical supergravity in AdS$_5 \times S^5$ \quad $\stackrel{\text{is dual to}}{\sim}$ \quad 
$\mathcal N = 4$ SYM theory with gauge group $U(N)$.   

SYM is short for super Yang-Mills. There is supersymmetry on each side of the correspondence, hence the words `super'. $\mathcal N$ is the number of corresponding supercharges. The field theory is a particular CFT that lives on the $(3+1)$-dimensional conformal boundary of AdS$_5$. The classical supergravity limit on the left hand side corresponds on the right hand side to the strongly coupled limit of the CFT combined with the planar limit of large $N$. Since this large $N$ CFT has a conjectured dual description that is higher-dimensional, it is called a \emph{holographic CFT}. 

The stronger statement is that:  

\hphantom{~~~~} \emph{Strong AdS/CFT}\\ 
\hphantom{~~~~} Super string theory (type IIB) in AdS$_5 \times S^5$ \quad $\stackrel{\text{is dual to}}{\sim}$ \quad 
$\mathcal N = 4$ SYM theory with gauge group $U(N)$. 

In this case there are no restrictions on the coupling constant and size of gauge fields of the CFT.

\section{Statement of the AdS/CFT correspondence}  



The Maldacena conjecture can be formulated in a more general way as follows. 

\hphantom{~~~~} \emph{Strong AdS/CFT}\\ 
\hphantom{~~~~} A theory of QG in asymptotically AdS$_{d+1} \times Q$ \quad $\stackrel{\text{is dual to}}{\sim}$ \quad  a CFT on the cylinder $\mathbb R \times S^{d-1}$ 

Here $Q$ is some (possibly trivial) compact manifold. 
The gravity side of the duality is commonly referred to as the \emph{bulk}, and the CFT side as the \emph{boundary}. 
In this formulation, the lower-dimensional CFT living on the conformal boundary of asymptotically AdS$_{d+1}$ is providing a \emph{definition} of a theory of quantum gravity in an asymptotically AdS universe. In the modern viewpoint, AdS/CFT can be seen as a stand-alone approach to quantum gravity (without string theory as origin). The challenge is to unravel the dictionary between the two sides of the duality. The AdS spacetime is an emergent low-energy description of the bulk QG theory, and the UV-completion of the bulk is to be understood in terms of degrees of freedom of the dual CFT, which is a perfectly well-defined and UV-complete theory. 

In practice, it is already very difficult to understand 
the duality between a (semi-)classical theory of gravity and a CFT, which is the more moderate version of AdS/CFT: 

\hphantom{~~~~} \emph{Moderate AdS/CFT}\\ 
\hphantom{~~~~} Classical gravity in asymptotically AdS$_{d+1} \times Q$ \quad $\stackrel{\text{is dual to}}{\sim}$ \quad  a \emph{holographic} CFT on 
$\mathbb R \times S^{d-1}$. 

The word \emph{holographic} refers to conditions on the field theory side  that make sure the dual theory is semi-classical. The main condition is that there is a parameter $N$ in the CFT that is large. It can be the size of the gauge fields of the theory. In 2D CFT's it is related to the central charge $c$ of the theory, $N = c^2$. 
We saw in section \ref{subsVirasoro} that the central charge of a 2D CFT is related to the gravitational parameters of an AdS$_3$ gravity theory by \eqref{BrHcentralcharge}. From that expression it indeed follows that large central charge $c$ in a holographic CFT corresponds to small gravitational coupling $G \ll \ell$ or $\ell_P \ll \ell$, consistent with the semi-classical limit \eqref{semicllP} in the bulk.  

Returning to the historical path of AdS/CFT, 
we consider in the rest of this chapter the original AdS/CFT dictionary by Witten and Gubser Klebanov Polyakov '98. It provides a practical recipe for computing large $N$ CFT correlation functions from tree diagrams in AdS gravity (when the length scale of AdS is large), which was used to obtain many tests of the duality. 
We will illustrate the dictionary by using it to calculate the CFT 2-point function from a gravitational bulk perspective.

\section{GKPW dictionary and calculation of CFT 2-point function}  \label{sect2ptfcalc}


Consider a $d$-dimensional CFT with action $S_{CFT}[\mathcal O]$ for fields $\mathcal O$, and a (Euclidean signature) path integral $Z = \int \mathcal D \mathcal O \, e^{-S_{CFT}[\mathcal O]}$.   
The main objects of study that contain physics are the correlation functions 
\ali{
	\langle \mathcal O(x_1) \cdots \mathcal O(x_n) \rangle := \frac{1}{Z} \int \mathcal D \mathcal O \,\, \mathcal O(x_1) \cdots \mathcal O(x_n) e^{-S_{CFT}[\mathcal O]} . 
}
The conformal invariance of the CFT imposes that these correlators obey the condition 
\ali{
	\langle \mathcal O \cdots \mathcal O \rangle = 	\langle \mathcal O' \cdots \mathcal O' \rangle
}
short for 
$\langle \mathcal O(x_1') \cdots \mathcal O(x_n') \rangle = 	\langle \mathcal O'(x_1') \cdots \mathcal O'(x_n') \rangle$. 
(For the derivation, see the video lecture.) 

A primary field $\mathcal O(x)$ with conformal dimension $\Delta$ transforms under a (global) conformal transformation to $\mathcal O'(x')$ given in \eqref{primaryfield}. Then the above condition on the 2-point correlator of primary fields imposes 
\ali{
    \langle O(x_1)O(x_2) \rangle = \left|\frac{\p x'}{\p x}\right|^{\Delta/d}_{x=x_1} \left|\frac{\p x'}{\p x}\right|^{\Delta/d}_{x=x_2} \langle O(x'_1)O(x'_2) \rangle.  
}
This has to be true for all $x \ra x'$ conformal transformations (translations, rotations, scale transformations and special conformal transformations). This requirement is enough to fix the 2-point function to 
\ali{
	\langle O(x_1)O(x_2) \rangle =  \frac{1}{|x_1-x_2|^{2\Delta}}  
}
independent of the details of the CFT and fixed entirely by conformal symmetry. The goal of this section is to rederive this result from a calculation in AdS gravity. 

The definition of the correlation function can be rewritten in terms of a so-called generating functional 
\ali{
	Z_{CFT}[\phi_0] \equiv \frac{1}{Z} \int \mathcal D \mathcal O \, e^{-S_{CFT}[\mathcal O] + \int d^d x \, \phi_0(x) \mathcal O(x)}  \quad (= \langle  e^{\int d^d x \, \phi_0(x) \mathcal O(x)} \rangle_{CFT} )   
}
as 
\ali{
	\langle \mathcal O(x_1) \cdots \mathcal O(x_n) \rangle := \left. \frac{\delta^n Z_{CFT}}{\delta\phi_0(x_1) \cdots \delta \phi_0(x_n) } \right|_{\phi_0=0}.  \label{correldef}
}
$Z_{CFT}[\phi_0]$ is the generating functional of correlation functions of operators $\mathcal O$ that couple to a source $\phi_0$. 
Writing 
\ali{
	Z_{CFT}[\phi_0] \equiv e^{-W_{CFT}[\phi_0]} 
}
defines $W_{CFT}[\phi_0]$ as the generating functional of \emph{connected} correlation functions of operators $\mathcal O$ that couple to a source $\phi_0$. For the 2-point function: 
\ali{
	\langle \mathcal O(x_1) \mathcal O(x_2) \rangle_{c} :=  -\left. \frac{\delta^2 W_{CFT}}{\delta\phi_0(x_1) \delta \phi_0(x_2) } \right|_{\phi_0=0} = \langle \mathcal O(x_1) \mathcal O(x_2) \rangle - \langle \mathcal O(x_1) \rangle \langle\mathcal O(x_2) \rangle . 
	\label{OO2ptf}
}

\paragraph{GKPW dictionary} 
Now we come to the original AdS/CFT dictionary known as GKPW dictionary [Gubser Klebanov Polyakov Witten '98]. The bulk theory is a QG theory (e.g.~string theory), or in a certain limit a classical gravity theory, with a field content that includes matter fields and gravitational fields (the metric). 
We will denote a collection of these fields 
as $\phi^i(Z,x)$. 
The boundary theory is a CFT with a field content that includes only matter fields. We use $\mathcal O^i(x)$ to denote the collection of CFT fields that match the bulk fields $\phi^i$ in the way prescribed below.    

The GKPW dictionary says that the equivalence of the bulk and boundary theory can be expressed by basically equating the (Euclidean) path integrals of each side, in such a way that the boundary conditions on the bulk fields provide source values for corresponding CFT operators: 
\ali{
	\Aboxed{ Z_{QG}[\phi^i \,\, | \,\, \phi^i_{\p AdS} = \phi^i_0] \quad = \quad Z_{CFT}[\phi^i_0] }  
}
with $Z_{CFT}$ the generating functional in the CFT that reduces to the CFT path integral $Z$ when $\phi_0 \ra 0$. 

The dictionary simplifies when the bulk is described by the gravitational sector and this gravitational sector can be described semi-classically through a saddle-point approximation, $Z_{QG} \ra Z_{grav} \approx \exp{-S_{grav}[\phi^i_*]}$: 
\ali{
	\Aboxed{ S_{grav}[\phi^i_* \,\, | \,\, \phi^i_{*,\p AdS} = \phi^i_0] \quad = \quad W_{CFT}[\phi^i_0] } 
}  
with $W_{CFT}$ the generating functional of connected correlators of CFT operators $\mathcal O^i(x)$ coupling to the boundary values $\phi^i_0(x)$ of bulk fields $\phi^i_*(Z,x)$ that solve the classical bulk EOM. 
The action evaluated on the classical solutions $S_{grav}[\phi^i_*]$ is called the on-shell action. 

Some examples of $\int \phi_0^i \mathcal O^i$ are: 
\begin{itemize} 
\item $\int \phi_0 \mathcal O$ for a scalar CFT operator $\mathcal O$ with conformal dimension $\Delta$ and spin $s = 0$, coupling to a scalar bulk field $\phi$ with mass $m$ and spin $s=0$. 
\item $\int A^\mu_0 \, J_\mu$ for a conserved CFT current $J_\mu$ with conformal dimension $\Delta=d-1$ and spin $s = 1$, coupling to a vector bulk field $A^\mu$ with mass $m=0$ and spin $s=1$. 
\item $\int g^\mn_0 T_\mn$ for a conserved CFT stress tensor $T_\mn$ with conformal dimension $\Delta=d$ and spin $s = 2$, coupling to a bulk graviton $g^\mn$ with mass $m=0$ and spin $s=2$.
\end{itemize} 

\paragraph{Application of GKPW dictionary} 


Consider now a scalar CFT operator $\mathcal O$ and its 2-point function \eqref{OO2ptf}. The dictionary tells us we should be able to calculate this correlator from an AdS gravity perspective as 
\ali{
	\langle \mathcal O(x_1) \mathcal O(x_2) \rangle_{c} =  -\left. \frac{\delta^2 S_{grav}^{onshell}}{\delta\phi_0(x_1) \delta \phi_0(x_2) } \right|_{\phi_0=0} \quad \stackrel{?}{=} \frac{1}{|x_1-x_2|^{2\Delta}} .  
}
Namely, we need to compute the classical action in AdS as a functional of the boundary conditions. 

The CFT operator $\mathcal O$ couples to a massive scalar field $\phi$ in the gravitational bulk theory \eqref{Sgrav}. As discussed in section \ref{sectionAdSgrav}, the set of bulk EOM for the gravitational action  \eqref{Sgrav} with matter action\footnote{
	$S_M$ is allowed to contain any higher order terms in $\phi$, e.g.~$b \, \phi^3 + \lambda \phi^4 + \cdots$, but as they won't contribute to the 2-point function (thanks to $\phi_0 \ra 0$ at the end of the prescription), they are left out from the start. 
} 
\ali{
	S_M = \frac{1}{2} \int d^{d+1} x \sqrt{g} \left( g_\ab \p^\alpha \phi \p^\beta \phi + m^2 \phi^2  \right)  
}
is 
\ali{
	&R_\ab - \frac{1}{2} R \, g_\ab + \Lambda g_\ab = 8 \pi G \, T_\ab^M(\phi) \\ 
	&\Box \phi - m^2 \phi = 0. 
}
Now we make the assumption that the bulk field $\phi$ is light enough so that there is no backreaction on the metric. Then the Einstein equation reduces to the vacuum Einstein equation and the set of equations to 
\ali{
	&R_\ab - \frac{1}{2} R \, g_\ab + \Lambda g_\ab = 0 \\ 
	&\Box \phi - m^2 \phi = 0 . 
}
We take as solution to the first equation the (Euclidean) AdS$_{d+1}$ metric 
\ali{
	ds^2 = g_\ab
	dx^\alpha dx^\beta = \frac{\ell^2}{Z^2} (dZ^2 + \delta_\mn dx^\mu dx^\nu) 
}
and we are left to determine the solution $\phi_*$ to the equation $\Box \phi - m^2 \phi = 0$ with $\Box \equiv \Box_{AdS}$, and evaluate the action $S_M[\phi_*(\phi_0)]$.  

By partial integration, the action can be rewritten as a part that will vanish on-shell and a boundary contribution\footnote{
	Some GR notation: 
	\ali{
		\nabla_\alpha \phi &= \p_\alpha \phi \\ 
		\nabla_\alpha p_\beta &= \p_\alpha p_\beta - \Gamma^\mu_{\ab} p_\mu \\ 
		\nabla_\al A^\al &= \frac{1}{\sqrt{g}} \p_\al (\sqrt{g} A^\al) \\ 
		\Box \phi &= \nabla_\al \nabla^\al \phi = \nabla_\al \p^\al \phi = \frac{1}{\sqrt{g}} \p_\al (\sqrt{g} g^\ab \p_\beta \phi) 	
		}
	} 
\ali{
	S_M = \frac{1}{2} \int d^{d+1} x \sqrt{g} \left( -\phi( \Box - m^2 ) \phi \right) + \frac{1}{2} \int d^{d+1} x \,\, \p_\alpha \left(\sqrt{g} g^\ab \phi \p_\beta \phi \right)  .  \label{SMaction}
}
To solve the AdS wave equation we start by exploiting the $d$-dimensional translation invariance $x^\mu \ra x^\mu + a^\mu$ of the background 
to Fourier decompose the scalar field, i.e.~write it in terms of plane waves (eigenmodes of the momentum operator) 
\ali{
	\phi(Z,x^\mu) = e^{i k_\mu x^\mu} f_k(Z).   
}
Then $f_k(Z)$ should satisfy 
\ali{
	\left( g^\mn k_\mu k_\nu - \frac{1}{\sqrt{g}} \p_Z(\sqrt{g} g^{ZZ} \p_Z) + m^2 \right) f_k(Z) = 0 . 
}
Or, with $g^\mn = \left( \frac{Z}{\ell} \right)^2 \delta^\mn$, 
\ali{
	\left(Z^2 k^2 - Z^{d+1} \p_Z(Z^{-d+1} \p_Z) + m^2 \ell^2 \right) f_k(Z) = 0. 
}
We could try a power law solution $f_k = Z^\Delta$: 
\ali{
	(k^2 Z^2 - \Delta(\Delta-d) + m^2 \ell^2 ) Z^\Delta = 0.  
	} 
Only when $Z$ is small, such a power law solution works, and we find that asymptotically (in the near-boundary limit $Z \ra 0$), $f_k = Z^\Delta$ gives a solution for $\Delta$ satisfying 
\ali{
	\Aboxed{ \Delta(\Delta - d) = m^2 \ell^2. }  \label{Deltaofm}
} 
There are two such values for the conformal dimension, 
\ali{
	\Delta_\pm = \frac{d}{2} \pm \sqrt{\left(\frac{d}{2}\right)^2 + m^2 \ell^2 } 
}
and the asymptotic behavior of the scalar field is given by 
\ali{
	\phi(Z,x^\mu) \,\, \stackrel{Z \ra 0}{\sim} \,\, Z^{\Delta_-} A(x^\mu) + Z^{\Delta_+} B(x^\mu) . 
}
A few comments: \begin{itemize} 
	\item The solution in $Z^{\Delta_-}$ is bigger near the boundary $Z \ra 0$ 
	\item $\Delta_+> 0$ for all $m$, $\Ra  Z^{\Delta_+}$ decays near the boundary 
		\item $\Delta_+ + \Delta_- = d$  
	\end{itemize}
It follows that a good boundary condition that will allow a solution is 
\ali{
	\lim_{Z \ra 0} \phi(Z,x) = Z^{\Delta_-} \phi_0(x) + \cdots 
}
or $\phi(Z=\epsilon,x) = \epsilon^{\Delta_-} \phi_0(x)$. 
A solution of the AdS wave equation with this boundary condition is given by [Witten '98]
\ali{
	\phi_*(Z,x) &= \int d^d x' \, K_{Witten}(Z,x;x') \phi_0(x') \\ 
	K_{Witten}(Z,x;x') &= c_d^{-1} \, \frac{Z^{\Delta_+}}{\left( Z^2 + |x-x'|^2 \right)^{\Delta_+}}, \qquad c_d = \frac{\pi^{d/2} \Gamma(\Delta_+ - \frac{d}{2})}{\Gamma(\Delta_+)}.    
}
It is left as an exercise to check this forms a solution, by checking that $K_{Witten}$ satisfies $(\Box - m^2) K_{Witten} = 0$ and $\lim_{Z\ra 0} K_{Witten} = Z^{\Delta_-} \delta^{(d)}(x-x')$. The function $K_{Witten}(Z,x;x')$ is called the \emph{boundary-to-bulk propagator} because it propagates a field $\phi_0(x')$ at the boundary to a field $\phi(Z,x)$ in the bulk. 

Once the imposed boundary condition is obeyed (and thus the leading asymptotic behavior of $K_{Witten}$ is known, by construction), the subleading behavior in $Z$ is fixed by the full solution. We can read off: 
\ali{
	\lim_{Z\ra 0} K_{Witten} = Z^{\Delta_-} \left(  \delta^{(d)}(x-x') + \mathcal O(Z^2) \right) + Z^{\Delta_+} \left( \frac{c_d^{-1}}{|x-x'|^{2\Delta_+}} + \mathcal O(Z^2) \right) 
	}
Now we can compute the on-shell action by plugging in the solution $\phi_*$ into \eqref{SMaction}.  The only contribution will come from the boundary term, which by Stokes theorem reduces to\footnote{See the derivation of Stokes theorem in \cite{Poisson} for details on the subtlety of the absence of  $\alpha,\beta = x$ contributions. Also, an alternative way of writing the Stokes theorem is $\int_M\sqrt{g} \nabla_\al J^\al = \int_{\p M} \sqrt{\gamma} \,  n_\al J^\al$, for conventions see also \cite{Poisson}.} 
\ali{
	S_M[\phi_*] &= \frac{1}{2} \int_{\p AdS} d^{d} x \,\, \left. \sqrt{g} g^{ZZ} \phi_*(Z,x) \p_Z \phi_*(Z,x) \right|_{Z = \epsilon \ra 0} \\
	&= \frac{1}{2} \int  d^{d} x_1  \int  d^{d} x_2 \phi_0(x_1) \phi_0(x_2) \int  d^{d} x \left. \frac{K_{Witten}(Z,x;x_1) \p_Z K_{Witten}(Z,x;x_2)}{Z^{d-1}} \right|_{Z = \epsilon \ra 0}
}
where in the second line we used $\sqrt{g} g^{ZZ} =  \left( \frac{\ell}{Z} \right)^{d-1}$ and then set $\ell=1$. 
From the asymptotic solution for $K_{Witten}$ we have 
\ali{
	&\int  d^{d} x  \frac{K_{Witten}(Z,x;x_1) \p_Z K_{Witten}(Z,x;x_2)}{Z^{d-1}} \nonumber \\ 
	&= \Delta_- Z^{2 \Delta_- - d} \delta(x_1-x_2) + \frac{d}{c_d} \frac{1}{|x_1-x_2|^{2 \Delta_+}} + \frac{\Delta_+}{c_d^2} Z^{2\Delta_+-d} \int \frac{d^d x}{|x-x_1|^{2\Delta_+} |x-x_2|^{2\Delta_+}} + \cdots . \label{eqforref}
}
Because $2 \Delta_+ - d > 0$, the last term will give a vanishing contribution for $Z \ra 0$, the second term gives a finite contribution and the first term an infinite contribution, since $2 \Delta_- - d < 0$. 
This divergence must be removed by renormalizing the bulk action by adding a counterterm $S_{ct} = -\frac{\Delta_-}{2} \epsilon^{\Delta_--d/2} \int d^d x \, \phi_0(x)^2$. The renormalized onshell action $S_{M,ren} = S_M + S_{ct}$ with slightly rescaled $\phi_0$ is finally given by 
\ali{
	S_{M,ren}[\phi_*(\phi_0)] = -\frac{1}{2} \int  d^{d} x_1  \int  d^{d} x_2 \frac{\phi_0(x_1) \phi_0(x_2)}{|x_1-x_2|^{2 \Delta_+}} . 
}
Then finally, 
\ali{
\langle \mathcal O(x_1) \mathcal O(x_2) \rangle_{c} =  -\left. \frac{\delta^2 S_{grav}^{onshell}}{\delta\phi_0(x_1) \delta \phi_0(x_2) } \right|_{\phi_0=0} = -\left. \frac{\delta^2 S_{M,ren}[\phi_*]}{\delta\phi_0(x_1) \delta \phi_0(x_2) } \right|_{\phi_0=0} =  \frac{1}{|x_1-x_2|^{2\Delta_+}}.  
}
The correct CFT result has followed from a gravity calculation! 

We conclude that a bulk scalar field $\phi$ with mass $m$ couples to a CFT operator $\mathcal O$ with scaling dimension 
\ali{
	\Delta = \Delta_+ = \frac{d}{2} + \sqrt{\left(\frac{d}{2}\right)^2 + m^2 \ell^2 } . 
}
The asymptotic behavior of the bulk field can be rewritten in terms of $\Delta$ as 
\ali{
	\Aboxed{ \phi(Z,x^\mu) \,\, \stackrel{Z \ra 0}{\sim} \,\, Z^{d-\Delta} \phi_0(x^\mu) + Z^{\Delta} \tilde \phi(x^\mu) } .	\label{phias}
}
This is an important ingredient of AdS/CFT duality: \emph{Local CFT operators $\mathcal O$ 
	map to light bulk fields $\phi$. } 

You can moreover check that taking the variation of $S_{grav}^{onshell}$ with respect to $\phi_0$ once extracts the field $\tilde \phi(x^\mu)$, such that by the AdS/CFT dictionary the expectation value of $\mathcal O$ is proportional to $\tilde \phi$,  
\ali{
	\vev{\mathcal O(x^\mu)} \sim \tilde \phi(x^\mu). 
}
To conclude, the leading asymptotic behavior of the bulk field $\phi$ gives the source $\phi_0$ that couples to $\mathcal O$, the subleading behavior gives $\vev{\mathcal O}$. 

\paragraph{UV/IR duality} 
The renormalization we had to include in the calculation is called \emph{holographic renormalization}. It just reflects the renormalization on the CFT side of short-distance divergences $x_1 \ra x_2$ in the correlators. Such short-distance divergences typically take the form of contact terms $\sim \delta(x_1-x_2)$, and are called \emph{UV divergences}. It is clear from the first term in \eqref{eqforref} that the contact term of the CFT is tied to the $Z \ra 0$ divergence in the bulk. But from the bulk perspective, the divergence as $Z \ra 0$ is a large-distance or \emph{IR divergence}, originating from the infinite size of the AdS spacetime. It is a feature of AdS/CFT that IR divergences in the bulk map to UV divergences in the boundary. This is sometimes referred to as \emph{UV/IR duality}.

\paragraph{Relation between $m$ and $\Delta$ from casimir} There is a faster way to derive or remember the important relation \eqref{Deltaofm} between the bulk field mass $m$ and the boundary field conformal dimension $\Delta$. 

We know the bulk theory is a QFT on a fixed AdS$_{d+1}$ background, since we have neglected backreaction of the light fields on the metric. Upon making this assumption, the full diff invariance of $S_M$ in $S_{grav}$ reduces to invariance under transformations that leave the background AdS$_{d+1}$ invariant. These are the isometries of AdS$_{d+1}$, with group $SO(d+1,1)$.  

Just as a QFT on a fixed Minkowski background has Poincar\'e symmetry with an associated casimir operator $c_2^{Poinc}$ that indicates the field content should be labeled by casimir eigenvalues mass $m$ and spin $s$ (section \ref{sectionCasimir}), our bulk QFT on fixed AdS$_{d+1}$ has a casimir operator $c_2^{AdS}$ that indicates how to label the bulk field $\phi$. We expect that label to be the bulk mass $m$ for a spinless field $\phi$, and we show this now.   

The casimir is  given by  $c_2^{AdS} = \frac{1}{2} J_{ab} J^{ab}$ 
with $J_{ab}$ the generators of the AdS isometry. When working on a field $\phi$, it can be worked out that $[c_2,\phi] = \Box \phi$ with $\Box \equiv \Box_{AdS}$. This is also true when the bulk is a classical theory, since the casimir is a mathematical concept associated with a Lie algebra, e.g.~given by the angular momentum operator squared $J^2$ in QM, but also applicable in differential geometry, where the quadratic casimir is defined as the $G$-invariant second order differential operator on a manifold $M$ with symmetry group $G$. This last definition immediately identifies the box operator $\Box$ as the casimir. In any case, we have on the bulk side that the action of the casimir (associated with the isometry group of AdS) on the bulk field $\phi$ gives $\Box \phi$ which by the EOM we know has to equal $m^2 \phi$. In conclusion 
\ali{
	[c_2^{AdS},\phi ] = m^2 \phi . 
}

From the point of view of the boundary CFT, the casimir associated with the isometry group of AdS$_{d+1}$ is the conformal casimir \eqref{c2conf} (consistent with the conclusions of section \ref{sectionAdSsymm}). It acts on scalar CFT operators with dimension $\Delta$ as 
\ali{
	[c_2^{conf},\mathcal O] = \Delta(\Delta-d) \mathcal O . 
}

This shows the casimir gives us the relation between bulk and boundary labels 
\ali{
	\Delta(\Delta - d) = m^2 .  
}
To fully derive this we would need the explicit relation between $\phi$ and $\mathcal O$, which is known as the HKLL representation of $\phi$ in the AdS/CFT program of \emph{bulk reconstruction}. But we will not cover this in the course and the relation between $\Delta$ and $m$ is already apparent without going into the further details.

\paragraph{Witten diagrams} 

The theories of the standard model typically use S-matrices to model the physics of scattering experiments. 
This is because these are QFT's with massive particles. 
In CFT's, massive particles are not allowed by symmetry, and in a sense the concept of a mass was replaced by the concept of a conformal dimension, as a labeling eigenvalue of the field content (as discussed around \eqref{c2conf}). The physics of a CFT is then described in terms of correlators rather than scattering matrices.    

We now see that the AdS/CFT duality allows to associate a mass to the CFT fields, by letting them couple to higher-dimensional, massive AdS fields! 

\begin{figure}[h!] 
	\centering \includegraphics[width=13cm]{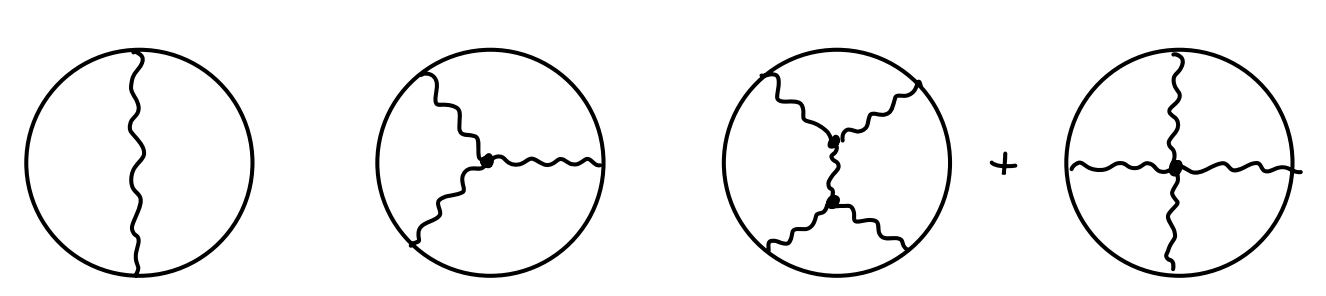} 
	\caption{
		An $n$-point correlator in the CFT is calculated from Witten diagrams with $n$ boundary vertices and thus $n$ boundary-to-bulk propagators. (We didn't discuss bulk-to-bulk propagators which arise in the $n>2$ case.) See e.g.~\cite{Kiritsis} chapter 14. 
	} \label{figKiritsis}
\end{figure}

Indeed, the duality also maps the physics content from correlators in the CFT to scattering diagrams in the bulk. The scattering diagrams that are used to calculate CFT correlation functions through the GKPW dictionary are called \emph{Witten diagrams}. Some of them are pictured in figure \ref{figKiritsis}. (They are tree diagrams. For some discussion on the inclusion of loop corrections to the Witten diagrams, see section 15.3 of \cite{HartmanQG}. That section also introduces the concept of \emph{multi-trace operators}, for interested students.)   

\paragraph{Other examples: global/gauge duality and dual of stress tensor} Coming back to the other examples of $\int \phi_0^i \mathcal O^i$, there are some comments to make. One, for the spin 1 case: a Noether current $J^\mu$ for any continuous boundary global symmetry is dual to a massless gauge field $A_\mu$ in the bulk. This reflects another feature of AdS/CFT sometimes referred to as \emph{global/gauge} duality, that a boundary global symmetry corresponds to a bulk gauge symmetry.  

Two, for the spin 2 case: the stress tensor of the CFT couples to the boundary value 
of the bulk metric $g_\ab$, mapping an essential component of the CFT to an essential component of bulk gravity. 
In FG gauge, using the notation of \eqref{FGgaugemetric}, the equivalent of \eqref{phias} becomes 
\ali{
	\tilde \gamma_\mn(Z,x^\mu) \,\, \stackrel{Z \ra 0}{\sim} \,\, \delta_\mn + Z^{d} \vev{T_\mn(x^\mu)}
}
with the subleading asymptotic behavior written in terms of the dual operator (see Exercise). 
This expression follows from solving the wave equation for the gravitational fluctuations $\tilde \gamma_\mn^{(2)}$ (which are massless, consistent with the masslessness of the graviton upon quantization) for the near-boundary behavior. See also Exercise 2 of this chapter.  
This means that the asymptotic gravitational stress tensor, by AdS/CFT, is equal to the expectation value of the CFT stress tensor 
\ali{
	T_\mn^{BY} \quad \stackrel{AdS/CFT}{=} \quad \vev{T_\mn} \label{BYdual}
}
as anticipated in the conclusions of section \ref{subsVirasoro}.



\newpage 
\section*{Exercises} 

\textbf{\underline{\smash{Exercise 1. Bulk-to-boundary propagator}}} 
 
 Show that Witten's bulk-to-boundary propagator 
 \ali{
 	K_{Witten}(Z,x;x') = \frac{1}{c_d} \frac{Z^{\Delta_+}}{(Z^2 + |x-x'|^2)^{\Delta_+}}, \qquad c_d = \frac{\pi^{d/2} \Gamma(\Delta_+ - d/2)}{\Gamma(\Delta_+)}  
 }	
 satisfies 
 \ali{
 	\lim_{Z \ra 0} K_{Witten} = Z^{\Delta_-} \delta^{(d)}(x-x')  
 }
 and 
 \ali{
 	(\Box - m^2) K_{Witten} = 0.  
 }
 You can use Mathematica to check for the Klein-Gordon equation.
 \\*
 \\*
 \\*
 \textbf{\underline{\smash{Exercise 2. Wave equation for metric fluctuations}}} 
 
 Consider a linearized fluctuation $h_{\mu\nu}$ of an AdS$_{d+1}$ spacetime in Poincar\'e coordinates and holographic gauge ($g_{ZZ}=g_{Z\mu}=0$): 
 \ali{ds^2=\frac{l^2}{Z^2}\left(dZ^2+g_{\mu\nu}dx^{\mu}dx^{\nu}\right)=\frac{l^2}{Z^2}\left(dZ^2+\left(\eta_{\mu\nu}+\frac{Z^2}{l^2}h_{\mu\nu}\right)dx^{\mu}dx^{\nu}\right).}
 
 To linear order in $h_{\mu\nu}$, the $ZZ$, $Z\nu$ and the trace of the $\mu\nu$ components of the vacuum Einstein's equations are given by:
 \ali{ZZ: \; \; \; \left(\partial_{Z}^2+\frac{3}{Z}\partial_Z \right) h=0,}
 \ali{Z\nu: \; \; \; \left(\partial_Z+\frac{2}{Z}\right)(\partial_{\mu}h^{\mu\nu}-\partial^{\nu}h)=0,}
 \ali{trace: \; \; \; \left(\partial_Z^2-\frac{2d-5}{Z}\partial_Z-\frac{4(d-1)}{Z^2}\right)h+2(\partial_{\mu}\partial^{\mu}h-\partial_{\mu}\partial_{\nu}h^{\mu\nu})=0.}
 
 Here $h=h_{\;\mu}^{\mu}$ and $\partial_Z^2=\partial_Z \partial_Z$.
 
 \textbf{a)} Show that 
 normalizable behavior at the boundary requires to impose $h=0$ and $\partial_{\mu}h^{\mu\nu}=0$.
 \\*

 Using tracelesness and conservation of the metric fluctuation, the $\mu\nu$ components of Einstein's equation simplify to:
 \ali{\left(\partial_{\alpha}^2+\partial_Z^2+\frac{5-d}{Z}\partial_Z-\frac{2(d-2)}{Z^2}\right)h_{\mu\nu}=0}
 
 \textbf{b)} Show that this implies that
 \ali{
 	\Box \phi_{\mu\nu} = \Box (Z^2 h_\mn) = 0,
 }
 meaning that each component of $\phi_{\mu\nu}$ satisfies a massless wave equation in $AdS_{d+1}$.
 	\\*
 	\\*
 	\textbf{c)} Argue that a consistent boundary behavior for the metric fluctuation is given by
 	
 	\ali{\phi_{\mu\nu}(x^{\rho},Z)\sim Z^d \langle T_{\mu\nu}(x^{\rho}) \rangle,}
 	
 	where $T_{\mu\nu}$ is the stress-energy tensor of the CFT on the boundary of AdS.

\chapter{Adding scales to AdS/CFT}  
 
\section{Finite temperature QFT}  \label{finiteTQFT}

A thermodynamic system in thermal contact with its environment at temperature $T$, and with 
volume $V$ and number of particles $N$ 
kept constant, is referred to as a canonical ensemble; its possible discrete quantum states as microstates. 
Denoting the energy of the system in microstate $s$ as $E_s$, the canonical partition function is given by  
\begin{equation} \label{Z}
	Z = \sum_s e^{-\beta E_s}, \quad \beta = \frac{1}{k_B T},
\end{equation}
with $k_B$ the Boltzmann constant which we will set equal to 1 from here on. 

Consider a system at temperature $T$ consisting of a field $\phi(\vec x, t)$ that is in the state $|\phi_a\rangle$ at time $t$. 
The partition function (\ref{Z}) can be rewritten for this system as 
\begin{align}
	Z &= \int d\phi_a \langle \phi_a | e^{-\beta \hat H} |  \phi_a \rangle = \text{Tr}\left(e^{-\beta \hat H}\right), \quad \beta = \frac{1}{T}. \label{partitiefunctie}
\end{align}
In the $T=0$ quantum field theory for $\phi(\vec x, t)$, the transition amplitude for a transition from an initial state $\phi_a(\vec x)$ at time $t$ to an end state $\phi_b(\vec x)$ at time $t'$ is given by the path integral 
\begin{equation}
	\langle \phi_b | e^{-i \hat H (t'-t)} | \phi_a \rangle  =  \int_{\phi_a(\vec x) = \phi(\vec x, t)}^{\phi_b(\vec x) = \phi(\vec x, t')} \mathcal D \phi \hspace{1mm} e^{i S[\phi]}.
\end{equation}
After performing a Wick rotation ($t \rightarrow -it_E$, $t'-t \rightarrow -i\beta$, $iS \rightarrow -S_E$) and identifying initial and end state, we find for the Euclidian path integral along a closed path: 
\begin{equation}
	\int d \phi_a \langle \phi_a | e^{-\beta \hat H } | \phi_a \rangle  =  \int_{\phi(\vec x, t_E+ \beta) = \phi(\vec x, t_E)} \mathcal D \phi \hspace{1mm} e^{- S_E[\phi]}.
\end{equation}
From comparison with (\ref{partitiefunctie}) we conclude that the Euclidean path integral along a closed path $\phi(\vec x, t_E+ \beta) = \phi(\vec x, t_E)$ is equal to the canonical partition function at temperature $T=\frac{1}{\beta}$:
\begin{equation}
	Z = \int_{\phi(\vec x, t_E+ \beta) = \phi(\vec x, t_E)} \mathcal D \phi \hspace{1mm} e^{- S_E[\phi]}.
\end{equation}
In other words: \emph{turning on temperature in the field theory corresponds to compactification of Euclidean time} 
\ali{
	t_E \sim t_E + \beta, \qquad \beta = \frac{1}{T}. \label{QFTbeta}
}

General reference: Polchinski \cite{Polchinski} Appendix A on path integrals.

\section{Euclidean Schwarzschild black hole} \label{sectSSbh}

Another instance where compactification of Euclidean time naturally appears is in the context of black holes. Let's consider the most familiar one first, the flat space Schwarzschild solution. Wick-rotated to Euclidean signature, it is given by 
\ali{
	ds^2 = \left( 1 - \frac{2m}{r} \right) dt_E^2 + \left( 1 - \frac{2m}{r} \right)^{-1} dr^2 + r^2 d\Omega^2   
}
with the parameter $m$ related to the mass $M$ of the black hole by $2m \equiv 2 G M$. 
We can do a coordinate transformation to $\tilde r = r - 2m$ and expand in $\tilde r$ to zoom in on the near-horizon region 
\ali{
	ds^2 = \frac{\tilde r}{2m} dt_E^2 + \frac{2m}{\tilde r} d\tilde r^2 + 4m(m+\tilde r) d\Omega^2 \, + \mathcal O(\tilde r^2)  
} 
or 
\ali{
	ds^2 = \left(\frac{\rho}{4m} \right)^2 dt_E^2 + d\rho^2 + (4m^2+ \frac{\rho^2}{2} ) d\Omega^2  \, + \mathcal O(\rho^4) 
}
in terms of yet another radius $\rho$ defined by $\rho^2 = 8 m \tilde r$.  

\begin{figure}
	\centering \includegraphics[width=10cm]{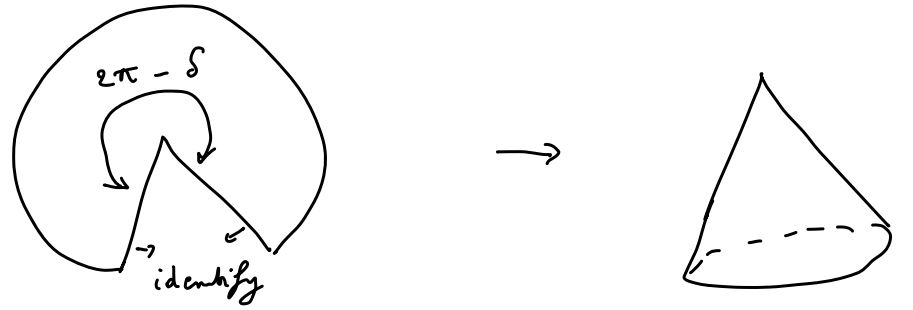} \caption{Conical singularity: Flat space in polar coordinates $dr^2 + r^2 d\theta^2$, $\theta \sim \theta + 2\pi$, would not be smooth but have a conical singularity in the origin $r=0$ if $\theta$ was identified with an angle smaller than $2\pi$. See e.g.~\cite{Nastase}. 
	}  \label{fig-consing} 
\end{figure}

To avoid a conical singularity at $\rho = 0$ ($r = 2m$) in the $(\rho, t_E)$ plane of the near-horizon Schwarzschild geometry, we require that $\frac{t_E}{4m}$ has period $2\pi$. See figure \ref{fig-consing}. 
Schwarzschild is a spherically symmetric solution of the vacuum Einstein equations and there is no source at the location of the horizon that could justify an infinite curvature in the form of a conical singularity there. Therefore we should require smoothness, which naturally imposes Euclidean time to be 
compactified 
\ali{
	t_E \sim t_E + 8 \pi G M. \label{SStE}
}
A QFT placed on a Euclidean Schwarzschild background geometry would then automatically feel a temperature 
\ali{
	T = \frac{1}{8\pi G M} . 
}

\section{Finite temperature AdS/CFT}

We start from a (Euclidean signature) AdS bulk, for historical reasons a 5-dimensional one:  
\ali{
	ds^2_{AdS} &= \cosh^2 \frac{\tilde \rho}{\ell} dt_E^2 + d \tilde \rho^2 + \ell^2 \sinh^2 \frac{\tilde \rho}{\ell} d\Omega_3^2 \\ 
	&=  \left( 1 + \frac{r^2}{\ell^2} \right) dt_E^2 + \frac{dr^2 }{ 1 + \frac{r^2}{\ell^2}} + r^2 d\Omega_3^2 . 
} 
The metric is given here in two different global AdS coordinates, which are related by $r = \ell \sinh \frac{\tilde \rho}{\ell}$. 
At $\tilde \rho$ or $r$ to infinity, this AdS$_5$ metric has a cylindrical conformal boundary  
\ali{
	\tilde \gamma^{(0)}_\mn dx^\mu dx^\nu =  dt_E^2 + \ell^2 d\Omega_3^2 \qquad (\mathbb R \times S_3)  \label{Dirichlet}
}
on which the dual 4-dimensional CFT lives. In the opposite reasoning, we could start from the CFT living on the flat metric $\tilde \gamma^{(0)}_\mn$ above, and determine the dual asymptotically AdS spacetime by solving the gravitational theory $S_{grav}$ \eqref{Sgrav} for a metric of the form \eqref{FGgaugemetric} with Dirichlet boundary condition \eqref{Dirichlet}. The 
lowest energy 
solution is the AdS$_5$ metric above.  

Now from section \ref{finiteTQFT} \eqref{QFTbeta} we know that we can turn on temperature in the CFT by compactifying the Euclidean time direction $t_E$ with a period $\beta$ equal to the inverse temperature. 
The finite temperature CFT then lives on  
\ali{
	\tilde \gamma^{(0)}_\mn dx^\mu dx^\nu =  dt_E^2 + \ell^2 d\Omega_3^2, \quad t_E \sim t_E + \beta \qquad (S_1 \times S_3) .  \label{DirichletT}
}
We can again determine the dual asymptotically AdS spacetime by solving the gravitational theory $S_{grav}$ \eqref{Sgrav} for a metric of the form \eqref{FGgaugemetric}, but this time with an altered Dirichlet boundary condition \eqref{DirichletT}. 

This time there are two solutions for the dual asymptotically AdS spacetime. One is the AdS$_5$ metric above but with compatified Euclidean time 
\ali{
	ds^2_{\text{thermal }AdS} &=  \left( 1 + \frac{r^2}{\ell^2} \right) dt_E^2 + \frac{dr^2 }{ 1 + \frac{r^2}{\ell^2}} + r^2 d\Omega_3^2 , \qquad t_E \sim t_E + \beta 
}
for arbitrary inverse temperature $\beta$. This is the thermal AdS background. 

The second solution is the AdS$_5$-Schwarzschild metric 
\ali{
	ds^2_{AdS\text{-}SS} &=  \left( 1 + \frac{r^2}{\ell^2} - \frac{\mu}{r^2} \right) dt_E^2 + \frac{dr^2 }{ 1 + \frac{r^2}{\ell^2} - \frac{\mu}{r^2}} + r^2 d\Omega_3^2 
}
with $\mu$ proportional to the mass $M$ of the black hole. 
It is a black hole solution with a horizon at $g_{t_Et_E}=0$ or $ 1 + \frac{r^2}{\ell^2} - \frac{\mu}{r^2} = 0$. The outer horizon radius is then  
\ali{
	r_+^2 = -\frac{\ell^2}{2} + \frac{\ell}{2} \sqrt{\ell^2 + 4 \mu} .  
}
To avoid a conical singularity at $r=r_+$ in the $(r,t_E)$ plane of the near-horizon AdS-Schwarzschild geometry, we require that time is periodically identified with a specific period $\beta$:  
\ali{
	t_E \sim t_E + \beta, \qquad \beta = \frac{4\pi}{g_{t_Et_E}'(r)|_{r=r_+}} = \frac{4 \pi \ell^2 r_+}{4 r_+^2 + 2 \ell^2} .  
}
This follows from a  completely analogous derivation as that of the periodicity of flat Schwarzschild Euclidean time \eqref{SStE}. 

\begin{figure}
	\centering \includegraphics[width=6.5cm]{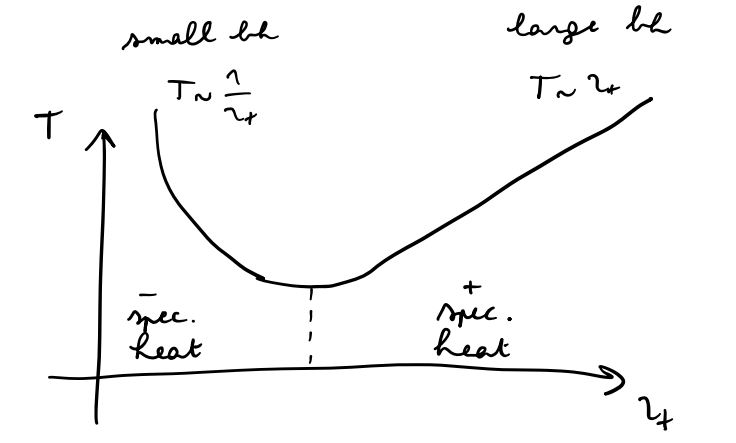} \qquad \qquad 
	\includegraphics[width=6.5cm]{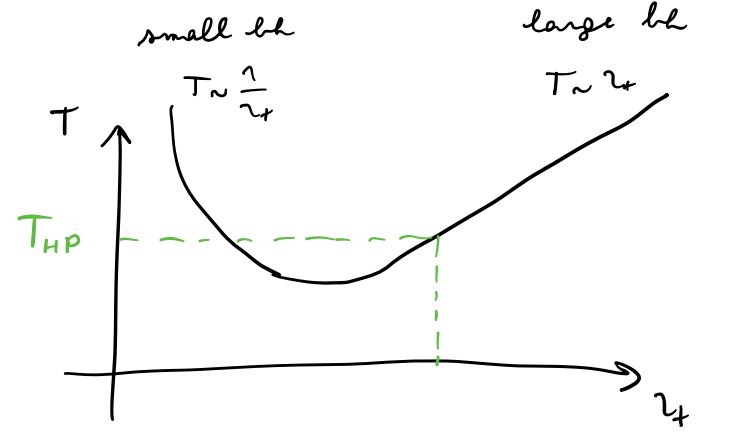}
	\caption{$T(r_+)$ profile for AdS black hole. 
	} \label{fig-TAdSbh} 
\end{figure}

From the point of view of observers on the AdS-Schwarzschild background, the AdS black hole thus radiates as a black body at a temperature 
\ali{
	T = \frac{4 r_+^2 + 2 \ell^2}{4 \pi \ell^2 r_+}. 
} 
This is the \emph{Hawking temperature} of the black hole.  
The profile of the horizon-dependence of the temperature is sketched in figure \ref{fig-TAdSbh}. The curve has a minimum at $\p T/\p r_+ = 0$ at 
\ali{
	r_{+,min} = \frac{\ell}{\sqrt{2}}, \qquad T_{min} = \frac{\sqrt{2}}{\pi \ell} = \frac{1}{\pi r_+} = \frac{d}{4\pi r_+} 
}
where I included the more general formula for general dimension $d$. 

As a thermal object, the AdS black hole has a specific heat defined as 
\ali{
	C = \frac{\p M}{\p T}. 
}
The sign of the specific heat is determined by the sign of $\p T/\p r_+$. (This is because $\p M/ \p r_+ > 0$ for $r_+ > 0$.) Namely, small black holes ($r_+ \ll \ell$ and $T \sim 1/r_+$) will have negative specific heat, while large black holes ($r_+ \gg \ell$ and $T \sim r_+$) will have positive specific heat. This means small AdS black holes are thermodynamically unstable, and will Hawking evaporate (note that Witten \cite{Witten98} remarks that ``The AdS/CFT correspondence gives apparently a completely unitary, QM description of the decay [of the small AdS black hole]"). The large AdS black holes are thermodynamically stable. 

There are two competing solutions and we have to calculate the on-shell Euclidean actions to determine which geometry dominates the gravitational path integral $Z = \int \mathcal D g e^{-S_E}$. The dominant geometry is the one with the smallest Euclidean action.   
The result for the difference in Euclidean actions is given by \cite{Witten98} 
\cite{HawkingPage}
\ali{
	\Delta S_E \, \equiv \, S_E(g_* = AdS\text{-}SS) -   S_E(g_* = \text{thermal }AdS) \quad = \frac{\pi r_+^2 (1 - r_+^2)}{1 + 3 r_+^2} . 
}
It is negative for sufficiently large $r_+$ (corresponding to sufficiently large $T$). We can conclude that the thermal AdS background dominates at low temperature while the large AdS black hole background dominates at high temperatures. The critical temperature at which $\Delta S_E = 0$ is called the \emph{Hawking-Page temperature} 
\ali{
	T_{HP} = \frac{3}{2\pi \ell} \qquad \text{(Hawking-Page temperature)}. 
}

\fbox{ \parbox{0.97\textwidth}{  For temperatures higher than $T_{HP}$, the \emph{Witten prescription} for turning on temperature in the CFT consists of \emph{placing a (large) black hole in the dual AdS bulk}. } }

\begin{figure}
\end{figure}

\section{RG and AdS/CFT} 


In the previous section we turned on $T$ in AdS/CFT, which you can think of as introducing a scale to the theory (namely a circumference $\beta = 1/T$ of $t_E$ or a horizon $r_+$ in the bulk).\footnote{
	This can actually be a bit confusing because the CFT at finite $T$ is still a scale invariant theory. The point is that in a scale invariant theory, there is no other scale present with which the temperature $T$ can be compared, so that all $T \neq 0$ are equivalent. The only 2 non-equivalent temperatures are $T = 0$ and $T \neq 0$. This translates in the bulk as the statement that $r_+$ can be eliminated from the metric by a coordinate transformation. This is deferred to an Exercise. 
} 
That scale $\beta$ was introduced by a Dirichlet boundary condition $\gamma_\mn^{(0)}(\beta)$ on the bulk metric, with $\int d^d x \,  \gamma_\mn^{(0)}(\beta) T^\mn$ a contribution to the CFT partition function.

There are other ways of introducing a scale to AdS/CFT by similarly introducing Dirichlet boundary conditions $\phi_0$ 
for other bulk fields that couple to CFT operators through $\int d^d x \, \phi_0 \mathcal O$. The idea is to keep the $\phi_0$ in order to deform the CFT (not send them back to zero at the end of a calculation, as in section \ref{sect2ptfcalc}). The source $\phi_0$ enters through \eqref{phias}, and to understand the deformation we first need to pause at the interpretation of the holographic direction $Z$ from the boundary perspective. 


\paragraph{What is $Z$?} 

In AdS 
\ali{
	ds^2 = \frac{\ell^2}{Z^2}(-dt^2 + d \vec x^2) \quad + \, \frac{\ell^2}{Z^2} dZ^2,  
}
the conformal factor $\frac{\ell^2}{Z^2}$ multiplying the conformal boundary part of the metric relates energies and distances measured by boundary observers (with $t$ and $\vec x$) to energies and distances measured by bulk observers (with $\frac{\ell}{Z} t$ and $\frac{\ell}{Z} \vec x$). 

\begin{figure}
	\centering \includegraphics[width=12cm]{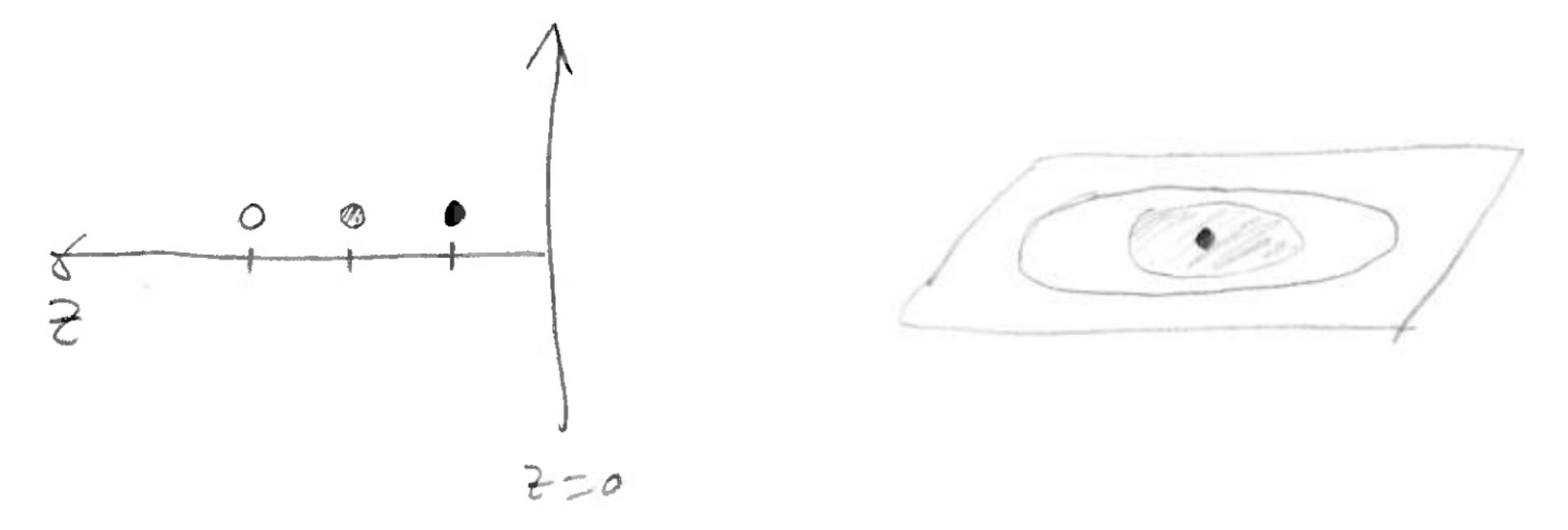} 
	\caption{UV/IR-correspondence: the same bulk object at different radial positions in the gravitational background corresponds to an object in the dual field theory of different sizes. 
	} 
	\label{fig-RG}
\end{figure}

A bulk object with size $d_{bulk}$ and energy $E_{bulk}$ corresponds to a boundary configuration with different size $d_{bdy}$ and energy $E_{bdy}$: 
\ali{
	d_{bulk} &= \frac{\ell}{Z} d_{bdy} \qquad \text{or} \quad d_{bdy} = \frac{Z}{\ell} d_{bulk} \\ 
	E_{bulk} &= \frac{Z}{\ell} E_{bdy} \qquad \text{or} \quad E_{bdy} = \frac{\ell}{Z} E_{bulk} . 
}
In particular, a bulk object close to the boundary $Z \ra 0$ corresponds to a boundary configuration of very small size: $Z \ra 0$ in bulk corresponds to probing the CFT in the UV-region, at very high energies or to very small distances (high resolution). This is a general 
explanation for the UV/IR correspondence in AdS/CFT that we first encountered in the context of holographic renormalization. It is illustrated in figure \ref{fig-RG} \cite{MaldacenaLectures}. 

As you move the bulk object deeper into the bulk, the corresponding region probed in the CFT becomes larger. This means the bulk direction $Z$ from the point of view of the boundary theory has the interpretation of an \emph{RG scale}: flowing into the bulk corresponds to an RG flow to the IR in the boundary theory.

\paragraph{RG theory} 

References: Hollowood lectures \cite{Hollowood}, David Tong \cite{Tongstat} Chapter 3.

We discuss in this section the basics of RG (renormalization group) theory in a $d$-dimensional flat background. In RG theory, one considers the space of theories. The theories, for a field $\varphi$, can be described by actions 
\ali{
	S[\varphi; g_i]  
}
that also parametrically 
depend on a set of coupling constants $\{g_i\}$ (an alternative notation for $S[\varphi; g_i]$ would be $S_{\{g_i\}}[\varphi]$), which you can think of as a set of coordinates spanning the space of theories. These parameters are `coupling constants' in the general sense, including e.g.~mass parameters.   

Now the coupling constants $g_i$ depend on the energy scale $\mu$ at which you resolve your theory; they are running couplings $g_i(\mu)$. 
The functions $g_i(\mu)$ define the \emph{RG flow} in the space of theories, which is conventionally 
thought of as a flow from UV to IR. 
Vanishing energy scale $\mu \ra 0$ corresponds to the IR limit, $\mu \ra \infty$ to the UV limit. 
\emph{Fixed points} of the flow $g_i = g_i^*$ have the property that 
\ali{
	\left. \frac{dg_i}{d\mu} \right|_{g_j^*} = 0. 
}

The action at a particular energy scale is the Wilsonian Effective Action $S[\varphi; \mu, g_i]$. This effective field theory action (EFT) is taken to be the most general action that one is allowed to write down based on symmetry considerations, including all possible couplings $g_i$ (infinitely many a priori) for allowed interaction terms of the fields e.g.~$\varphi^n, \varphi^n \p_\mu \varphi \p^\mu \varphi$, etc. Writing $\mathcal O_i(x)$ for such linear combinations of the fields and their derivatives, the action takes the form of a kinetic term for $\varphi$ plus `deformation terms' (we'll come back to this name later).    
\ali{
	S[\varphi; g_i] = \int d^d x \, \left( \frac{1}{2} \p_\mu \varphi \p^\mu \varphi + \sum_i g_i \mathcal O_i(x) \right). 
}
The $\mu$-dependence can be made explicit by extracting a factor $\mu^{d-\Delta_i}$ from $g_i$ for $\Delta_i$ the classical scaling dimension of $\mathcal O_i(x)$, in order to make $g_i$ dimensionless: $S[\varphi; \mu, g_i] = \int d^d x \, \left( \frac{1}{2} \p_\mu \varphi \p^\mu \varphi + \sum_i \mu^{d-\Delta_i} g_i \mathcal O_i(x)  \right)$. But we will use the notation with the dimensionful coupling. 

For e.g.~a scalar field theory, examples of `deformation terms' are $\frac{1}{2} m^2 \varphi^2$ and $\frac{1}{4!} \lambda \varphi^4$. That is, $g_2 = m^2$ and $g_4 = \lambda$.  
In the IR limit $\mu \ra 0$, mass scales decouple as $m/\mu \ra \infty$ except for $m=0$: only massless particles remain, and moreover, more generally speaking, no dimensionful parameters are left.  
This leaves a scale invariant theory (with no mass or other scales) described by a CFT. This is referred to as a \emph{CFT fixed point} in the IR.  
A curve $g_i(\mu)$ forms a `renormalized trajectory' if it connects the IR CFT fixed point to a UV CFT fixed point (called ``asymptotic safety" by Weinberg), defining a theory on all length scales.   
\emph{CFT's, as fixed points, play a very special role in RG theory as the mileposts to and from which RG trajectories flow.}  
From the point of view of the CFT fixed points, the $\sum g_i \mathcal O_i$ terms in the action reintroduce scales to the theory, deforming us 
away from scale invariance. Hence the name `deformation terms'. 

In the neighborhood of a fixed point, the RG flow can be linearized $g_i(\mu) = g_i^* + \delta g_i$ (for details see \cite{Hollowood}), leading to a classification of couplings $g_i$ into \emph{relevant, marginal and irrelevant}, according to the scaling dimension $\Delta$ of $\mathcal O_i$: 
\ali{
	&\Delta < d \qquad \text{relevant} \\
	&\Delta = d \qquad \text{marginal} \\
	&\Delta > d \qquad \text{irrelevant} . 
}
A relevant coupling has the property that it grows, i.e.~becomes relevant in the IR. An irrelevant coupling vanishes, i.e.~becomes weaker or `irrelevant' in the RG flow to the IR. Said otherwise, a relevant perturbation vanishes in the UV, while an irrelevant perturbation blows up in the UV. It is the sign of $d-\Delta$ that is important for this classification.  

It follows that the behavior of theories in the IR is determined by only a small number of couplings, namely the relevant couplings, instead of by the infinite initial set of couplings $g_i$. In the scalar field theory example, $g_2=m^2$ is relevant in $d \geq 2$ and $g_4 = \lambda$ is relevant in $d=2,3$ but irrelevant in $d>4$. This justifies why we in practice only considered a few couplings $g_2$ and $g_4$ from the start.  

\paragraph{RG in AdS/CFT} 

Now let's say we start from a CFT, e.g. $S[\varphi] = \frac{1}{2} \int d^d x \,  \p_\mu \varphi \p^\mu \varphi$, and deform it with a term 
\ali{
	\int \phi_0 \, \mathcal O 
}  
with constant (i.e.~$x$-independent) coupling constant $\phi_0$,  
and $\mathcal O$ an operator of dimension $\Delta$. 

Assuming the CFT is holographic, i.e.~has a good semi-classical AdS dual, $\phi_0$ represents the boundary value of a bulk field $\phi$ that is dual to the CFT operator $\mathcal O$. This is put into a formula in equation \eqref{phias}: 
\ali{
	\phi(Z,x^\mu) \,\, \stackrel{Z \ra 0}{\sim} \,\, Z^{d-\Delta} (\phi_0  + \tilde \phi_0(x^\mu)) 
	+ Z^{\Delta} \tilde \phi(x^\mu). \label{phiasrepeated}
} 
Correlators of $\mathcal O$ in the deformed theory are calculated by setting $\tilde \phi_0 = 0$ in \eqref{correldef}.

The value of $\phi$ in the bulk as $Z$ is varied corresponds to the running of the coupling $\phi_0$ with energy scale from the perspective of the field theory. 

For $d-\Delta > 0$, i.e.~for a relevant deformation of the CFT, $Z^{d-\Delta} \phi_0 $ vanishes as $Z \ra 0$, corresponding to a vanishing coupling $\phi_0$ in the UV from the boundary point of view. 

For $d-\Delta < 0$, i.e.~for an irrelevant deformation of the CFT, $Z^{d-\Delta} \phi_0 $ blows up as $Z \ra 0$, corresponding to a coupling $\phi_0$ that blows up in the UV from the boundary point of view. 


Consistent with our previous discussion of the interpretation of the $Z$-direction in AdS as an RG direction in the boundary theory, it is the power of $Z$ in the holographic description of the deformation that contains the information about the type of deformation. \emph{The holographic direction makes the RG direction of the boundary theory explicit in a geometric way.}

For applications in AdS/CFT (domain wall solutions), I refer to \cite{dHoker}.

The $T\bar T$ deformation, briefly discussed in class, is a modern research topic. 

\newpage 
\section*{Exercises} 

	\textbf{\underline{\smash{Exercise 1. Conical singularity}}}

Show that for a metric of the form 
\ali{
	ds^2 = F(r) dt_E^2 + \frac{dr^2}{F(r)} 
}
which, at $r=r_+$ where $F(r_+) = 0$, can be written in the form of a plane metric in polar coordinates, a conical singularity is avoided only if the Euclidean time direction $t_E$ is periodic with period:
\ali{
	\beta = \left. \frac{4\pi}{F'(r)}\right|_{r=r_+}. 
}
\\*
\\*
\\*
\textbf{\underline{\smash{Exercise 2. AdS black hole}}}

Write the metric we used in class for the AdS$_{d+1=5}$ black hole in the general dimensional form 
\ali{
	ds^2 = \frac{l^2}{r^2} (-f(r) dt^2 + \frac{dr^2}{f(r)} + dx^i dx^i ), \quad f(r) = 1 - \left(\frac{r}{r_+}\right)^d.   
}	
Write down the simple coordinate transformation by which $r_+$ can be eliminated from the metric. 
This expresses that there are only two non-equivalent temperatures in the scale invariant theory: zero and non-zero (since there is no other scale to compare the temperature to).

\chapter{Black hole entropy and its statistical interpretation} 



\section{Black hole thermodynamics} 


Black hole mechanics laws formally look like laws of thermodynamics. Including quantum effects at the semi-classical level (Hawking calculation in QFT on curved space), the black hole is no longer completely black but radiates at a Hawking \emph{temperature} and has a Bekenstein-Hawking \emph{entropy} $S_{BH}$. This allows to interpret the black hole laws physically, not just formally, as black hole \emph{thermodynamics} laws for the black hole as a thermal object: 
\ali{
	&\text{zeroth law} \qquad  T \text{ constant} \\
	&\text{first law}  \qquad \quad \delta M = T \delta S_{BH}  \label{bhfirstlaw}  \\ 
	&\text{second law} \qquad \delta S_{gen} \geq 0 
}
where $M$ is the mass of the black hole, $T$ the Hawking temperature, $S_{BH}$ the black hole entropy or Bekenstein-Hawking entropy, and $S_{gen}$ the generalized entropy consisting of $S_{BH}$ plus the entropy of matter on the outside-horizon region.


In section \ref{finiteTQFT}, we saw that in a finite temperature QFT, the 
canonical partition function is given by the Euclidean path integral along a closed path of length $\beta= 1/T$, 
\ali{
	Z(\beta) = \sum_s e^{-\beta E_s} = \int_{\phi(\vec x, t_E+ \beta) = \phi(\vec x, t_E)} \mathcal D \phi \hspace{1mm} e^{- S_E[\phi]}, \qquad \beta = \frac{1}{T}. 
}  
Similarly, in a quantum theory of gravity, the canonical partition function for a system of gravitational and matter fields in thermal equilibrium at temperature $T$ is at least formally\footnote{
	Not knowing how to evaluate the path integral exactly corresponds to not knowing the QG microstates.  
} given by  
\ali{
	Z(\beta) = \sum_{\text{QG states } s \, \in \, \mathcal H_{QG}} e^{-\beta E_s} = 
	\int_{t_E \sim t_E + \beta \, \text{at } \p M} \mathcal D g \mathcal D \phi \hspace{1mm} e^{- S_{E,grav}[g_\ab,\phi]}, \qquad \beta = \frac{1}{T} 
}
with periodicity of $t_E$ imposed as a boundary condition on the metric $g_\ab$ at `infinity' $\p M$, and the Euclidean gravitational action given in \eqref{SEgrav}.   

In the semi-classical approximation, 
\ali{
	Z(\beta) \approx e^{- S_{E,grav}[g^*_\ab,\phi^*]} 
}
for $g^*_\ab$ and $\phi^*$ classical solutions. 

Once the thermal partition function is identified with the Euclidean path integral, you can use the known formulas of statistical physics for obtaining the free energy, entropy and energy (with $F = E - T S$): 
\ali{
	F &= - \frac{1}{\beta} \log Z(\beta) \approx \frac{1}{\beta} S_{E,grav}[g_*] \\
	S &= (1 - \beta \p_\beta) \log Z(\beta) \\ 
	E &= - \p_\beta \log Z .  
}  
This allows to calculate the entropy of a black hole solution. As a first example, consider the Schwarzschild black hole of section \ref{sectSSbh}. Its horizon $r_h = 2 M$ (in units $G=1$) in Euclidean signature becomes the origin $\rho = 0$ of a polar coordinate system with angular coordinate $t_E \sim t_E + 8 \pi M$. This means the Hawking temperature of this black hole is $T = 1/(8 \pi M)$. The Euclidean action can be calculated to be $S_E = 4 \pi M^2 = \beta M/2$. It follows that $Z(\beta) = \exp{\{-\beta^2/(16\pi)\}}$, which allows to derive the thermodynamic properties of the Schwarzschild black hole: 
\ali{
	S &= 4 \pi M^2 \\ 
	E &= M. 
}
Unsurprisingly, the energy is given by the mass of the black hole. The entropy can be written in terms of the surface area $A = 4 \pi r_h^2 = 4 \pi (4 M^2)$ of the black hole horizon as 
\ali{
	S = \frac{A}{4}. 
}
This is the famous \emph{Bekenstein-Hawking} entropy 
\ali{
	S_{BH} = \frac{A}{4G\hbar},   \label{BHentropy}
	}
which combines gravity ($G$) and quantum mechanics ($\hbar$). While derived here for a specific black hole, it is a much more general result \cite{HartmanQG}. 

The fact that the black hole entropy scales as the surface area of the horizon rather than the volume, has been interpreted as revealing a \emph{holographic} nature of black holes, and by extension of \emph{gravity} itself, according to the holographic principle of 't Hooft and Susskind '93-'94. The AdS/CFT correspondence conjectured by Maldacena in '97 is the most concrete implementation of that holographic principle. So we could have also started the course with black hole thermodynamics. (This is what Hartman does in \cite{HartmanQG}.)

Big question: is there a statistical, i.e.~quantum gravitational, interpretation of $S_{BH}$? 
That is, can we write $S_{BH} = k_B \log \Omega$ for an $\Omega$ describing the total number of microscopic states (i.e.~QG black hole microstates) available to the system? 

One comment before moving on in the next sections to that question and an AdS/CFT answer to it, concerns the question of including quantum loop corrections in $Z(\beta)$. Reference for more information: Jacobson's lectures \cite{Jacbhthdyn}. Including corrections $g = g_* + \tilde g$, $\phi = \phi_* + \tilde \phi$, the partition function takes the form 
\ali{
	Z(\beta) = \int \mathcal D \tilde g \mathcal D \tilde \phi \,  e^{-S_{E,grav}[g_* + \tilde g, \phi_*+ \tilde \phi; G]} \approx e^{-S_{E,grav}^{eff}[g_*,\phi_*; G_{ren}]} = \text{tr} \, e^{-\beta H_0[\tilde g, \tilde \phi]} 
}  
with $G_{ren}$ a renormalized gravitational constant, and $H_0$ the evolution operator for fluctuations in the background $(g_*, \phi_*)$. 
You can again apply the statistical physics formulae to obtain the entropy $S$, this time from the \emph{effective} action $S_{E,grav}^{eff}$. Without providing any details, let me mention that the correction $S'$ to the entropy that follows from this procedure is itself divergent, and its interpretation is not resolved. Here, $S'$ is the thermal entropy of quantum fields which gives a 1-loop 
contribution to the full black hole entropy.



\paragraph{Black hole thermodynamics of BTZ} 

Consider the AdS black hole in 3 dimensions known as the BTZ black hole 
\ali{
	ds^2 = -\frac{r^2 - r_+^2}{\ell^2} dt^2 +  \frac{\ell^2}{r^2 - r_+^2} dr^2 + r^2 d\phi^2, \qquad \phi \sim \phi + 2 \pi 
}
with mass and angular momentum 
\ali{
	M = \frac{r_+^2}{8 G \ell^2} = \frac{1}{\ell}(L_0+\bar L_0), \qquad J = 0 = L_0 - \bar L_0 
}
given in terms of the Virasoro generator modes of the Brown-York stress tensor. For the more general rotating BTZ: $M = \frac{r_+^2 + r_-^2}{8 G \ell^2}$ and $J = \frac{r_+ r_-}{4G \ell}$. 

By a similar analysis as for the flat space black hole [see the Exercise and \cite{Krauslectures}], it has a temperature 
\ali{
	T = \frac{r_+^2-r_-^2}{2\pi \ell^2 r_+} \quad \stackrel{J=0}{=} \frac{r_+}{2\pi \ell^2}
	\label{BTZtemp}
}   
and entropy 
\ali{
	S = \pi \sqrt{\frac{\ell (\ell M + J)}{2G}} + \pi \sqrt{\frac{\ell (\ell M - J)}{2G}} \quad \stackrel{J=0}{=} 2\pi \sqrt{\frac{\ell^2 M}{2G}}. \label{BTZentropy} 
}
Through AdS/CFT, there is an interpretation of $\Omega$ when writing $S = \log \Omega$! Namely in terms of CFT states, by comparison to the Cardy formula in 2D CFT. 

Comment: Formally, you can go from the BTZ metric to the AdS metric by setting $r_+^2 = -\ell^2$, which corresponds to $M = -\frac{1}{8G}$. This has a direct interpretation 
in the CFT, see later 
in section \ref{sectholCardy}.  


\section{Cardy formula in 2D CFT} \label{sectCardyCFT}


Main reference: \cite{HartmanQG} chapter 25. Further reading: \cite{difran} chapter 10. 

In this section  we discuss 2D CFT at finite temperature, more specifically the torus partition function and Cardy formula. Only in the next section \ref{sectholCardy} will we go back to a bulk gravity perspective. \\

\underline{1) \, CFT on cylinder (at $T=0$) } 

\begin{figure}[t]
 \centering	\includegraphics[width=14cm]{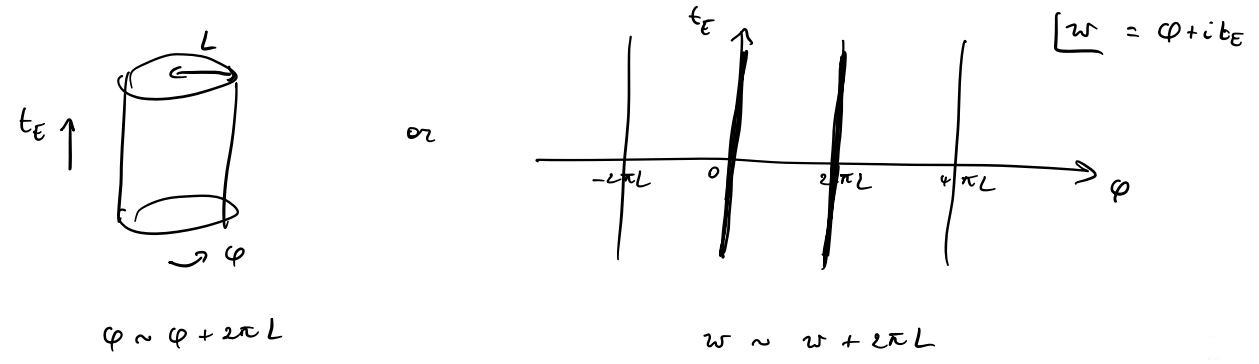} 
 \caption{Set-up of 2D CFT on cylinder }   \label{figCFToncyl} 
\end{figure}

We consider a CFT on a cylinder, pictured in two different ways in figure \ref{figCFToncyl}. The first representation is in terms of the $(t_E,\phi)$ coordinates with $\phi $ the spacelike compactified coordinate, with identification $\phi \sim \phi + 2 \pi L$ for $L$ the radius of the cylinder, which will sometimes be set to 1 for simplicity of the formulas. 
The second representation is in terms of the complex coordinate $w = \phi + i t_E$, with identification $w \sim w + 2 \pi L$ inherited of course directly from the $\phi$ identification. In this rolled out picture of the cylinder, the strip between the bold lines $\phi = 0$ and $\phi = 2\pi L$ 
is the fundamental domain which corresponds to one wrapping of the cylinder upon identification of the bold lines, and all vertical lines in the picture $(\phi = n \pi L)$ are to be identified. 
The metric on the cylinder is $ds^2_{cyl} = d\phi^2 + dt_E^2 = dw d\bar w$ but has to be supplemented by the identifications of the coordinates, $\phi \sim \phi + 2 \pi L$ or $w \sim w + 2 \pi L$. 

The compactification of the spacelike direction $\phi$ leads to a Casimir contribution $\sim \frac{1}{L}$ to the energy, more precisely the energy in the vacuum state is 
\ali{
	E_{cyl}^{vac} = -\frac{c}{12} \frac{1}{L}. \label{Evac}
}
To arrive at this result, use the transformation behavior of the CFT stress tensor under general conformal transformations $z \ra z'$ in \eqref{CFTstresstensortransffull} 
and apply it to the plane to cylinder conformal transformation $z \ra w$ discussed in \eqref{planetocyltransf}, $z = e^{i w}$ or $z = L e^{i w/L}$ (where we see $w$ automatically has the correct identifications because it appears in the exponent and $e^{2 \pi i}=1$). It follows that 
\ali{
	T_{cyl}(w) = -\frac{z^2}{L^2} T_{plane}(z) - \frac{c}{24} \frac{1}{L^2} 
}   
where the second term comes precisely from the anomalous ($=$ non-tensorial) contribution to the transformation law of the stress tensor, namely the Schwarzian contribution in \eqref{CFTstresstensortransffull} proportional to the central charge $c$. The energy is defined as $H = \int d\phi \, T_{t_E t_E} = \int d\phi (T_{ww} + T_{\bar w \bar w})$ and has to be evaluated in the vacuum state to find indeed $E_{cyl}^{vac} = \int_0^{2\pi L} d\phi \vev{T_{t_E t_E}^{cyl}(t_E=0,\phi)}_{vac} = (2 \pi L) (-\frac{c}{12 L^2}) = -\frac{c}{12 L}$. 

Now, the goal will be to derive the partition function for the CFT on the cylinder at finite temperature $T$. In the canonical ensemble $(T,V,N)$, knowledge of the partition function allows to derive the entropy and free energy: 
\ali{
	Z(T,V,N) \quad \leadsto \quad S(T,V,N), \,\,\, F(T,V,N) = E - T S 
}
where $Z = \sum_{states} e^{-\beta E}$. 
To turn on temperature, we 
periodically identify Euclidean time with period $\beta = 1/T$. This procedure implies that we need to discuss the CFT on a torus. \\

\underline{2) \, CFT on a torus }

\begin{figure}[t]
	\centering	\includegraphics[width=14cm]{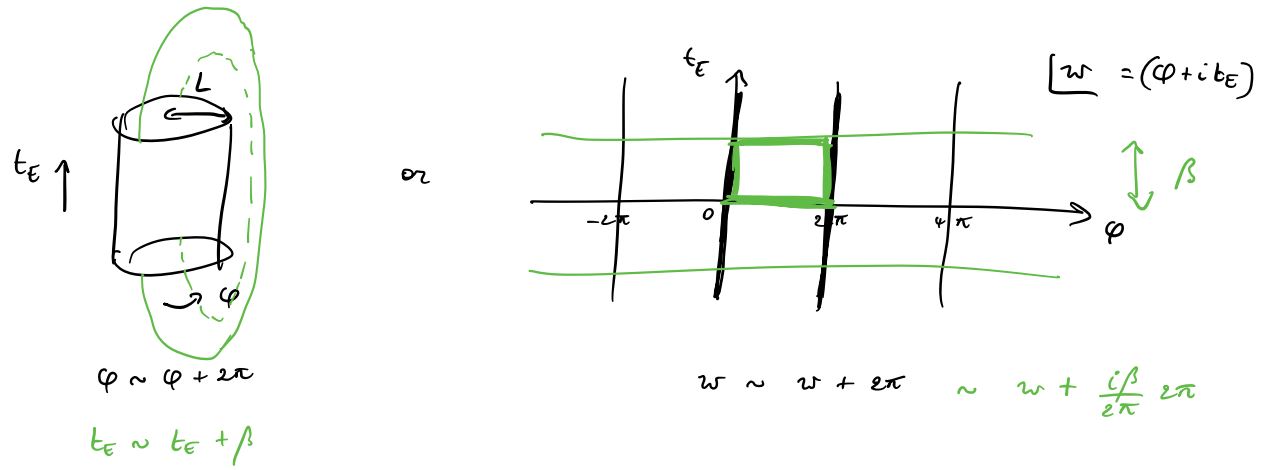} 
	\caption{Set-up of 2D CFT on torus }   \label{figCFTontorus} 
\end{figure}

\begin{figure}[t]
	\centering	\includegraphics[width=12cm]{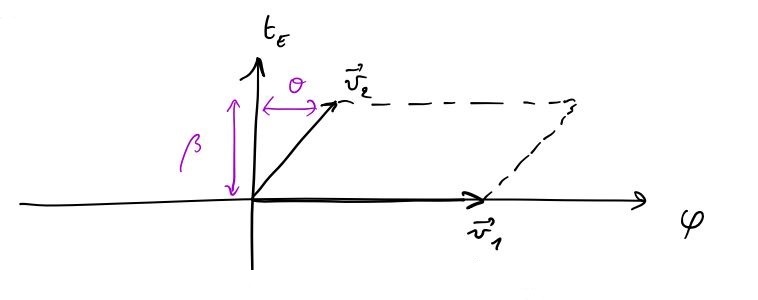} 
	\caption{Torus definition }   \label{figtorusdef} 
\end{figure}

Compactification of Euclidean time direction $t_E$ with periodicity $\beta$ creates a torus, pictured in two ways in figure \ref{figCFTontorus} with the additional coordinate identifications (compared to the cylinder) in green. The fundamental domain is a rectangle (rather than the strip) of size $2 \pi \times \beta$. In fact, the more general fundamental domain for a torus is  a parallellogram, specified by real numbers $\beta$ and $\theta$ as pictured in figure \ref{figtorusdef}. Alternatively, a torus is defined by the specification of two vectors $\vec v_1 = (2\pi,0)$ and $\vec v_2 = (\theta, \beta)$. In figure \ref{figtorusdef}, the torus has already been `normalized' by rotation of a general given $\vec v_1$ to align with the $\phi$-axis, and rescaling to $|\vec v_1| = 2\pi$. This is possible without loss of generality when the theory is rotational invariant 
and scale invariant, which is true for our 2D CFT. 

A torus is fully characterized by its \emph{modulus} $\tau$ 
\ali{
	\tau = \frac{\theta + i \beta}{2\pi}.  
}

Given the fundamental domain in figure \ref{figtorusdef}, the identifications required to cover the torus (once) are 
\ali{
	(t_E,\phi) \sim (t_E + \beta, \phi + \theta) \sim (t_E, \phi + 2\pi) 
}
or 
\ali{
	w \sim w + 2\pi \sim w + 2 \pi \tau . 
}
The full lattice pictured in figure \ref{figCFTontorus} describes the multiple coverings of the torus as 
\ali{
	w \sim w + (m + n \tau) 2 \pi, \qquad m,n \in \mathbb Z . 
}

\begin{figure}[t]
	\centering	\includegraphics[width=5.8cm]{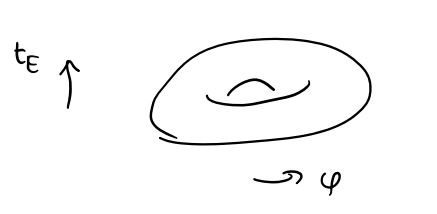}  \qquad \qquad  \qquad  \includegraphics[width=6.5cm]{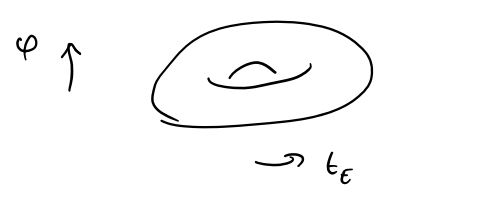}
	\caption{Two possible choices to label the torus directions }   \label{figtorusx2} 
\end{figure}

Now, to obtain the partition function of the finite temperature CFT on a cylinder, we need to write down the path integral on the 
torus.  
In order to do this, there are two equally valid choices of the Euclidean coordinates spanning the torus, pictured in figure \ref{figtorusx2}. First, choosing $t_E$ as Euclidean `time' and $\phi$ as `space' direction, one writes the path integral by taking a state on the $t_E=0$ surface $\phi:0\ra 2\pi$, which is an element of the Hilbert space $\mathcal H_{(2\pi,0)}$, and evolving it in the direction of $\vec v_2$, i.e.~along the vector $\beta \p_{t_E} + i \theta \p_\phi$, to obtain 
\ali{
	Z(\tau,\bar \tau) = \tr_{\mathcal H_{(2\pi,0)}} e^{-\beta H + i \theta J}.  
}
From now on, we will consider again the special case $\theta = 0$ (rectangular lattice cell for the torus) for simplicity (it can be included without problem). 

Second, one can instead choose $\phi$ as Euclidean `time' and $t_E$ as `space' direction, and write the path integral by taking a state on the $\phi=0$ surface $t_E:0\ra \beta$, which is an element of the Hilbert space $\mathcal H_{(0,\beta)}$, and evolving it in the direction of $\vec v_1$, i.e.~along the vector $2\pi \p_\phi$, to obtain 
\ali{
	Z(\tau,\bar \tau) = \tr_{\mathcal H_{(0,\beta)}} e^{-2\pi J}.  
}
Since both choices are valid, we require equality of 
\ali{
	\tr_{\mathcal H_{(2\pi,0)}} e^{-\beta H} = \tr_{\mathcal H_{(0,\beta)}} e^{-2\pi J}. \label{consistencyZ}
	}

\begin{figure}[t]
	\centering	\includegraphics[width=16cm]{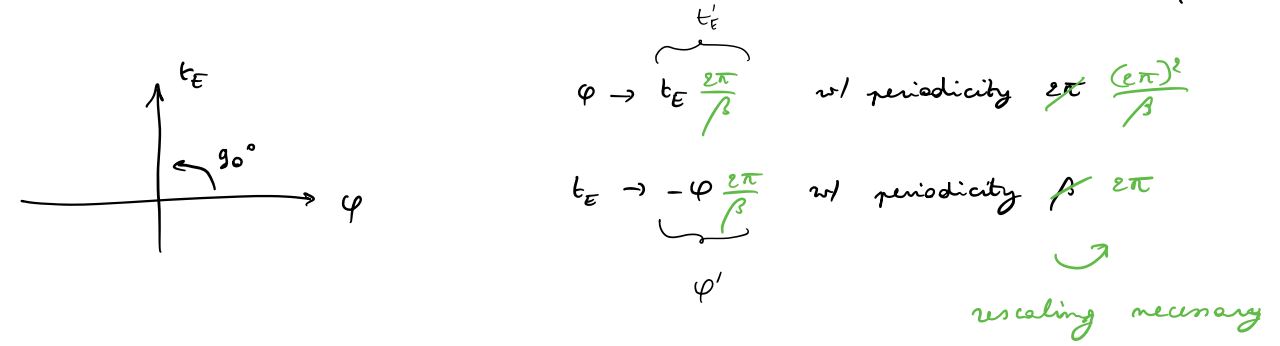}  
	\caption{Required conformal transformation to connect $\mathcal H_{(2\pi,0)}$ and $\mathcal H_{(0,\beta)}$ Hilbert spaces. 
	}   \label{figtorusrotation} 
\end{figure}

In a general QFT, there is no relation between the Hilbert spaces $\mathcal H_{(2\pi,0)}$ and $\mathcal H_{(0,\beta)}$, but in CFT there is! Namely, we can transform between the 2 choices of `time' and `space' by a conformal transformation: a rotation over 90 degrees plus a rescaling by $2\pi/\beta$. This is pictured in figure \ref{figtorusrotation}. The rescaling is necessary for `normalization' of the torus to $\vec v_1 = (2\pi,0)$, i.e.~imposing $\Delta \phi' = 2\pi$ after the rotation. It follows that the other coordinate is rescaled with the same factor, leading to a periodicity $\Delta t_E' = (2\pi)^2/\beta$. By conformal equivalence, it then follows that 
\ali{
	\tr_{\mathcal H_{(0,\beta)}} e^{-2\pi J} = \tr_{\mathcal H_{(2\pi,0)}} e^{-\frac{(2\pi)^2}{\beta} H}. \label{confeq}
}

Making use of \eqref{confeq}, the 
condition \eqref{consistencyZ} becomes 
\ali{
	\tr_{\mathcal H_{(2\pi,0)}} e^{-\beta H} = \tr_{\mathcal H_{(2\pi,0)}} e^{-\frac{(2\pi)^2}{\beta} H}. 
} 
In this rewritten form, the traces left and right are over the same Hilbert space, so that the heavy notation can be replaced simply by a trace $\tr$. It is an equality condition between the path integral for traveling a distance $\beta$ and $4 \pi^2/\beta$ in the (Euclidean) time direction 
\ali{
	Z(\beta) = Z(\frac{4\pi^2}{\beta}) \qquad \text{high $T$ - low $T$ \emph{duality}}
	. \label{highTlowT} 
} 
It expresses a duality between high temperature and low temperature phases of the theory. 

What the conformal transformation $(t_E,\phi) \ra \frac{2\pi}{\beta} (-\phi, t_E)$ seems to have done is to relate $Z$ on the torus $\tau = \frac{i \beta}{2\pi}$ with $Z$ on the torus $\tau' = \frac{i}{2\pi} \frac{4\pi^2}{\beta} = - \frac{1}{\tau}$. The transformation $\tau \ra - \frac{1}{\tau}$ between tori 
is called an $S$-transformation. It leaves the definition of the torus invariant: 
\ali{
	w \sim w + (m + n \tau) 2 \pi \quad \stackrel{S: \, \tau \ra -1/\tau}{\longrightarrow} \quad &w \sim w + (m - \frac{n}{\tau} 2\pi \\
	&w \tau \sim w \tau + (m \tau - n) 2\pi \\ 
	&w' \sim w' + (n' \tau + m') 2\pi \quad;  
	} 
that is, the $(m,n)$ lattice is transformed into an $(m',n') = (-n,m)$ lattice, which describes the \emph{same} torus.  
The only thing that has changed is how you count the lattice cells (with $m,n \in \mathbb Z$). 

Similarly, the so-called $T$-transformation $\tau \ra \tau + 1$ leaves the torus invariant, with $(m,n) \ra (m+n,n)$. With repeated $T$- and $S$- transformations one can build a $PSL(2,\mathbb Z)$ transformation 
\ali{
	\tau \ra \frac{a \tau + b}{c \tau + d}, \qquad \bar \tau \ra \frac{a \bar \tau + b}{c \bar \tau + d} 
	} 
with $a,b,c,d \in \mathbb Z$ and $a d - b c = 1$. We have arrived at the $PSL(2,\mathbb Z)$ or \emph{modular invariance} of a 2D CFT on a torus, expressed by 
\ali{
	Z(\tau,\bar \tau) = Z\left( \frac{a \tau + b}{c \tau + d}, \frac{a \bar \tau + b}{c \bar \tau + d} \right) 
}
of which the high $T$ - low $T$ duality \eqref{highTlowT} is one example. \\

\underline{3) \, Cardy formula } 

Now we are ready to follow Cardy's derivation of entropy of a 2D CFT on a cylinder. The torus path integral $Z(\beta)$ is interpreted as the partition function for the system at temperature $1/\beta$. 
Let us discuss this partition function $Z(\beta) = \sum_{states} e^{-\beta E}$ in two opposite temperature regimes. 

At low temperatures, $\beta \ra \infty$, the sum over states is dominated by the lowest energy or vacuum energy contribution: 
\ali{
	\text{low $T$ or $\beta \ra \infty$} \qquad Z(\beta) = \sum_{states} e^{-\beta E} \stackrel{\beta \ra \infty}{\approx} e^{-\beta E_{vac}} = e^{\beta \frac{c}{12} \frac{1}{L}}  
} 
where we used \eqref{Evac}. 
Instead, at high temperatures $\beta \ra 0$, the partition function $Z(\beta) = \sum_{states} e^{-\beta E}$ is dominated by high energy states and is usually impossible to evaluate. In this case, however, we can make use of the powerful modular invariance \eqref{highTlowT} to write the partition function as 
\ali{
	\text{high $T$ or $\beta \ra 0$} \qquad Z(\beta) = \sum_{states} e^{-\frac{4 \pi^2 L^2}{\beta} E} \stackrel{\beta \ra 0}{\approx} e^{-\frac{4 \pi^2 L^2}{\beta} E_{vac}} = e^{\frac{4 \pi^2}{\beta} \frac{c}{12}L}  
}
where the modular invariant rewritten form is dominated in the $\beta \ra 0$ limit by the vacuum energy contribution again. This only works because of the high temperature - low temperature duality. 

The free energy defined by $Z(\beta) = e^{-\beta F}$ can be written for high temperature as $F = -\frac{\pi^2}{3} c T^2 \frac{L}{2\pi}$, proportional to the central charge $c$ (a measure for the number of degrees of freedom). 
The entropy $S(\beta) \equiv (1 = \beta \p_\beta) \log Z = 2 \pi^2 c/(3\beta)$ and  the energy $E(\beta) \equiv -\p_\beta \log Z = \pi^2 c/(3 \beta^2)$ from which it follows that the entropy as a function of energy is given by 
\ali{
	S(E) = 2\pi \sqrt{ \frac{c}{3} E L} \qquad \quad \text{(Cardy formula)}. \label{SCardy} 
}
This is the famous Cardy formula for the entropy of high energy states of the 2D CFT on a cylinder of radius $L$. 
Correspondingly, one can associate to it a density of states $\rho(E)$ defined by $S = \log \rho$ (called $\Omega$ in a previous section), 
\ali{
	\rho(E) = e^{2 \pi \sqrt{ \frac{c}{3} E L}}.  
}  
This is the Cardy formula for the density of states at high energy. It gives the number of CFT states with energy $E$, for $E$ high, of said otherwise, the degeneracy of high-$E$ CFT states.

The energy of a state on the cylinder is the conformal dimension $\Delta$ because $H = L_0 + \bar L_0 = D$, so the Cardy formulas above can alternatively be written in terms of the dilation eigenvalue $\Delta$ instead of the energy eigenvalue $E$: entropy $S(\Delta) = 2 \pi \sqrt{\frac{c}{3} \Delta}$ and density of states $\rho(\Delta) = \exp{\{ 2 \pi \sqrt{\frac{c}{3} \Delta} \}}$.

\section{Holographic interpretation of Cardy formula}  \label{sectholCardy}


AdS$_3$ gravity is dual to a CFT$_2$ with $c = \frac{3l}{2G}$. The bulk theory is semi-classical for $\hbar G$ small of $G \ll l$, with the dimension of $G$ being a length to the power (bulk dimension$- 2$). This regime corresponds to large central charge $c \gg 1$ from the boundary perspective. 
According to the Witten prescription, a large black hole in the bulk corresponds to high temperature in the dual CFT. 
We consider in 3 bulk dimensions the BTZ black hole. It has a temperature $T$ and entropy $S$ given in \eqref{BTZtemp} and \eqref{BTZentropy}, and a mass $M$ given in terms of the (classical) bulk Virasoro generators in \eqref{BTZmass} as $M = \frac{L_0 + \bar L_0}{l}$. 

By AdS$_3$/CFT$_2$, the Virasoro generators of the asymptotic symmetry of the 3D bulk geometry (in the gravity theory with parameters the AdS radius $l$ and gravitational constant $G$) map to Virasoro generators of the conformal symmetry of the 2D boundary CFT with a central charge given by the Brown-Henneaux result $c = 3l/(2G)$. The mass of BTZ, $M l = L_0 + \bar L_0$,  then has a dual interpretation as the eigenvalue of the $L_0 + \bar L_0$ operator in the CFT, which is $\Delta$ or $E l$: 
\ali{
	\frac{3l}{2G} \ra c, \qquad M l \ra 
	\Delta . 
}   
Making these holographic replacements in the formula for the BTZ Bekenstein-Hawking entropy \eqref{BTZentropy} gives 
\ali{
	S_{BTZ} = 2\pi \sqrt{\frac{\ell^2 M}{2G}} \stackrel{AdS/CFT}{=} 2\pi \sqrt{\frac{c}{3}\Delta} = S_{Cardy}(\Delta) 
}
meaning the black hole entropy has a dual interpretation as Cardy entropy of high-energy states of the dual CFT! This allows a statistical counting of black hole microstates as high-$\Delta$ CFT states which describe quantum gravity holographically: 
\ali{
	S_{BTZ} \stackrel{AdS/CFT}{=} \log \rho_{Cardy}(\Delta).  
} 
This is a famous result by Strominger '97 and one of the major accomplishments of the holographic approach to quantum gravity, providing a statistical interpretation of the black hole entropy. It is not satisfactory yet in the sense that the number of black hole microstates can be counted, but are not yet identified (in a bulk perspective).

\begin{figure}[t]
	\centering	\includegraphics[width=16cm]{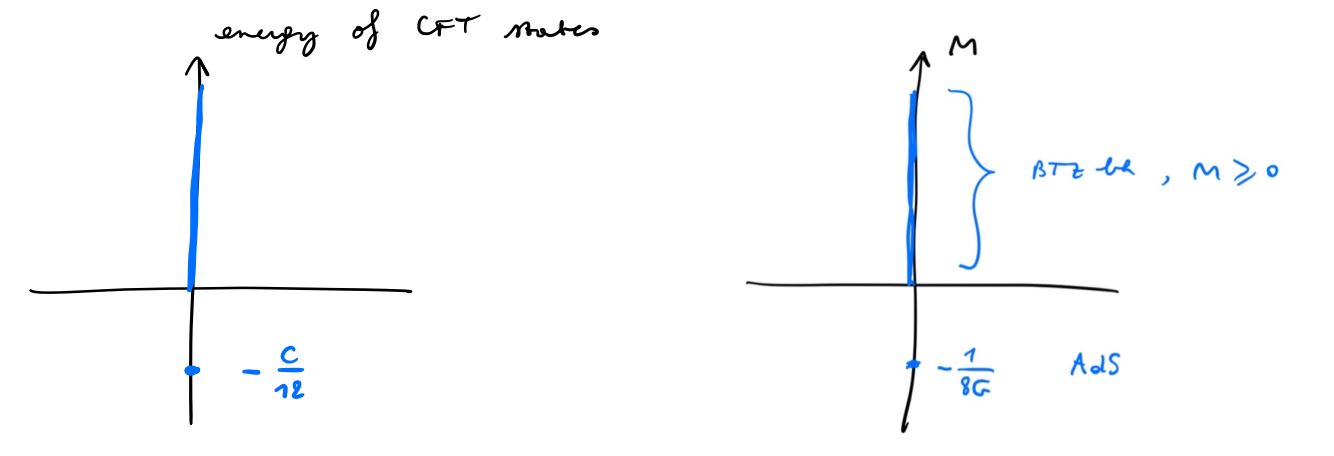}  
	\caption{
	}   \label{figBTZstates} 
\end{figure}

The limits that go into the derivation of the result (semi-classical gravity for large mass black holes) can be summarized in the order of limits condition $\Delta \gg c \gg 1$. 
The vacuum state of 2D CFT on the cylinder maps to empty AdS on the bulk side, separated with a gap from the BTZ solutions. This is illustrated in figure \ref{figBTZstates}. 

Before this holographic result, the highlight in the quest for a statistical interpretation of black hole entropy came from the string theory approach to quantum gravity. In a famous string theory result, Strominger and Vafa '96 provided a microscopic state counting of a 5D extremal Reissner-Nordstrom black hole in supergravity, using the Cardy formula. 
In modern language, this has a similar AdS/CFT interpretation as the BTZ case discussed above.   

\newpage 
\section*{Exercises} 

\textbf{\underline{\smash{Exercise 1. Entropy of a Schwarzschild black hole from the Euclidean path integral}}}


We saw in class that the Euclidean geometry of a 4-dimensional Schwarzschild black hole, described by the metric 
\ali{ds^2=\left(1-\frac{2M}{r}\right)d\tau^2+\left(1-\frac{2M}{r}\right)^{-1}dr^2+r^2d\Omega^2,}
is free of a conical singularity, thus a classical solution of Einstein's equations, only when one periodically identifies the Euclidean time: $\tau \sim \tau +8\pi M$.

Compute the entropy of the Schwarzschild black hole using that the entropy is given by $S=(1-\beta\partial_{\beta})\log Z$. Use also that the partition function for this system can be described as a Euclidean path integral which can be solved using a saddle-point approximation, where the Euclidean Schwarzschild metric is taken to be the solution to the equations of motion of the Einstein-Hilbert action plus Gibbons-Hawking boundary term:
\ali{S_E= -\frac{1}{16\pi}\int d^4x \sqrt{g}R-\frac{1}{8\pi}\int d^3x \sqrt{h}K.}
Here $h_{ab}$ is the induced metric on the boundary geometry (which must be taken initially at fixed $r=r_0$) and $K=h^{ab}K_{ab}$ is the trace of its extrinsic curvature. After evaluating the on-shell Euclidean action, notice that the result is divergent once one takes the limit $r_0\rightarrow\infty$. Therefore, regularize it by adding a counterterm to the action of the form
\ali{S_{ct}=\frac{1}{8\pi}\int d^3x \sqrt{h}\tilde{K},}
where $\tilde{K}$ is the trace of the extrinsic curvature of a boundary geometry with the same intrinsic geometry as $h_{ab}$, but embedded in flat spacetime. Find an expression for the regularized on-shell Euclidean action and use it to compute the entropy in the saddle-point approximation.

What is the black hole entropy you find in terms of the area of the event horizon?
	\\*
\\*
\\*
\textbf{\underline{\smash{Exercise 2. BTZ and Cardy}}}

Check that the transformation 
\ali{
	\phi &= \frac{2\pi}{\beta} t_E' \\
	t_E &= -\frac{2\pi \ell^2}{\beta} \phi' \\ 
	r &= \frac{\ell}{r_+} \sqrt{r'^2 - r_+^2} 
} 
with $\beta = \frac{2\pi \ell^2}{r_+}$ 
takes you from Euclidean AdS$_3$ 
\ali{
	ds^2_{EAdS_3} &= \left(1 + \frac{r^2}{\ell^2} \right) dt_E^2 + \frac{dr^2}{1 + \frac{r^2}{\ell^2}} + r^2 d\phi^2 
}
to Euclidean BTZ 
\ali{
	ds^2_{BTZ} &= \frac{r'^2 - r_+^2}{\ell^2} dt_E'^2 + \frac{\ell^2}{r'^2-r_+^2} dr'^2 + r'^2 d\phi'^2 \, . 
}
This transformation is a bulk extension of the boundary conformal transformation that interchanges the (Euclidean) time and space directions of the torus 
\ali{
	\phi &= \frac{2\pi}{\beta} t_E' \\
	t_E &= - \frac{2\pi}{\beta} \phi' . 
}
Write this conformal transformation as a transformation between complex coordinates $w = \phi + i t_E$ and $w' =  \phi' + i t_E'$ (and $\bar w$ and $\bar w'$), in terms of the torus modulus 
\ali{
	\tau = \frac{i \beta}{2\pi} . 
}

As discussed in class, this conformal transformation has the same effect as the $S$-transformation of the torus 
\ali{
	\tau = - \frac{1}{\tau'}.  
} 
The bulk coordinate transformation between EAdS$_3$ and EBTZ then describes the $S$-transformation of the \emph{solid} tori pictured below. Namely, they are related by exchanging the $\phi$ and $t_E$-cycle. 
\begin{figure}
	\centering \includegraphics[width=12cm]{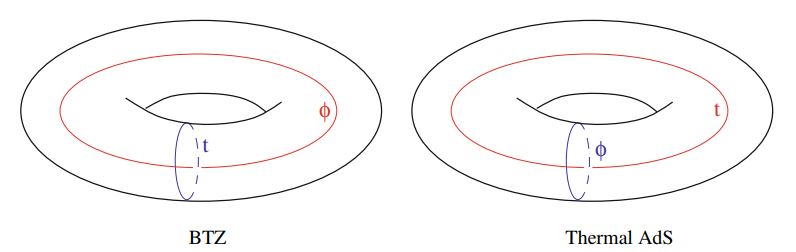}
\end{figure}
In EAdS, which of the two cycles of the torus is contractible in the bulk? 
Same question for EBTZ. 

Taking as a given that the on-shell Euclidean action for the AdS solution is given by (for a derivation see e.g.~Kraus' lectures on black holes and AdS$_3$/CFT$_2$ \cite{Krauslectures} (2.43))
\ali{
	S_E = \frac{i \pi}{12} (c \tau - \bar c \bar \tau) ,  
} 
write down the on-shell Euclidean action for the BTZ solution. 
Which of the two solutions dominates at high temperature $\beta \ra 0$? 

Calculate the Bekenstein-Hawking entropy of BTZ from the previous answer. 
We have followed the same steps of the derivation of the Cardy formula, but on the bulk side now. 

Above, we assumed the torus modulus  only has an imaginary component, its real component $\theta$ was set to zero. If we kept it, what would $\theta$ correspond to on the gravity side? 


\chapter{Entanglement in AdS/CFT} 

\section{Entropy} 


References: Holzhey \cite{Holzhey}, Tong statistical physics \cite{Tongstatphys} 

The \emph{von Neumann entropy} 
\ali{
	S = - \text{tr } (\rho \log \rho)  
}
gives a fundamental 
definition for entropy as a measure for randomness/uncertainty of a statistical ensemble described by $\rho$, where $\rho$ is the probability distribution (classically) or density matrix (QM). $S$ gives an observer-dependent measure for the indeterminacy or lack of resolution of your system.  

Let us check that this definition reduces to the notions of entropy we have used earlier in the course. 
In the microcanonical ensemble, every microstate is equally probable, i.e.~for an observer in a closed system $(E,V,N)$, $\rho = \frac{1}{\Omega(E)}$ with $\Omega$ the accessible phase space volume (classically) 
or the number of available microstates (QM). 
Applying the von Neumann definition gives the entropy $S = k_B \log \Omega(E)$, 
which depends on the energy.    
$S$ quantifies the lack of information (not knowing which microstate is occupied) about your system.

In the canonical ensemble $(T,V,N)$ for an open system observer, the Boltzmann probability distribution $\rho = \frac{e^{-\beta H}}{Z(\beta)}$ 
gives a von Neumann entropy $S(T) = (1 - \beta \p_\beta) \log Z(\beta)$, which depends on the temperature. In the thermodynamic limit $N \ra \infty$, it becomes equal to the microcanonical entropy (for the average energy).

\section{Geometric entropy or entanglement}

The von Neumann entropy 
can be applied in the context of 
$d$-dimensional CFT to define the concept of `geometric entropy' or  entanglement.

\begin{figure}
	\centering \includegraphics[width=4.5cm]{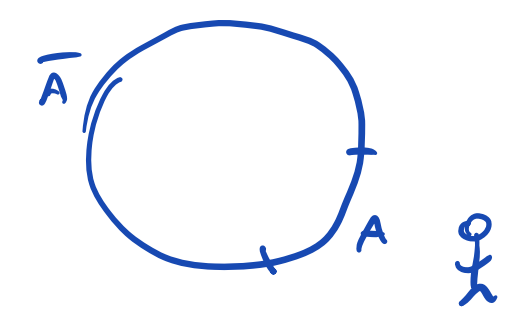} 
\caption{The picture shows the spacelike  (compact) direction $\theta$ of a $(1+1)$-dimensional CFT at some constant time $t$, bipartitioned into two intervals. 
} \label{fig-regionA}
\end{figure}

\paragraph{Entanglement} 
Consider a 
$d$-dimensional CFT in a pure state $|\psi\rangle$ described by the density matrix $\rho = |\psi\rangle \langle \psi|$. We consider the constant time slice pictured in figure \ref{fig-regionA} where we geometrically bipartition the system (assuming the Hilbert space can be factorized $\mathcal H = \mathcal H_A \bigotimes \mathcal H_{\bar A}$) 
into a spatial region $A$ and its complement $\bar A$. In $d>2$ the region $A$ is a ball, in $d=2$ it is an interval. 

An observer that only has access to region $A$ will measure a different density matrix, called the reduced density matrix 
\ali{
	\rho_A = \tr_{\bar A} \rho.  
	}
It is obtained from $\rho$ by tracing out degrees of freedom in $\bar A$. 
The observer's lack of information about the full system can be quantified by the von Neumann entropy of $\rho_A$: 
\ali{
	S_A = - \tr (\rho_A \log \rho_A ) \qquad \text{(entanglement entropy)} . 
}
This is by definition the \emph{geometric entropy} or \emph{entanglement entropy} associated with region $A$. 
%


To understand its interpretation, we consider some limits in which $S_A$ vanishes. 
The von Neumann entropy will vanish when $\rho$ is pure. You can see this for example by using that $\tr \rho^2 = \tr \rho$ 
for a pure state. Now the reduced density matrix $\rho_A$ is in general no longer pure, but mixed, $\rho_A = \sum_\alpha p_\alpha |\psi_\alpha \rangle \langle \psi_\alpha|$, with $\sum_\alpha p_\alpha = 1$ or $\tr \rho_A = 1$. It will be pure in the limit that the region $A$ becomes the full time slice ($\bar A$ vanishes), because then $\rho_A \ra \rho$, which was pure by assumption. This shows that $S_A$ is a measure for the \emph{amount of missing information} from the point of view of the observer in $A$, namely $S_A \ra 0$ when this observer has access to the full system. 
   
Another instance in which $\rho_A$ will be pure, is when the pure state $\rho$ of the CFT is separable, $|\psi \rangle = |\psi_A \rangle|\psi_{\bar A} \rangle$, meaning there is no quantum entanglement between degrees of freedom in $A$ and degrees of freedom in $\bar A$. In that case $\rho_A = |\psi_A\rangle \langle \psi_A|$ and $S_A = 0$. This is consistent with the interpretation of $S_A$ as a measure for the \emph{amount of entanglement} between dof in $A$ and dof in $\bar A$.

Properties: \\
$S_A = S_{\bar A}$ because $\rho$ is pure \\ 
$S_{A_1} + S_{A_2} \geq S_{A_1 \cup \, A_2} $ (subadditivity) \\ 
$S_{ABC}  + S_B \leq S_{AB} + S_{BC}$ (strong subadditivity or SSA)

\paragraph{Modular Hamiltonian} Because $\rho_A$ is a positive semi-definite matrix ($x' \rho_A x \geq 0$ for any vector $x \neq 0$), one can write it as 
\ali{
	\rho_A = \frac{e^{-H_{mod}}}{\tr e^{-{H_{mod}}}}  \label{Hmoddef}
}
for some Hermitian matrix $H_{mod}$ in the exponent. This is reminiscent of the thermal density matrix taking the form $\rho_{thermal} = e^{-\beta H}/\tr e^{-\beta H}$ in terms of the Hamiltonian $H$, and therefore in analogy $H_{mod}$ is called the \emph{modular Hamiltonian} or \emph{entanglement Hamiltonian} associated with the region $A$: 
\ali{
	H_{mod,A} = -\log \rho_A + c' 
	}
where $c'$ is a normalization constant. It is in general a very non-local operator, but we will soon discuss a case where a closed expression for $H_{mod,A}$ is known.

\paragraph{First law of entanglement}  

Under a change of the state $\rho_A' = \rho_A + \delta \rho_A$, the entanglement changes to $S'_A = S_A + \delta S_A$ with $\delta S_A = -\tr (\delta \rho_A \log \rho_A) + \mathcal O(\delta \rho_A^2)$ (using $\tr \delta \rho_A= 0$). 
On the other hand, the change in the expectation value of the modular Hamiltonian $\delta \vev{H_{mod,A}} \equiv \vev{H_{mod,A}}_{\rho_A'} -  \vev{H_{mod,A}}_{\rho_A}$ is also equal to $-\tr (\delta \rho_A \log \rho_A)$. It follows that the entanglement fluctuations and modular Hamiltonian fluctuations are related by 
\ali{
	\delta S_A = \delta \vev{H_{mod,A}} .  \label{firstlawent}
}
This is known as the \emph{`first law' of entanglement}, again in analogy with the thermodynamic first law relating changes in entropy with changes in energy, $dE = T dS$, in the thermal state.


\section{Ryu-Takayanagi}

\begin{figure}
		\centering \includegraphics[width=7cm]{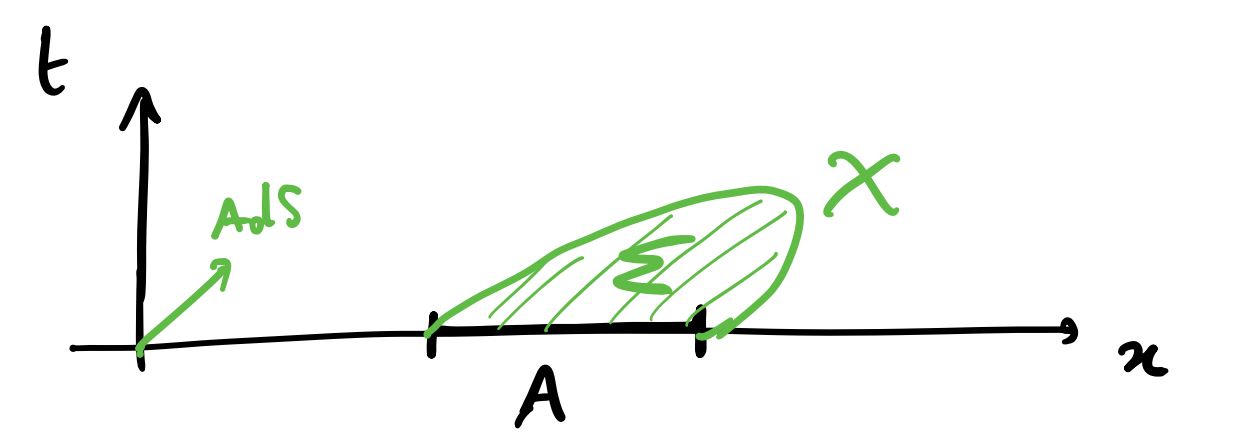} 
	\caption{The RT formula. 
	} \label{fig-RT}
\end{figure}

\paragraph{RT prescription} The interpretation of the CFT entanglement $S_A$ in the dual AdS theory is provided by the Ryu-Takayanagi formula or \emph{RT formula}  
\ali{
	S_A \quad \stackrel{AdS/CFT}{=} \quad \frac{\mathcal A(\chi)}{4G} 
}
where $\chi$ is the RT surface $\equiv$ co-dimension 2 surface of minimal area $\mathcal A$ that is homologous to the boundary region $A$. In AdS$_3$/CFT$_2$ that means the RT surface is a boundary-anchored geodesic (see figure \ref{fig-RT}) and $\mathcal A(\chi)$ is the length of that geodesic, see \eqref{RTfirstencounter}. The homology condition can be written as the condition $\p \Sigma = \chi \cup A$, 
with $\Sigma$ the area between the RT surface $\chi$ and the boundary region $A$ that is bounded by $\chi$ and $A$. (If there was a hole in the spacetime region $\Sigma$ for example, that hole would contribute to $\p \Sigma$ and the homology condition would not hold.) 

According to RT, a \emph{quantum} connection in the CFT has an interpretation as a \emph{geometric} connection in the AdS bulk theory. 

The RT formula resembles closely the Bekenstein-Hawking entropy formula \eqref{BHentropy}. However, the first involves the RT surface $\chi$ while the second involves a black hole horizon $h$. To understand the Bekenstein-Hawking entropy inspiration for the RT formula, we first consider a special choice of region $A$. 

\begin{figure}
	\centering \includegraphics[width=7cm]{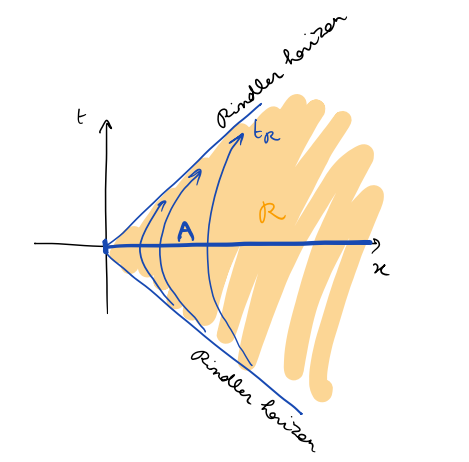} 
	\caption{The Rindler wedge $\mathcal R$.} \label{fig-Rindler} 
\end{figure}

\paragraph{Special case of $A$} 
Consider the special case of $A$, where $A =$ half-line $x \geq 0$ in CFT$_2$. It is the region whose domain of dependence $\mathcal D(A)$, defined as the set of points that are causally connected to  $A$, is the Rindler wedge $\mathcal R$. The Rindler wedge is the region of Minkowski space $x\geq 0, |t| < x$ spanned by the Rindler time coordinate $t_{\mathcal R}$ and Rindler space coordinate $R$: 
\ali{
	ds^2_{\mathcal R} &= -dt^2 + dx^2 , \qquad x\geq 0, |t| < x \\
		&= dR^2 - R^2 dt_{\mathcal R}^2 , \qquad R \geq 0, \text{any } t_{\mathcal R}  
}  
with 
\ali{
	x = R \cosh t_{\mathcal R}, \qquad t = R \sinh t_{\mathcal R} . 
}
The Rindler metric has a horizon at $R = 0$. This is the Rindler horizon, which is the edge of the Rindler wedge in the Minkowski spacetime. A Rindler observer, moving along a line of constant $R$ as it evolves in  Rindler time $t_{\mathcal R}$ (see figure \ref{fig-Rindler}), does not have access to the region of Minkowski space behind the Rindler horizon. 

To avoid a conical singularity at the Rindler horizon in Euclidean signature, Euclidean Rindler time should be periodically identified with $\beta = 2\pi$, such that a Rindler observer measure a temperature 
\ali{
	T = \frac{1}{2\pi}. 
}
The more precise statement is the following (reference:  \cite{HartmanQG} (5.8) for a very nice derivation from path integral pictures). If a Minkowski observer, moving along a line of constant $x$ as it evolves in time $t$, measures a vacuum state $\rho$, then an observer confined to $A$, i.e.~a Rindler observer moving along a line of constant $R$ as it evolves in time $t_{\mathcal R}$, measures a thermal state 
\ali{
	\rho_A= \frac{e^{-2\pi H_{t_{\mathcal R}}}}{\tr e^{-2\pi H_{t_{\mathcal R}}}}  . 
}
Here $ H_{t_{\mathcal R}}$ is the Rindler time Hamiltonian. From the definition of the modular Hamiltonian in \eqref{Hmoddef}, we read off that 
\ali{
	H_{mod,A} = 2 \pi \, H_{t_{\mathcal R}} . \label{HmodRindler}
}
That is, in the special case of $A$ given by the half-line, the modular Hamiltonian is equal to $2\pi$ times the 
generator of translations in $t_{\mathcal R}$ (i.e.~Minkowski boosts).  
The entanglement $S_A$ associated with the half-line becomes a thermal entropy.  


To the Rindler observer, its time $t_{\mathcal R}$ is just regular time, and 
its full spacelike region it the Minkowski half-line $A$. Since it measures a temperature $T$, its bulk dual (called the AdS-Rindler spacetime) will contain a black hole (Witten prescription) with a black hole horizon. 
The Bekenstein-Hawking entropy of this black hole is holographically dual to the thermal entropy of the CFT state $\rho_A$ 
\ali{
	S_A = \frac{\mathcal A(h)}{4G}   \label{SAdual}
}
with $\mathcal A(h)$ the area of the AdS-Rindler black hole horizon. It means that the RT surface $\chi$ of the half-line $A$ wraps the AdS-Rindler black hole horizon.   
The above reasoning explains the \emph{Bekenstein-Hawking form} of the RT formula, at least for the special case that $A$ is the half-line. What about more general $A$? 

There is a conformal transformation [Casini Huerta Myers '11] that maps any interval $A$ to the half-line $A'$ (in general dimensional language, any ball $A$ to the half-plane $A'$). Since the transformation is conformal, the reasoning of the previous paragraph still applies.

\section{Gravity from entanglement}

\begin{figure}
	\centering \includegraphics[width=14cm]{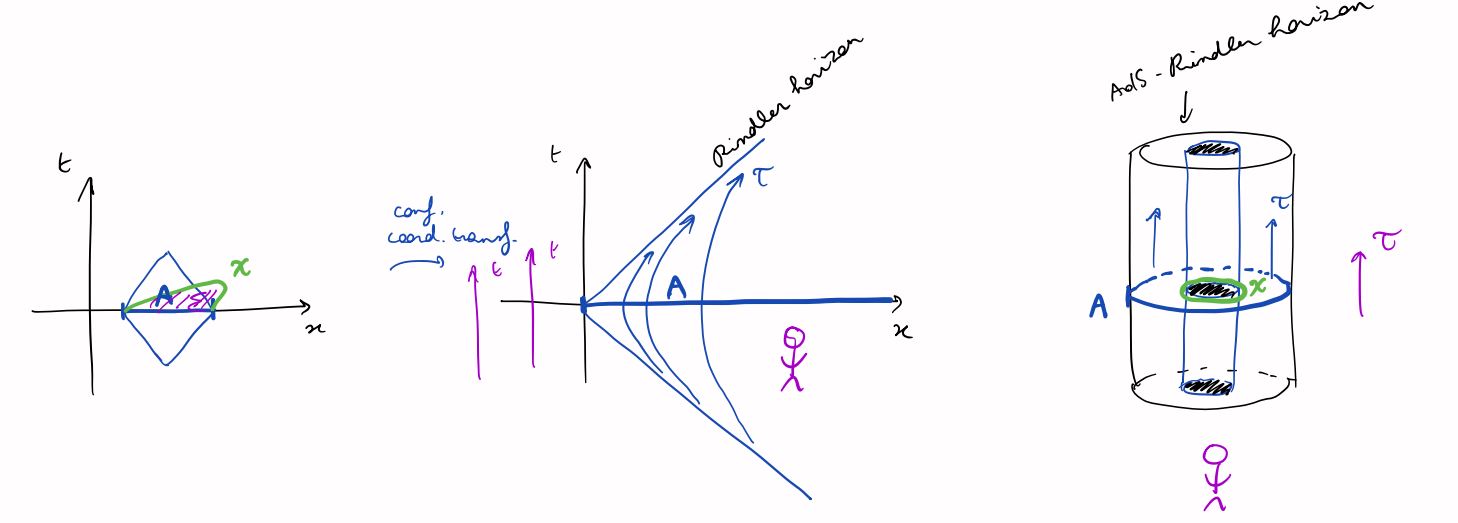} 
	\caption{The Bekenstein-Hawking inspiration for the RT formula.} \label{fig-BHinspiration} 
\end{figure}

Consider the first law of black hole thermodynamics \eqref{bhfirstlaw} applied to the AdS-Rindler black hole 
\ali{
	\delta M = \frac{1}{2\pi} \, \delta S_{BH}.     \label{AdSRindlerfirstlaw}
}
It expresses how a small change of mass of a black hole is related to its change in entropy.  Said otherwise, it expresses how two black hole \emph{solutions} with slightly different mass are related to each other, namely the black hole first law is an \emph{on-shell} expression. 
%
Relating two solutions of the Einstein equations (one with $M$ and one with $M + \delta M$), it is natural that 
the first law is equivalent to 
the linearized Einstein equations. 
(This is made precise in the Iyer-Wald formalism.) 

First consider the left hand side. The black hole mass is calculated from the Brown-York stress tensor, \eqref{QBY}. Schematically, $\delta M = \int \delta T_{t_{\mathcal R} t_{\mathcal R}}^{BY}$. By AdS/CFT, the Brown-York stress tensor is identified with the expectation value of the CFT stress tensor \eqref{BYdual}, so that $\delta M = \int \delta \vev{T_{t_{\mathcal R} t_{\mathcal R}}} = \delta \vev{ H_{t_{\mathcal R}}}$. Making use of \eqref{HmodRindler}, we arrive at $\delta M = \frac{1}{2\pi} \delta \vev{ H_{mod,A}}$ for $A$ the half-line. 

Now for the right hand side. The formula for the Bekenstein-Hawking entropy tells us that $\delta S_{BH} = \delta \mathcal A(h)/4G$. By AdS/CFT, this thermal entropy equals the entanglement entropy $S_A$ for $A$ the half-line \eqref{SAdual}. So we have $\delta S_{BH} = \delta S_A$. 

Having used the AdS/CFT dictionary on each side of the black hole first law \eqref{AdSRindlerfirstlaw}, it now reads 
\ali{
	\delta \vev{ H_{mod,A}} = \delta S_A. \label{rewriting}
}   
In first instance for $A$ the half-line, but since the half-line is mapped to any interval by a conformal transformation, to which the AdS/CFT physics is insensitive, it applies to any interval or ball $A$. 

We recognize this dual CFT rewriting \eqref{rewriting} of the black hole first law \eqref{AdSRindlerfirstlaw} as the first law of entanglement \eqref{firstlawent}. The conclusion is that a gravitational first law in the bulk corresponds to an entanglement first law in the CFT. It follows that, from the CFT perspective, the entanglement first law imposes linearized Einstein equations in the dual bulk theory. This is a famous result by van Raamsdonk, and is often summarized in statements that \emph{gravity emerges from entanglement} in AdS/CFT!

\newpage 
\section*{Exercises} 

\textbf{\underline{\smash{Exercise 1. Ryu-Takayanagi formula}}}

Work out the length of a geodesic in AdS$_3$ that is homologous to a boundary interval $A$ of length $L_A$, and check the Ryu-Takayanagi formula. You can use that boundary-anchored geodesics in AdS$_3$ 
\ali{
	ds^2 = \frac{\ell^2}{Z^2} (dZ^2 - dt^2 + dx^2) 
}
are given by half-circles of diameter $L$  
\ali{
	x^2 + Z^2 = \left(\frac{L}{2}\right)^2, \qquad (Z>0) .   
}
The UV cutoff $\epsilon$ that appears in the entanglement of the CFT  
\ali{
	S_A = \frac{c}{3} \log \frac{L_A}{\epsilon} 
}
has acquired a geometric interpretation in the bulk. Briefly discuss this geometric interpretation in the context of the UV/IR correspondence in AdS/CFT. 
\\*
\\*
\\*
\textbf{\underline{\smash{Exercise 2. Geometric interpretation of SSA}}} 

Consider 3 adjacent intervals $A, B$ and $C$ in the $x$-direction of a 2D CFT. In a $(Z,x)$ diagram with $Z$ the AdS direction, draw the geometric interpretation of the entanglement measures $S_{AB}$, $S_{BC}$, $S_{ABC}$ and $S_B$. Now argue why the strong subadditivity (SSA) property should hold.

\end{document}